\documentclass[12pt]{article}
\pdfoutput=1
\usepackage{amsmath}
\usepackage{amsfonts}
\usepackage{float}
\usepackage{url}
\usepackage{hyperref}
\usepackage{mathtools}
\usepackage{bbold}
\usepackage{slashed}
\usepackage{simplewick}
\usepackage{tikz}
\usepackage{graphicx}
\usepackage{epstopdf}
\usepackage{subfigure}
\usepackage[labelfont=bf,font={small}]{caption}
\usepackage{pgfplots}
\usepackage{amssymb}
\usepackage{color}
\usepackage{mathrsfs}
\usepackage{fancybox}
\usepackage{lipsum}
\usepackage{eurosym}
\usepackage{tcolorbox}
\usepackage{physics}
\usepackage{tensor}
\usepackage{empheq}

\newcommand{\sltr}{\text{SL}(2,\text{R})}
\newcommand{\dpi}{\mathcal{D}}

\newcommand{\mo}{\mathcal{O}}

\newcommand{\cas}{\mathcal{C}}

  \newcommand{\sj}[6]{ \begin{Bmatrix}
   #1 & #2 & #3 \\
   #4 & #5 & #6 
  \end{Bmatrix}}

\newcommand{\sinc}{\text{sinc}}
\newcommand{\average}[1]{\left\langle #1 \right\rangle}

\numberwithin{equation}{section}
\begin{document}
\begin{titlepage}

\setcounter{page}{1} \baselineskip=15.5pt \thispagestyle{empty}

%\begin{flushright}
%hep-th/13mmnnn\\
%\end{flushright}
\vfil

${}$
%\vspace{1cm}

\begin{center}
\def\thefootnote{\fnsymbol{footnote}}
\begin{changemargin}{0.05cm}{0.05cm} 
\begin{center}
{\Large \bf Dissecting the ensemble in JT gravity
}
\end{center} 
\end{changemargin}
\vskip 1cm
\textbf{Andreas Blommaert}
\\[0.5cm]
{\normalsize { \sl Department of Physics and Astronomy, Ghent University,\\
Krijgslaan 281-S9, 9000 Ghent, Belgium}}

\vspace{0.2cm}

\medskip

\end{center}

\vskip1cm

\begin{center} 
\textbf{Abstract}
\end{center} 
We calculate bulk and boundary correlators in JT gravity by summing over geometries. The answers are reproduced by computing suitable ensemble averages of correlators of chaotic quantum systems. We then consider bulk correlators at large spatial separations and find that semiclassical decay eventually makes way for erratic oscillations around some nonzero answer. There is no cluster decomposition because of wormholes connecting distant regions. 

We construct more microscopic versions of JT gravity which are dual to a single quantum system by including a set of branes in the gravitational theory the data of which describes the Hamiltonian of the dual system. We focus on the bulk description of eigenstates which involves end of the world branes and we explain how observables factorize due to geometries connecting to these branes.

%\vspace{0.3cm}
%\hrule
%\vfil
%\begin{flushleft}
%\today
%March 20, 2013
%\end{flushleft}

\end{titlepage}

\newpage
\tableofcontents
\vspace{0.5cm}
\noindent\makebox[\linewidth]{\rule{\textwidth}{0.4pt}}
\vspace{1cm}
%\newpage

\addtolength{\abovedisplayskip}{.5mm}
\addtolength{\belowdisplayskip}{.5mm}

\def\plus{\raisebox{.5pt}{\tiny$+$\smpc}}

\addtolength{\parskip}{.6mm}
\def\spc{\hspace{1pt}}

\def\nspc{{\hspace{-2pt}}}
\def\ff{\rm\smpc f\smpc} 
\def\fff{\mbox{Y}}
\def\ww{{\rm w}}
\def\smpc{{\hspace{.5pt}}}

\def\zz{{\spc \rm z}}
\def\xx{{\rm x\smpc}}
\def\xxi{\mbox{\footnotesize \spc $\xi$}}
\def\jj{{\rm j}}
\addtolength{\baselineskip}{-.1mm}

\renewcommand{\Large}{\large}

\setcounter{tocdepth}{2}
\addtolength{\baselineskip}{0mm}
\addtolength{\parskip}{-0.6mm}
\addtolength{\abovedisplayskip}{1mm}
\addtolength{\belowdisplayskip}{1mm}

\setcounter{footnote}{0}

\section{Introduction}\label{sect:1}
One of the more important lessons of holography is that realistic models of quantum gravity in AdS have a discrete spectrum with random matrix level statistics \cite{bhrm}. The spectrum is discrete because AdS quantum gravity has a dual description as a field theory on the compact boundary of AdS.

The spectrum has random matrix level statistics because black holes are chaotic quantum systems \cite{bhrm,bhrm2}. Semiclassical chaos is diagnosed by exponential sensitivity to changes in initial conditions, the butterfly effect. In a quantum theory this can be probed by computing out of time ordered correlators \cite{ss,ssmultiple,ssstringy,stanfordroberts}. For chaotic systems these grow exponentially fast at early times. There is a similarly universal appearance of exponential growth in scattering experiments near black hole horizons. The exponential growth corresponds to exponential redshifts between static and infalling coordinates close to the horizon \cite{ss,ssmultiple,ssstringy,stanfordroberts}. There is a second universal characteristic of chaos in quantum systems \cite{haake}. Energy levels of chaotic quantum systems repel. More in particular the statistics of neighbouring energy levels of chaotic quantum systems is described by random matrix theory \cite{mehta}.

We see that discreteness and random matrix level statistics are universal features of quantum black holes and hence of quantum gravity in AdS. However it is not obvious how to explain these universal features from the bulk gravitational point of view. One way to phrase this in a particularly sharp way is to consider a version of the information problem due to Maldacena \cite{maldainfo} stating that boundary correlators cannot decay to zero at late times. Take the two point function of a chaotic system with an $L$ dimensional Hilbert space
\begin{equation}
    \Tr(e^{-\beta_1 H}\mo e^{-\beta_2 H}\mo) =e^{-2S_0}\int_0^\infty d E_1\,e^{-\beta_1 E_1}\int_0^\infty d E_2\,e^{-\beta_2 E_2}\,\rho(E_1,E_2)\,\mo\mo(E_1,E_2)\nonumber\,.
\end{equation}
Here we take
\begin{equation}
    \rho(E_1,E_2)=\sum_{i=1}^L \delta(E_1-\lambda_i)\sum_{j=1}^L \delta(E_2-\lambda_j).\label{spikes}
\end{equation}
We set the typical level spacing proportional to $e^{-S_0}$. At early times $t\ll e^{S_0}$ the two point function is self averaging which implies it is well approximated by averaging the matrix elements over a set of $e^{S_0}$ neighboring energy levels \cite{bhrm}. The result is that we can effectively take $\mo\mo(E_1,E_2)$ to be a smooth function of the energies. This is a version of ETH \cite{eth1,eth2}. Furthermore the self averaging property of the two point function implies that at early times it is well approximated by suitably averaging $\rho(E_1,E_2)$ to obtain a smooth function. 

At such early times the two point function decays exponentially. At exponentially late times we see the effects of the individual levels and neither $\rho(E_1,E_2)$ nor $\mo\mo(E_1,E_2)$ can be approximated by smooth functions. The result is an erratically oscillating curve which at asymptotic times oscillates around a constant so called plateau value \cite{maldainfo,bhrm}. It is nontrivial to explain this late time behavior from the bulk gravitational path integral.
\\~\\
As a first and important step one might hope to explain the averaged features of the two point function at late times. From the boundary analysis it turns out the averaged late time behavior of correlators of a chaotic quantum system is accurately captured by averaging the correlator over an ensemble of random matrices \cite{bhrm} 
\begin{equation}
    \average{\Tr(e^{-\beta_1 H}\mo e^{-\beta_2 H }\mo)}=\frac{1}{\mathcal{Z}}\int d H\,P(H)\Tr(e^{-\beta_1 H}\mo e^{-\beta_2 H }\mo)\,.\label{1.2}
\end{equation}
In particular the ensemble averages the matrix elements over a set of $e^{S_0}$ neighboring energy levels and it treats $\lambda_1\dots \lambda_L$ as the eigenvalues of an $L$ dimensional random matrix. This is a more refined way to coarse grain $\rho(E_1,E_2)$ which introduces correlations when $E_1$ and $E_2$ are close together. These correlations capture level repulsion in the original chaotic system.

This strips of the erratic oscillations \cite{bhrm}. There is much recent progress on reproducing this averaged curve from the gravitational path integral within JT gravity \cite{bhrm,sss,sss2,paper5,phil}. This is a theory of AdS quantum gravity in two dimensions. For early work on the model see \cite{J,T,alpol,jensen,malstanyang,ads2}. Within this model the averaged late time behavior of the boundary two point function \eqref{1.2} is captured by including contributions due to Euclidean wormholes or baby universes in the gravitational path integral. Upon accounting for those contributions one in fact finds that JT gravity is manifestly identical to a particular random matrix ensemble \cite{sss2,phil}. Related recent work includes \cite{paper6,clifford,joatho,maxjoa,olga}.

This identification of JT gravity as an ensemble raises several fundamental questions. It suggests that perhaps the gravitational path in general wants to compute ensemble averages of observables. This would imply that pure gravity is at best to be interpreted as an effective description of the real world. Pure gravity would not be a microscopic theory but rather a mesoscopic one like hydrodynamics. To address this question we can return to the boundary two point function. If there are microscopic theories of pure gravity which are not dual to an ensemble then these must be able to capture the details of the late time erratic oscillations. We would like to understand if pure gravity is rich enough to capture this microstructure. If so then what comes of the Euclidean wormholes which seem to be consequences of statistical correlations in ensemble averages \cite{sss,sss2,rw1,malmaoz,soo,jdb}?

There is recent evidence that there do exist more microscopic theories of pure gravity. Part of the erratic oscillations at late times have been explained within pure JT gravity \cite{paper6}. The point is that more microscopic definitions of gravity are obtained by including certain spacetime branes in the theory which can emit and absorb baby universes. This suggests that perhaps branes and wormholes are enough to capture microstructure in pure gravity.
\\~\\
Despite recent efforts many mysteries still surround the role of the ensemble in gravity. In this work we focus on JT gravity. In this model in particular we lack a bulk description of eigenstates of the Hamiltonian matrix of the dual system. This is important to capture the late time microstructure of correlation functions. There also lacks a satisfactory bulk description of the averaged properties of higher point functions at late times. In this work we address the following topics.
\begin{enumerate}
    \item The identification of JT gravity as a matrix integral in \cite{sss2} indicates there is a natural averaging over eigenvalues in gravity. To capture the averaged properties of correlators we further require a suitable average over eigenstate components. It is not obvious if and how JT gravity captures this type of averaging. Progress on this was made in \cite{phil} where it is explained for the boundary two point function on a disk with a handle that summing over different particle trajectories on this topology reproduces predictions from ETH. In section \ref{sect:2} we refine ETH to accurately capture Lyapunov growth \cite{ethotoc}. We reproduce the resulting averaged description of boundary correlators by summing over all possible particle trajectories on all possible bulk geometries. We explain how to include contributions due to wormholes to bulk correlators in JT gravity by applying bulk operator reconstruction \cite{paper5,paper201,wopjordan}.
    \item Much like boundary correlators at late times we point out that bulk correlators cannot decay to zero at large spatial separations in finite entropy quantum gravity. In other words there can be no strict notion of cluster decomposition. This is due to wormholes connecting distant regions in spacetime \cite{phil}. At large distances the correlators oscillate erratically around an analogue to the plateau. As we discuss in section \ref{sect:4} the averaged analogue to the plateau is captured by the ensemble averaged effective description of JT gravity.
    \item In section \ref{sect:5} we construct more microscopic versions of JT gravity which describe one quantum system instead of an ensemble. The construction boils down to introducing a specific set of branes in the gravitational path integral. In particular we focus on the role of random eigenstate components. Combined with eigenbranes this construction defines alpha states \cite{paper6,maxfieldmarolf}. Alpha states are particularly useful to understand why the gravitational path integral without branes computes ensemble averages. We highlight from the bulk geometrical point of view how observables factorize due to geometries connecting to this specific set of branes. When this work was nearing completion we learned of an independent investigation of factorization in JT gravity \cite{wopphilshunyusteve}.
\end{enumerate}
The sections can be read more or less independently. Throughout sections \ref{sect:2} and \ref{sect:4} we ignore branes and consider the ensemble averaged description of JT gravity. Let us introduce each section in a bit more detail.
%%%%%%%%%%%%%%%%%%%%%%%%%%%%%%
\subsection{Correlation functions}
In section \ref{sect:2} we consider boundary matter correlators in JT gravity such as the time ordered $4$ point function
\begin{equation}
    \average{\Tr(e^{-\beta_1 H}\mo_1e^{-\beta_2 H}\mo_1e^{-\beta_3 H}\mo_2 e^{-\beta_4 H}\mo_2)}\,.
\end{equation}
We will often use this correlator as an example as it is both tractable enough to facilitate an elegant presentation as well as rich enough to capture all relevant features of the problem. JT gravity is a theory of hyperbolic Riemann surfaces ending on the thermal boundary circle \cite{malstanyang,sss2}. The inverse temperature sets the regularized length of the circular boundary of the hyperbolic Riemann surfaces \cite{malstanyang}. The JT gravity action comes with a topological term $S_0\chi$ which penalties higher genus contributions by their Euler character. The result is a genus expansion of for example the partition function
\begin{align}
    \nonumber\average{Z(\beta)}&=\sum_{g=0}^\infty e^{-2g S_0}\,Z_g(\beta)=\quad \raisebox{-5mm}{\includegraphics[width=23mm]{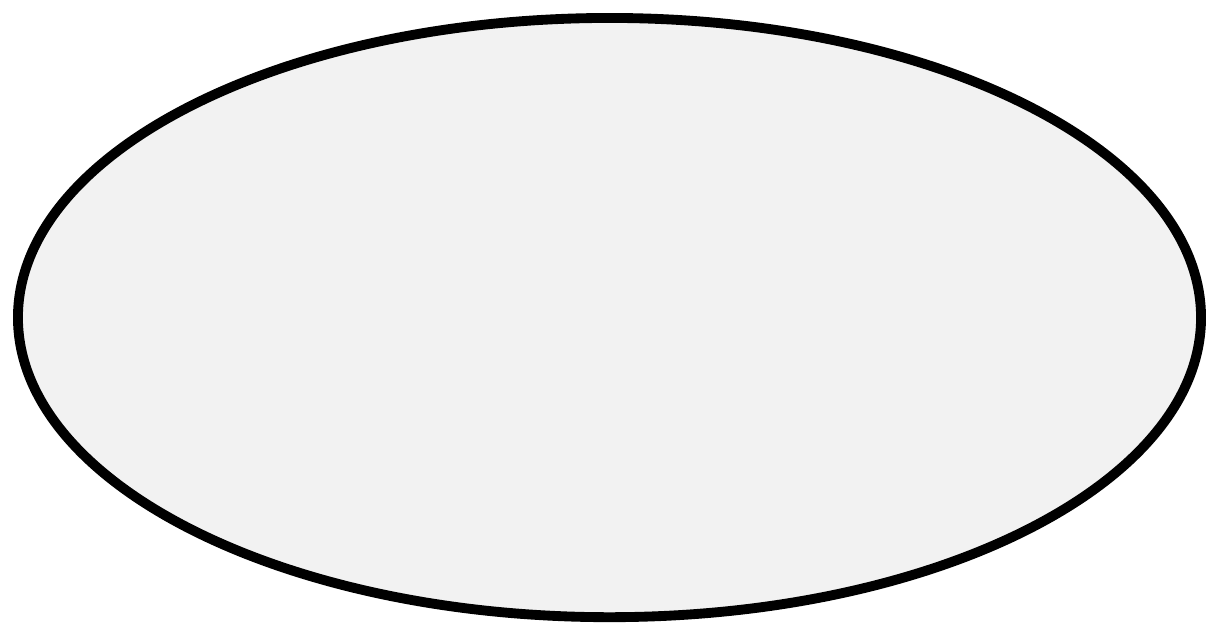}}\quad +\quad\raisebox{-5mm}{\includegraphics[width=23mm]{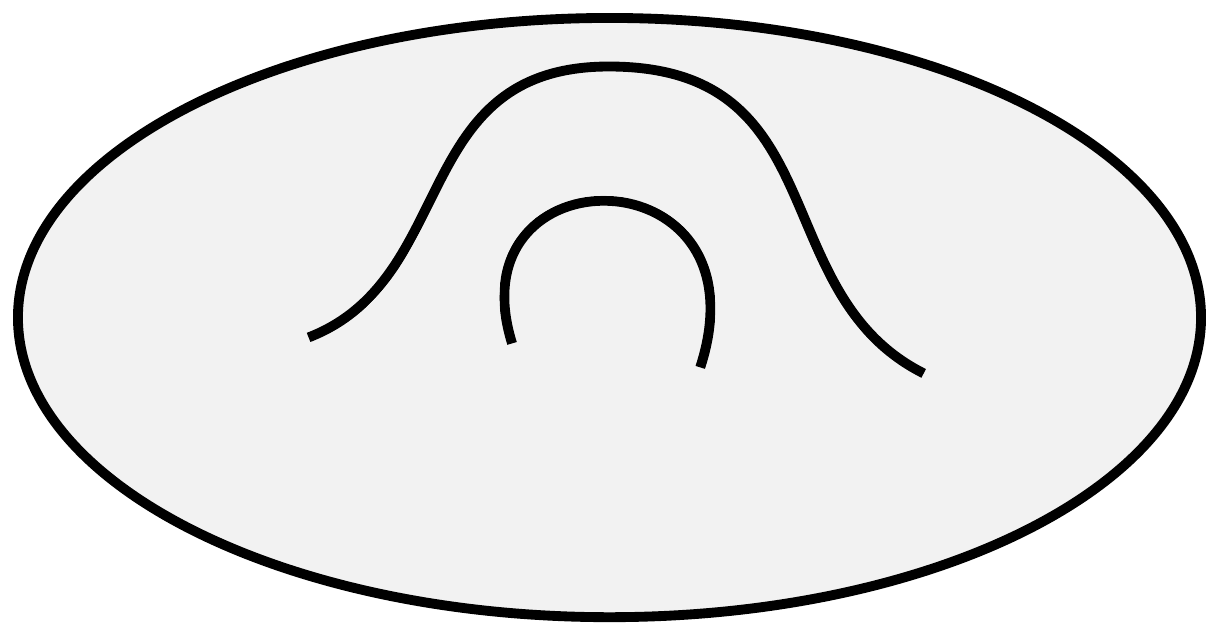}}\quad+\dots\\&=\int_\cas d E\,e^{-\beta E}\,\rho(E)\,.\label{1.4}
\end{align}
For more details including the specification of the energy contour see \cite{sss2}. Multi boundary correlators are computed by summing over all hyperbolic geometries that end on the union of all boundaries. These may be connected or disconnected. For example
\begin{align}
       \nonumber &\average{Z(\beta_1)Z(\beta_2)}_\text{conn}=\int_\cas d E_1\, e^{-\beta_1 E_1}\int_\cas d E_2\,e^{-\beta_2 E_2}\,\rho(E_1,E_2)_\text{conn}.\\
       &\qquad= \quad \raisebox{-5mm}{\includegraphics[width=50mm]{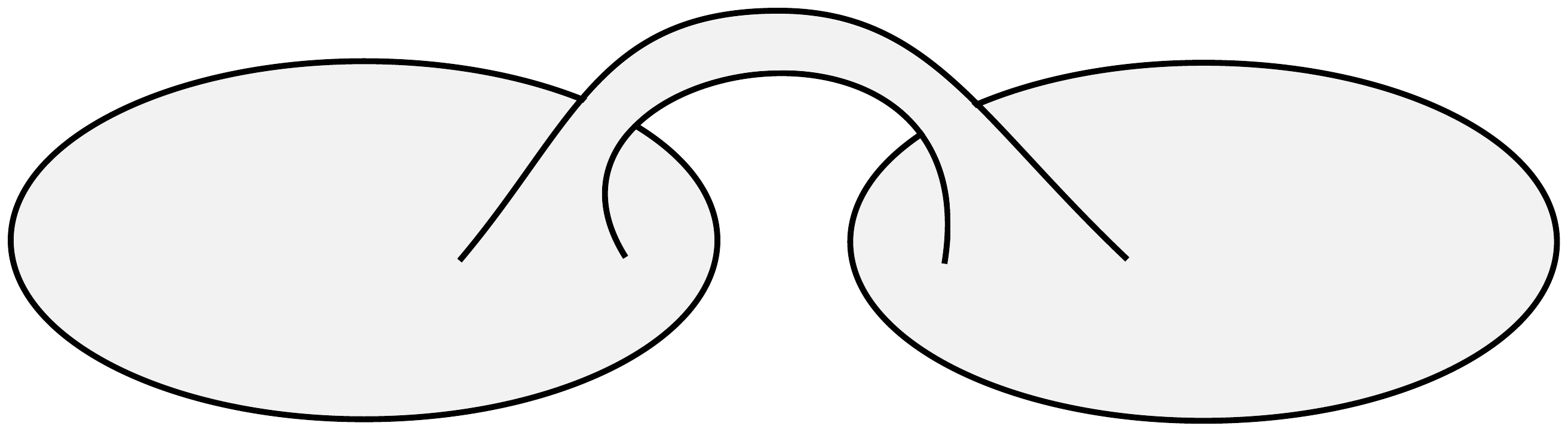}}\quad+\quad\raisebox{-5mm}{\includegraphics[width=50mm]{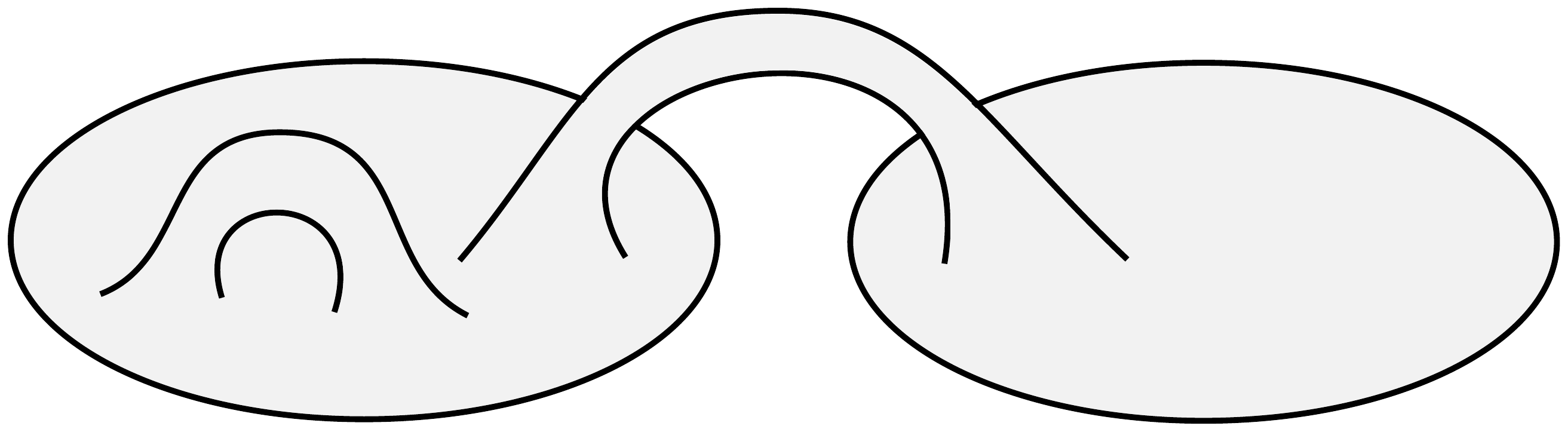}}\quad +\dots\label{2.35}
\end{align}

The gravitational genus expansion of the inverse Laplace transformed multi boundary correlation functions $\rho(E_1\dots E_n)$ matches to the genus expansion of the multi level spectral density $\rho(E_1\dots E_n)$ of a particular double scaled matrix integral \cite{sss2}. The double scaling boils down to zooming in on low energies where $e^{S_0}$ controls the local density of eigenvalues. The perturbative expansion of $\rho(E_1\dots E_n)$ is asymptotic so JT gravity is perturbatively ill defined. The full matrix integral answer is well behaved and known universally. This urges us to define JT gravity nonperturbatively as this double scaled matrix integral \cite{sss2}.
\\~\\
Free boundary matter correlators in JT gravity are computed as boundary to boundary propagators of free  massive quantum mechanical probe particles. JT gravity has a first order formulation as an $\sltr$ BF theory \cite{paper3,paper4,sss2,lucahermanpufu}. In this formulation the boundary to boundary propagator of a quantum mechanical probe particle reduces to a gravitational Wilson line anchored with both endpoints on the boundary. Intuitively these are related because the basis of the Hilbert space of $\sltr$ BF on an interval diagonalizes the geodesic length $d$ along the trajectory of the particle. In this basis the Wilson line matrix element is \cite{paper3,zhenbin,phil}
\begin{equation}
    \bra{d}\mo\mo\ket{d}=\exp(-\ell d).
\end{equation}
This is related to the on shell action of a heavy probe particle with $m^2=\ell(\ell-1)$ and with $\ell$ the weight of the boundary operator. We will use the first order formulation throughout as computing Wilson line amplitudes in $\sltr$ BF theory is quite straightforward \cite{witten2dgt,paper3}.

At early times the higher genus geometries do not contribute significantly to correlation functions. Correlators are then effectively calculated by computing the path integral of JT gravity on a disk with several Wilson lines inserted. For example for the two point function
\begin{equation}
    \average{\Tr(e^{-\beta_1 H}\mo e^{-\beta_2 H}\mo)}\supset\quad\raisebox{-10mm}{\includegraphics[width=40mm]{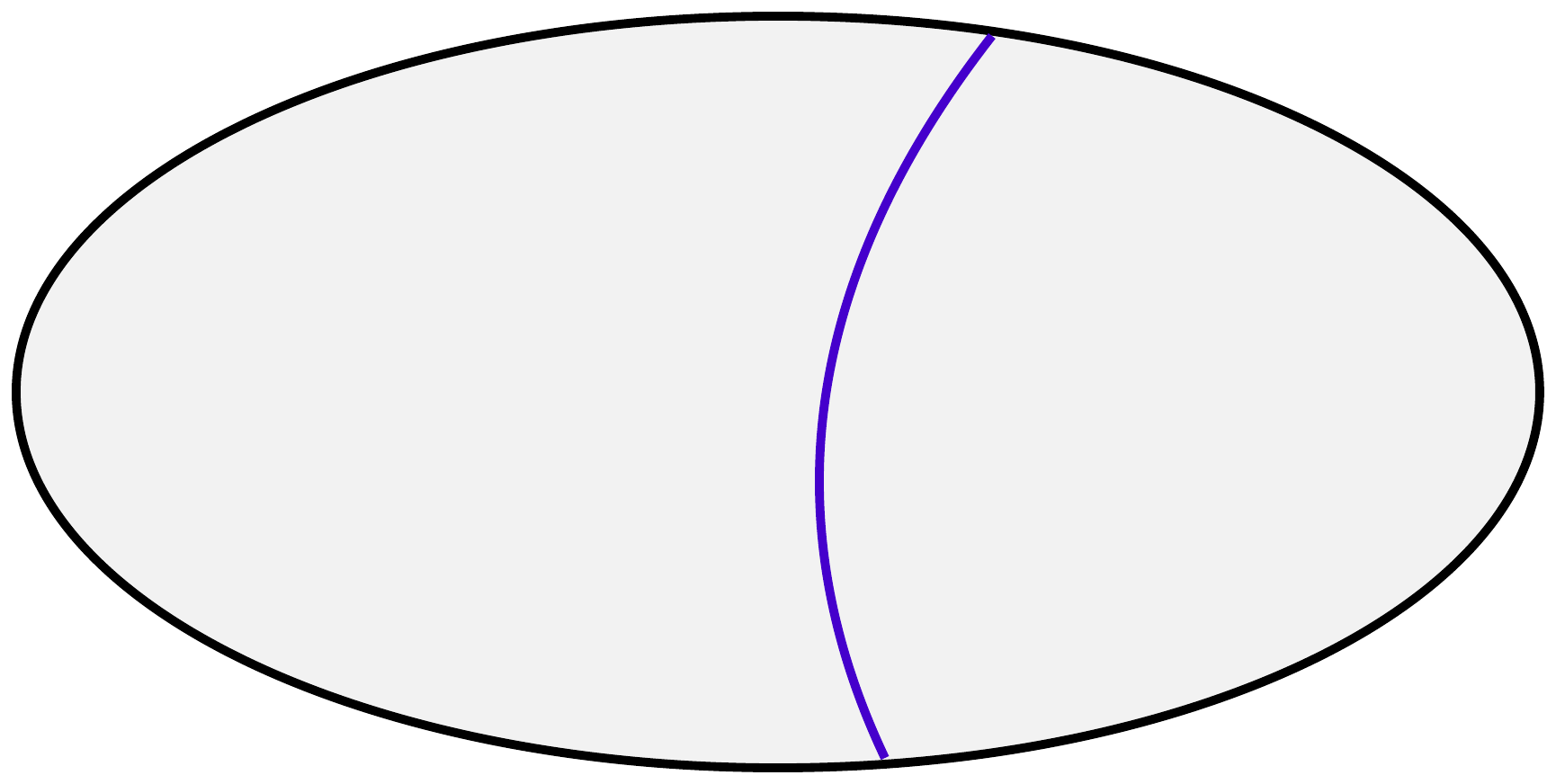}}\quad.\label{1.7}
\end{equation}
The Wilson line is represented by a single representative path of the particle through the gravitational bulk. Various labels will be left implicit in pictures. Another example is the time ordered or $s$ channel $4$ point function
\begin{equation}
    \average{\Tr(e^{-\beta_1 H}\mo_1e^{-\beta_2 H}\mo_1e^{-\beta_3 H}\mo_2 e^{-\beta_4 H}\mo_2)}\supset\quad\raisebox{-10mm}{\includegraphics[width=50mm]{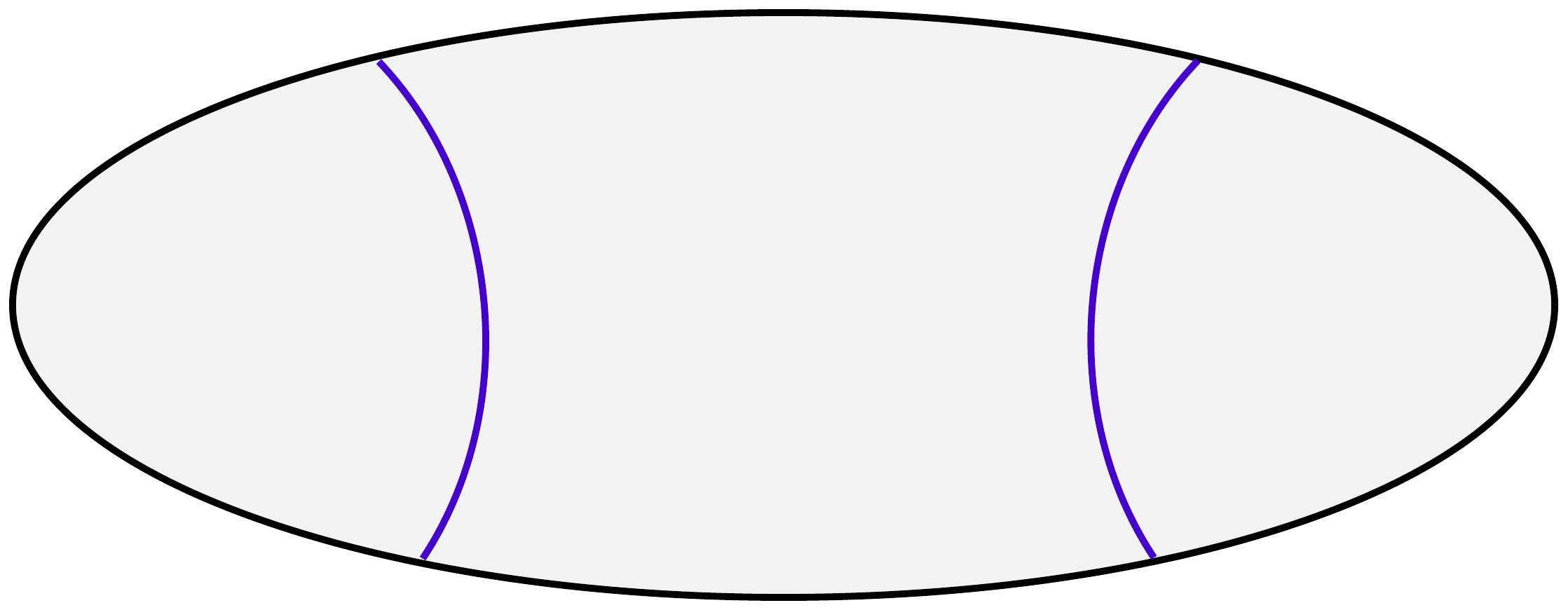}}\quad.\label{49}
\end{equation}
For the $4$ point function with all operators identical there is a sum over $s, t$ and $u$ channel diagrams. Each of these disk amplitudes can be computed in an elementary manner \cite{paper3,paper4,paper5}. See also \cite{altland,altland2,schwarzian,origins,kitaevsuh,zhenbin,lucahermanpufu}. For example the time ordered $4$ point function is
\begin{align}
\displaybreak[0]
     \nonumber&\average{\Tr(e^{-\beta_1 H}\mo_1e^{-\beta_2 H}\mo_1e^{-\beta_3 H}\mo_2 e^{-\beta_4 H}\mo_2)}\\\nonumber&\qquad\supset e^{-4S_0}\int_0^\infty d E_1\,e^{-(\beta_1+\beta_3) E_1}\int_0^\infty d E_2\,e^{-\beta_2 E_2}\int_0^\infty d E_4\,e^{-\beta_4 E_4}\\&\qquad\qquad\qquad\qquad\qquad\qquad\qquad\rho_0(E_1)\rho_0(E_2)\rho_0(E_4)\,\mo_1\mo_1(E_1,E_2)\mo_2\mo_2(E_1,E_4)\,.\label{2.39}
\end{align}
Here we introduce the genus zero spectral density \cite{remarks,malstanyang,bhrm,stanfordwitten} and the matrix element
\begin{equation}
    \rho_0(E)=\frac{e^{S_0}}{2\pi^2}\sinh 2\pi \sqrt{E}\quad,\quad \mo\mo(E_1,E_2)=e^{S_0}\,\frac{\Gamma(\ell\pm i \sqrt{E_1}\pm i \sqrt{E_2})}{\Gamma(2\ell)}\,.
\end{equation}

At later times or lower temperatures the higher genus geometries in the gravitational bulk can no longer be neglected. On each gravitational topology the boundary to boundary propagators decompose into an infinite sum of topologically inequivalent contributions. For example for the two point function on a disk with a handle the probe particle can wind any number of times around the handle \cite{phil}
\begin{equation}
    \average{\Tr(e^{-\beta_1 H}\mo e^{-\beta_2 H}\mo)}
    \supset \quad \raisebox{-9mm}{\includegraphics[width=36mm]{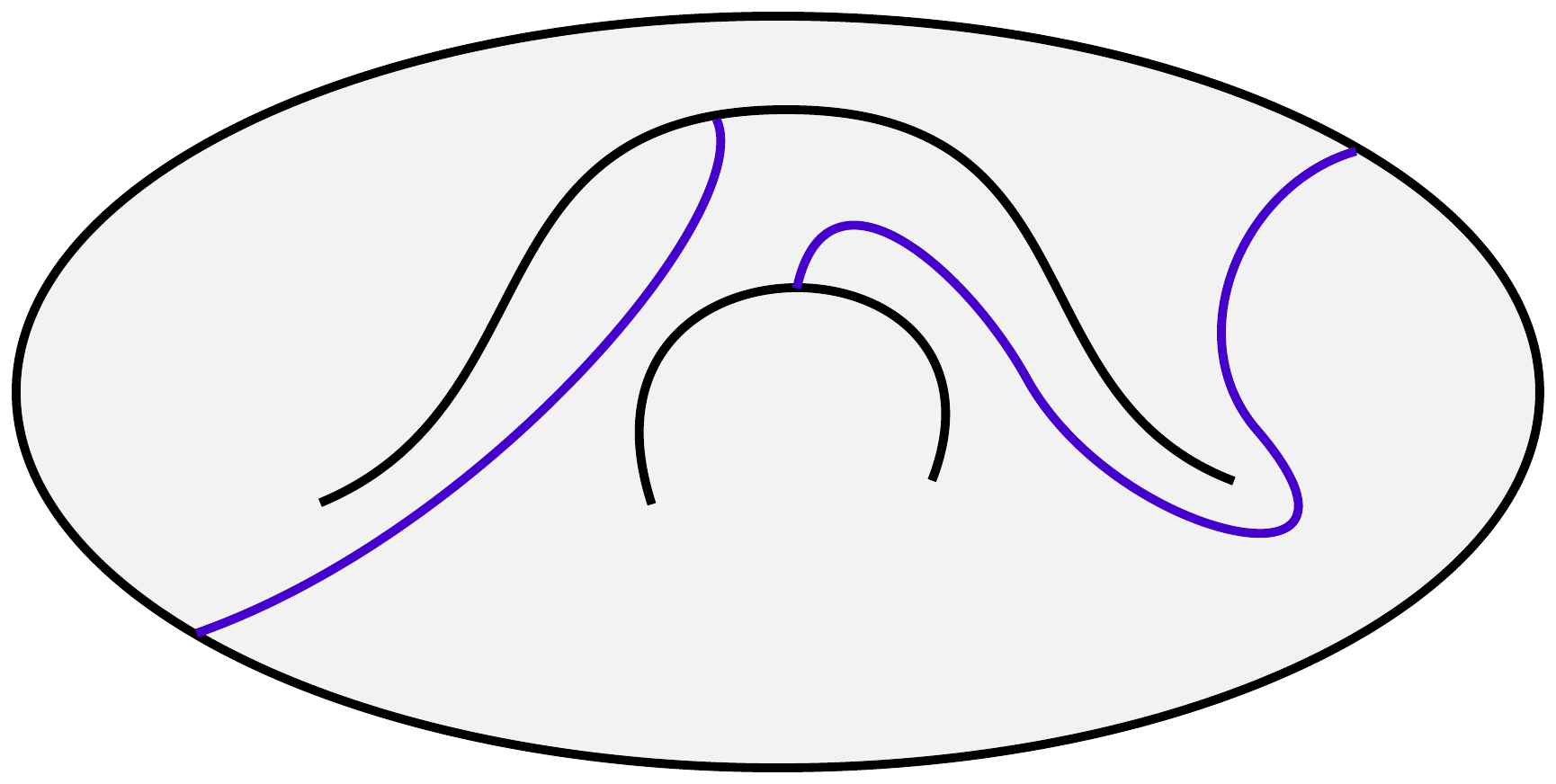}}\quad.\label{138}
\end{equation}
For any fixed number of windings the path integral of the massive quantum mechanical probe particle reduces to a boundary anchored gravitational Wilson line of the same topology. We see that on each surface of fixed topology $\Sigma$ that boundary correlators in JT gravity decompose into a sum of amplitudes. Different amplitudes are distinguished by inequivalent or non homologous trajectories $\gamma$ of the probe particles. For example for the time ordered $4$ point function
\begin{align}
    &\average{\Tr(e^{-\beta_1 H}\mo_1e^{-\beta_2 H}\mo_1e^{-\beta_3 H}\mo_2 e^{-\beta_4 H}\mo_2)}\nonumber\\ &\qquad\qquad\qquad\qquad=\sum_\Sigma \sum_{\gamma_1\,\gamma_2}\average{\Tr(e^{-\beta_1 H}\mo_1e^{-\beta_2 H}\mo_1e^{-\beta_3 H}\mo_2 e^{-\beta_4 H}\mo_2)}_{\Sigma,\gamma_1,\gamma_2}\,.\label{53}
\end{align}
In section \ref{sect:2} we bring order to this zoo of contributions specified by topologically inequivalent trajectories $\gamma$ and calculate the sum of all contributions. We find that the leading answer at any time scale is obtained by replacing in \eqref{2.39} the product of genus zero spectral densities by a random matrix correlator
\begin{align}
     \nonumber&\average{\Tr(e^{-\beta_1 H}\mo_1e^{-\beta_2 H}\mo_1e^{-\beta_3 H}\mo_2 e^{-\beta_4 H}\mo_2)}\\\nonumber&\qquad\supset e^{-4S_0}\int_\cas d E_1\,e^{-(\beta_1+\beta_3) E_1}\int_\cas d E_2\,e^{-\beta_2 E_2}\int_\cas d E_4\,e^{-\beta_4 E_4}\\&\qquad\qquad\qquad\qquad\qquad\qquad\qquad\rho(E_1,E_2,E_4)\,\mo_1\mo_1(E_1,E_2)\mo_2\mo_2(E_1,E_4)\,.
\end{align}
This corresponds to summing over all Riemann surfaces which end on the genus zero diagram \eqref{49} as claimed in \cite{paper5}. We match this leading contribution as well as various subleading contributions to predictions from random matrix theory.
%%%%%%%%%%%%%%%%%%%%%%%%%%%%%
\subsection{Cluster decomposition}
In section \ref{sect:4} we discuss cluster decomposition in JT gravity. The cluster decomposition principle asserts that connected correlators of quantum matter on some fixed background eventually decay to zero when the separation between any two operators is taken to be large. In finite entropy quantum gravity this decay cannot go on forever. This is a bulk translation of the version of the information problem due to Maldacena \cite{maldainfo}. Similar intuition was used to question the existence of an extrapolate dictionary in dS quantum gravity in \cite{susskindlindesay}.

In particular for the two point function of massive fields cluster decomposition implies that $\average{\phi_{m^2}(0,z_1)\phi_{m^2}(0,z_2)}$ decays to zero when the geodesic separation $d(z_1,z_2)$ is taken to $\infty$. In quantum gravity we need to account for contributions due to geometries with wormholes that connect two points on the parent universe
\begin{equation}
    \average{\phi_{m^2}(0,z_1)\phi_{m^2}(0,z_2)}=\average{\phi_{m^2}(0,z_1)\phi_{m^2}(0,z_2)}_\text{parent}+\average{\phi_{m^2}(0,z_1)\phi_{m^2}(0,z_2)}_\text{wormhole}+\dots \nonumber
\end{equation}
The firs term decays to zero when we take the separation to infinity but the second term in general does not. %The Lorentzian setup on the parent universe is (operators are black dots)
%\begin{equation}
%    \raisebox{-10mm}{\includegraphics[width=73mm]{sunday11_73.pdf}}\label{top1}
%\end{equation}
%However the contribution from the wormhole geometry in general does not.
%\begin{equation}
%    \raisebox{-10mm}{\includegraphics[width=45mm]{sunday12_45.pdf}}\label{top2}
%\end{equation}
The wormhole can create a shortcut which effectively brings distant regions close together and so we should not expect cluster decomposition to hold in quantum gravity. This is perhaps not too surprising as the cluster decomposition principle strictly speaking only needs to hold for local quantum field theories. In quantum gravity we generically sum over nonlocal quantum field theories on different topologies. At minimum we get a nonlocal quantum field theory on the parent universe by integrating out the wormholes \cite{coleman,gidstro1,gidstro2}.

For an independent argument consider massless matter. There is an infinite amount of conformal space near the semiclassical horizon and so we can probe cluster decomposition by investigating conformal matter near the horizon. For conformal matter we can trust the Penrose diagram for intuition
\begin{equation}
    \raisebox{-10mm}{\includegraphics[width=27mm]{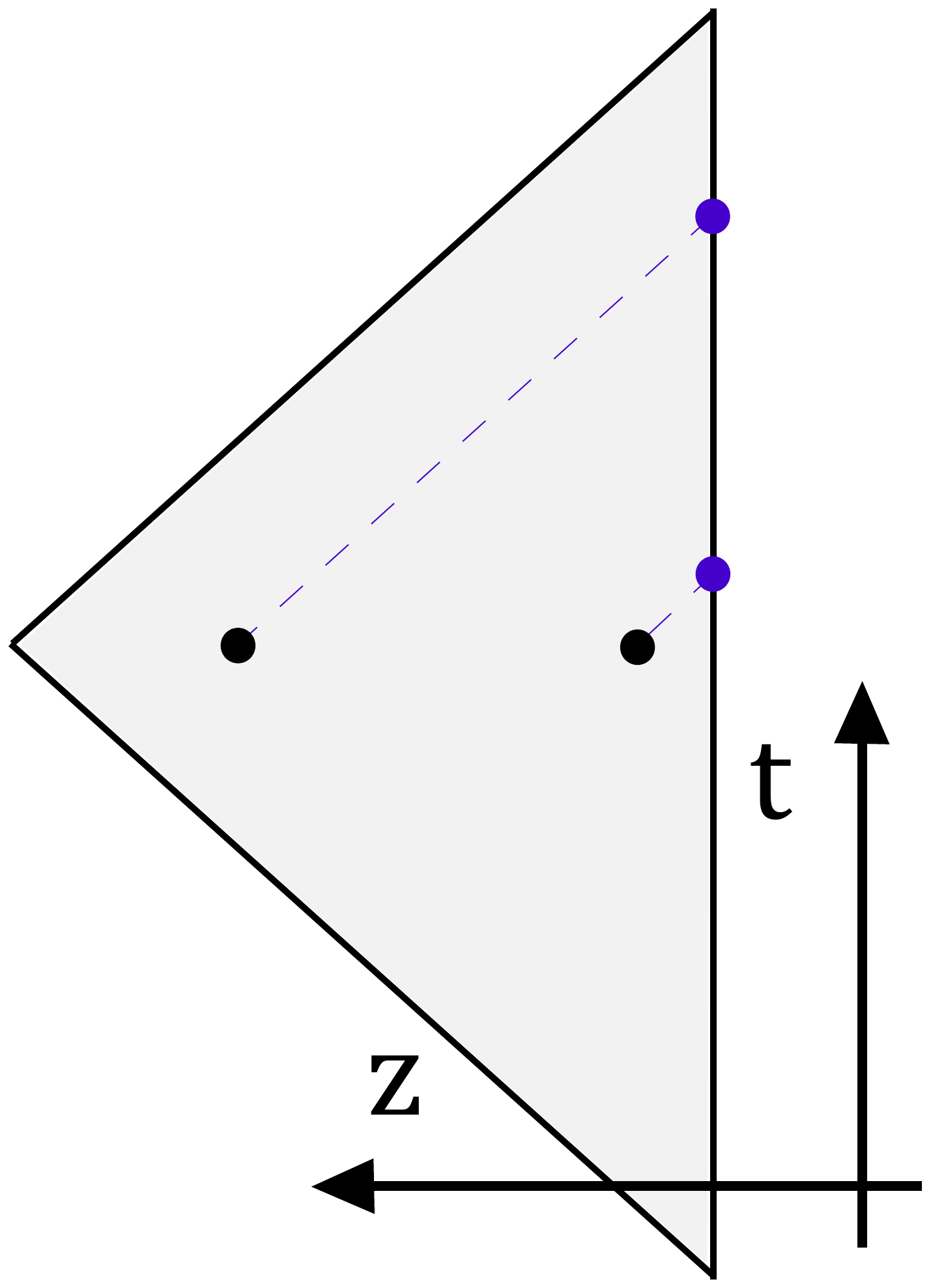}}
\end{equation}
The decay of the boundary two point function (blue dots) makes way at late times for erratic oscillations around a plateau \cite{maldainfo}. The Penrose diagram suggests to expect similar behavior for the bulk two point function (black dots).

In section \ref{sect:2} we explain how one can include contributions due to Euclidean wormholes to bulk correlators in JT gravity. We reverse engineer bulk operator reconstruction in the Lorentzian geometry to write the bulk correlator in terms of boundary correlators \cite{paper5,thomassingle,paper201,wopjordan} for which we know how to include contributions due to Euclidean wormholes. In section \ref{sect:4} we use this definition of bulk correlators to probe cluster decomposition. In particular we consider conformal matter close to the semiclassical horizon. We then further investigate the large distance behavior of the two point function of a massive field on an extremal black hole. In this case we can probe cluster decomposition due to the infinite throat. We identify an analogue to the plateau in both examples as expected of ensemble averages of large distance correlators in a finite entropy system.
%%%%%%%%%%%%%%%%%%%%%%%%%%%%%%%%%%%%%%%%%
\subsection{Factorization}
In section \ref{sect:5} we discuss factorization and the role of the ensemble in JT gravity. There is a sense in which gravitational path integrals naturally compute statistical ensemble averages of observables \cite{coleman,gidstro1,gidstro2}. This was recently emphasized and realized in a toy model of two dimensional quantum gravity \cite{maxfieldmarolf} with comments in \cite{joaquingiddings,vafa}. For the general argument we can work within the approximation of free baby universes for which we can model the gravitational system as \cite{coleman,gidstro1,gidstro2,joaquingiddings}
\begin{equation}
    H(\phi)=H_0(\phi)+(a+a^\dagger)\,V(\phi)\,.\label{theory}
\end{equation}
Here $H_0(\phi)$ is the Hamiltonian of the gravitational theory on the parent universe. The operators $a$ and $a^\dagger$ annihilate and create baby universes. To diagonalize the Hamiltonian we first construct an orthonormal basis of eigenstates of $(a+a^\dagger)$
\begin{equation}
  (a+a^\dagger)\ket{\alpha}=\alpha\ket{\alpha}\,.  
\end{equation}
These so called alpha states are coherent multi baby universe states \cite{joaquingiddings}. We have
\begin{equation}
    H(\phi)\ket{\alpha}=(H_0(\phi)+\alpha V(\phi))\ket{\alpha}\,.\label{hamphi}
\end{equation}
This means we have tuned certain coupling constants in the bare theory of the parent universe. For example we could imagine $V(\phi)=\phi^4$. If we allow for more and more species of baby universes we can ultimately tune all coupling constants of the parent universe by changing between different alpha states. Within each alpha state we can solve the parent theory with fixed coupling constants. So the baby universe eigenvalues act as superselection sectors
\begin{equation}
    \mathcal{H}=\int_\oplus d\alpha \ket{\alpha}\otimes \mathcal{H}_\alpha\,.\label{hamphi}
\end{equation}
This is under the assumption that observers in the parent universe cannot interact with the baby universes to change the coupling constants in the parent universe. This seems sensible from an anthropological point of view. 
\\~\\
In the context of holography we do not typically average over data of the CFT. Any given boundary theory with specific data $\alpha$ is dual to one particular superselection sector of bulk quantum gravity. On the other hand naive calculations using the gravitational path integral correspond to observations in the no boundary state. The no boundary state has a nonzero overlap with all alpha states. This turns out to imply that the gravitational path integral naturally computes ensemble averages of observables \cite{maxfieldmarolf}.

Let us explain this in more detail. In quantum field theory in general we may interpret boundary conditions in the Euclidean path integral as defining states in a Hilbert space. For example in JT gravity we may read the total amplitude with $2$ circular fixed energy boundaries as defining an inner product on single boundary states
\begin{equation}
    \rho(E_1,E_2)=\bra{\rho(E_1)}\ket{\rho(E_2)}=\bra{\text{HH}}\ket{\rho(E_1)\rho(E_2)}=\bra{\text{HH}}\rho(E_1)\rho(E_2)\ket{\text{HH}}=\dots\,.\label{4.13}
\end{equation}
We may equivalently read this as the inner product of the no boundary state and a state with two boundaries or as the matrix element of two boundary creating operators in the state with no boundaries. Diagonalizing the boundary creating operators in this multi boundary Hilbert space gets the basis of alpha states of \cite{maxfieldmarolf}. 

In quantum field theory there is in general a stand alone notion of an inner product on states. This often manifests in the form of a local cutting and gluing relation of amplitudes. With this independent notion of an inner product, path integrals in quantum field theory can be understood as computing transition amplitudes between normalized states. In quantum gravity we may not have this luxury. The hope of some local cutting and gluing relation in the bulk perishes upon contemplating the required large diff invariance of quantum gravity. Therefore there may not be an independent notion of an inner product on states in quantum gravity. If so then the best we can do is to use the path integral itself to define the inner product on asymptotic states as in \eqref{4.13}. In case of dS quantum gravity this was argued for in \cite{wittends}. See however recently \cite{cotleremergent,wopjordan}. For our purposes and those of \cite{maxfieldmarolf} the notion of using the path integral to define an inner product as in \eqref{4.13} suffices.

Say now we compute the partition function in JT gravity as specified in \eqref{1.4}. This introduces a probe boundary operator in our gravitational theory. There are no additional boundaries in \eqref{1.4} on which geometries may end so in the language of \cite{maxfieldmarolf} we are probing the no boundary state. By introducing a complete set of alpha states we find
\begin{equation}
    \average{Z(\beta)}=\bra{\text{HH}}Z(\beta)\ket{\text{HH}}\,=\int d\alpha\,\rvert \bra{\text{HH}}\ket{\alpha}\rvert^2\,Z_\alpha(\beta).\label{4.12}
\end{equation}
Here $Z_\alpha(\beta)$ corresponds in the context of holography to the partition function of a dual boundary theory with data $\alpha$. For JT gravity the boundary dual is a quantum mechanical system and the data is the Hamiltonian matrix. From \eqref{4.12} we see that the gravitational path integral naturally computes ensemble averages. The statistical nature of gravitational calculations in the no boundary state is emphasized if we consider two partition functions
\begin{equation}
    \average{Z(\beta_1)Z(\beta_2)}=\bra{\text{HH}}Z(\beta_1)Z(\beta_2)\ket{\text{HH}}\,=\int d\alpha\,\rvert \bra{\text{HH}}\ket{\alpha}\rvert^2\,Z_\alpha(\beta_1)Z_\alpha(\beta_2).
\end{equation}
Due to the ensemble averaging there is a nontrivial correlation between the two operators
\begin{equation}
    \average{Z(\beta_1)Z(\beta_2)}= \average{Z(\beta_1)}\average{Z(\beta_2)}+\average{Z(\beta_1)Z(\beta_2)}_\text{conn}\,.
\end{equation}
From the gravitational point of view this is due to geometries connecting the two boundaries
\begin{align}
\displaybreak[0]
       \nonumber &\average{Z(\beta_1)Z(\beta_2)}_\text{conn}\\&\qquad=\quad \raisebox{-5mm}{\includegraphics[width=50mm]{intro3.pdf}}\quad+\quad\raisebox{-5mm}{\includegraphics[width=50mm]{intro4.pdf}}\quad+\dots
\end{align}
Clearly the product of two CFT partition functions with the same data $\alpha$ factorizes. The total connected contribution vanishes. In the gravity calculation this factorization maps to the property that the alpha states diagonalize the boundary creating operators 
\begin{equation}
    \average{Z(\beta_1)Z(\beta_2)}_\alpha = \frac{\bra{\text{HH}}Z(\beta_1)Z(\beta_2)\ket{\alpha}}{\bra{\text{HH}}\ket{\alpha}}=Z_\alpha(\beta_1)Z_\alpha(\beta_2)\,.
\end{equation}
Our goal in section \ref{sect:5} is to understand what an alpha state geometrically looks like for JT gravity. Furthermore we explain how observables factorize in an alpha state by summing over geometries. In case of $Z(\beta_1)Z(\beta_2)$ this was largely addressed in \cite{paper6}. There is however an important class of observables for which there is an additional factorization issue. This is because $Z(\beta_1)Z(\beta_2)$ is not sensitive to eigenvector correlations whilst correlation functions in general are.

The explicit realization of the ensemble for JT gravity is the matrix integral definition itself \cite{sss2}. Consequently we find in section \ref{sect:5} that the alpha states are labeled by a set of energies $E_1\dots E_L$ and a matrix $Z$ of eigenstate components. An alpha state is constructed by defining the gravitational path integral to include a well chosen set of spacetime branes. Similar results were obtained in the toy model of \cite{maxfieldmarolf}. For JT gravity these branes include eigenbranes \cite{paper6} with data $E_1\dots E_L$ and additionally eigenvector branes with data $Z$ which are essentially exponentials of EOW branes \cite{rw1,malkou}. When this work was nearing completion we learned of an independent investigation of factorization of observables in JT gravity \cite{wopphilshunyusteve}.
%%%%%%%%%%%
% SECTION %
%%%%%%%%%%%
\section{Correlation functions}\label{sect:2}
We compute correlation functions in random matrix theory assuming local typicality. This is a refinement of ETH that captures for example Lyapunov growth \cite{ethotoc}. We then reproduce the resulting averaged description of correlators in JT gravity by summing over all possible trajectories of probe matter particles on all possible bulk geometries.
%%%%%%%%%%%%%%%%%%%%%%%%%%%%%%
\subsection{Correlation functions in random matrix theory}\label{ss21}
We consider correlation functions in $L\gg 1$ random matrix theory. The ensemble is defined in textbooks as
\begin{equation}
    \mathcal{Z}=\int d H\,P(H)= \int \prod_{a=1}^L d\lambda_a\,e^{-L V(\lambda_a)}\prod_{b<a}^L (\lambda_a-\lambda_b)^2\int d U\,.\label{2.1}
\end{equation}
Here $d U$ is the Haar measure on $L$ dimensional unitary basis transformations between the eigenstates $\ket{a}$ of the Hamiltonian $H$ and some reference basis
\begin{equation}
    \ket{i}=\sum_{a=1}^L U_{i a}\ket{a}\,.
\end{equation}
For example for a system consisting of multiple fermions $\ket{i}$ could be fixed spin states. We want to calculate the ensemble average of a generic correlator
\begin{equation}
    \Tr(e^{-\beta_1 H}\mo_1\dots e^{-\beta_n H}\mo_n)\,.\label{12}
\end{equation}
Here $\beta_1\dots \beta_n$ in general have complex parts denoting time evolution of the operators. We assume knowledge of the nondynamical data $\Tr(\mo_1\dots \mo_n)$. We then have
\begin{align}
    &\Tr(e^{-\beta_1 H}\mo_1\dots e^{-\beta_n H}\mo_n)\nonumber \\&=\sum_{a_1\dots a_n=1}^L e^{-\beta_1 \lambda_{a_1}}\dots e^{-\beta_n\lambda_{a_n}}\sum_{i_1\dots i_n=1}^L\sum_{I_1\dots I_n=1}^L \mo^1_{I_1 i_1}\dots\mo^n_{I_n i_n}\,U_{i_n a_1}U_{I_1 a_1}^*\dots U_{i_{n-1} a_n}U_{I_n a_n}^*\,. \label{14}
\end{align}
To compute the ensemble average we first take the integral over unitaries. Correlation functions of the Haar random ensemble are known in terms of Weingarten functions $\text{Wg}\,(\sigma,L)$. In general we have a sum of permutations $\alpha$ and $\beta$ \cite{wein,gu}
\begin{align}
    &\int d U\,U_{i_1 j_1}U_{k_1 l_1}^*\dots U_{i_n j_n}U_{k_n l_n}^*\nonumber\\&\qquad\qquad=\sum_{\alpha}\delta_{i_1 k_{\alpha(1)}}\dots \delta_{i_n k_{\alpha(n)}}\sum_{\beta}\delta_{j_1 l_{\beta(1)}}\dots \delta_{j_n l_{\beta(n)}}\,\text{Wg}\,(\beta\cdot \alpha^{-1},L)\,. \label{16}
\end{align}
The Weingarten functions are known in closed form \cite{wein,gu}. If we denote by $m$ the number of cycles in $\sigma$ then $\text{Wg}\,(\sigma,L)$ goes like $L^{m-2n}$ for $L\gg 1$. Therefore the dominant terms in the sum are those with maximal number of cycles $\alpha=\beta$ and in this case the prefactor in the leading behavior of the Weingarten functions is one. So to leading order there is just a single sum over permutations $\alpha$
\begin{equation}
    \int d U\,U_{i_1 j_1}U_{k_1 l_1}^*\dots U_{i_n j_n}U_{k_n l_n}^*=\frac{1}{L^n} \sum_{\alpha}\delta_{i_1 k_{\alpha(1)}}\dots \delta_{i_n k_{\alpha(n)}}\delta_{j_1 l_{\alpha(1)}}\dots \delta_{j_n l_{\alpha(n)}}\,.\label{17}
\end{equation}
This approximation is accurate for not too complex correlators such as \eqref{12} with $n\ll L$. However is breaks down when we would compute for example the purity of some density matrix as in \cite{rw1}.

We see that to leading order the correlators of the Haar random ensemble are just the Wick contractions of an ensemble of independent Gaussian random complex variables
\begin{equation}
    \mathcal{Z}=\int \prod_{i=1}^L\prod_{a=1}^L d U_{i a}\, d U_{i a}^*\,e^{-L\rvert U_{i a}\rvert^2}\,.\label{18}
\end{equation}
It is indeed a well known property of large Haar random unitary matrices that any relatively small subset of their matrix elements can be approximated as independent Gaussian random complex variables. JT gravity has a dual description as a matrix integral. From the gravitational side of the duality by summing over geometries we will find no indications of the subleading contributions due to the Weingarten functions. For the moment it seems that JT gravity is dual to an ensemble where the eigenstate components are independent Gaussian random complex variables and not Haar random unitaries. This is slightly surprising. We comment further on this in section \ref{sect:disc}.

After computing the ensemble average over unitaries in \eqref{14} we can trace over $i_1 \dots i_n$ and compute the ensemble average over eigenvalues. For clarity of presentation we focus on two simple examples and we henceforth consider the Gaussian ensemble \eqref{18}.
\\~\\
\textbf{\emph{Two point function}}
\\~\\
Consider the two point function of two identical operators
\begin{equation}
    \Tr(e^{-\beta_1 H} \mo e^{-\beta_2 H} \mo)=\sum_{a\,b=1}^Le^{-\beta_1\lambda_a}\,e^{-\beta_2\lambda_b}\sum_{i\,j=1}^L\sum_{I\,J=1}^L \mo_{I i}\mo_{J j}\,U_{i a}U_{j b}U_{J a}^*U_{I b}^*\,.\label{19}
\end{equation}
We consider traceless operators $\Tr(\mo)=0$ and furthermore take two different operators to have a vanishing two point function $\Tr(\mo_1\mo_2)=0$. This choice is inspired by the results of the gravity calculations later on. Taking Wick contractions we find
\begin{equation}
    \sum_{i\,j=1}^L\sum_{I\,J=1}^L \mo_{I i}\mo_{J j}\average{U_{i a}U_{j b}U_{J a}^*U_{I b}^*}=\frac{1}{L^2}\Tr(\mo\mo)\,.
\end{equation}
Here we used that the operators are traceless. This factors out of \eqref{19} and we recognize the remainder as the product of two macroscopic loop operators $Z(\beta_1)Z(\beta_2)$. Averaging over eigenvalues we find
\begin{equation}
    \average{\Tr(e^{-\beta_1 H}\mo e^{-\beta_2 H}\mo)}=\frac{1}{L^2}\Tr(\mo\mo)\average{Z(\beta_1)Z(\beta_2)}\,.\label{2.10}
\end{equation}
Clearly this is not an accurate description of the two point function of a chaotic quantum system. Otherwise every such two point function would be proportional to the spectral form factor at all times which is not the case. We will improve on this further on.
\\~\\
\textbf{\emph{Four point function}}
\\~\\
Consider now the $4$ point function with all operators identical
\begin{align}
   \label{23} &\Tr(e^{-\beta_1 H}\mo e^{-\beta_2 H}\mo e^{-\beta_3 H}\mo e^{-\beta_4 H}\mo)\\\nonumber&=\sum_{a\,b\,c\,d=1}^Le^{-\beta_1 E_a}\dots \,e^{-\beta_4 E_d}\sum_{i\,j\,k\,l=1}^L\sum_{I\,J\,K\,L=1}^L\mo_{I i}\mo_{J j}\mo_{K k}\mo_{L l}\,U_{i b}U_{j c}U_{k d}U_{l a}U_{I a}^*U_{J b}^*U_{K c}^*U_{L d}^*\,.
\end{align}
We sum over Wick contractions and trace over $i,j,k$ and $l$ using the fact that there are no one point functions. This limits the contractions which contribute to the correlator to $9$ out of a possible $24$. Two examples are
\begin{align}
   \sum_{i\,j\,k\,l=1}^L\sum_{I\,J\,K\,L=1}^L\mo_{I i}\mo_{J j}\mo_{K k}\mo_{L l}\,\big\langle
    \contraction[4mm]{}{U_{i b}}{U_{j c}U_{k d}U_{l a} U_{I a}^* U_{J b}^* U_{K c}^*  }{U_{L d}^*}
    \contraction[3mm]{U_{i b}}{ U_{j c} }{U_{k d}U_{l a} U_{I a}^* U_{J b}^*}{U_{K c}^*}
    \contraction[2mm]{U_{i b}U_{j c}}{U_{k d}}{U_{l a} U_{I a}^*}{U_{J b}^*}
    \contraction[1mm]{U_{i b}U_{j c}U_{k d}}{U_{l a}}{}{U_{I a}^*}
    U_{i b}U_{j c}U_{k d}U_{l a}U_{I a}^*U_{J b}^*U_{K c}^*U_{L d}^*\big\rangle &=\frac{1}{L^4}\Tr(\mo \mo)^2\,\delta_{b d}\nonumber \\
    \sum_{i\,j\,k\,l=1}^L\sum_{I\,J\,K\,L=1}^L\mo_{I i}\mo_{J j}\mo_{K k}\mo_{L l}\,\big\langle
    \contraction[4mm]{}{U_{i b}}{U_{j c}U_{k d}U_{l a} U_{I a}^*}{U_{J b}^*}
    \contraction[3mm]{U_{i b}}{ U_{j c} }{U_{k d}U_{l a} U_{I a}^* U_{J b}^*}{U_{K c}^*}
    \contraction[2mm]{U_{i b}U_{j c}}{U_{k d}}{U_{l a} U_{I a}^*U_{J b}^* U_{K c}^*}{U_{L d}^*}
    \contraction[1mm]{U_{i b}U_{j c}U_{k d}}{U_{l a}}{}{U_{I a}^*}
    U_{i b}U_{j c}U_{k d}U_{l a}U_{I a}^*U_{J b}^*U_{K c}^*U_{L d}^*\big\rangle &=\frac{1}{L^4}\Tr(\mo \mo\mo\mo)\,.\label{24}
\end{align}
All $9$ terms are
\begin{align}
    \label{215}&\sum_{i\,j\,k\,l=1}^L\sum_{I\,J\,K\,L=1}^L\mo_{I i}\mo_{J j}\mo_{K k}\mo_{L l}\,\average{U_{i b}U_{j c}U_{k d}U_{l a}U_{I a}^*U_{J b}^*U_{K c}^*U_{L d}^*}\\&=\frac{1}{L^4}\Tr(\mo\mo)^2(\delta_{a c}+\delta_{b d}+\delta_{a b c d})+\frac{1}{L^4}\Tr(\mo\mo\mo\mo)(1+\delta_{a c}\delta_{b d}+\delta_{a b c}+\delta_{a b d}+\delta_{a c d}+ \delta_{b c d})\,.\nonumber
\end{align}
Here a delta with multiple labels is one only if all labels are the same. We can now term by term ensemble average over eigenvalues. For example the first term in \eqref{24} identifies two of the energies that are being summed over in the calculation of the four point function \eqref{23}. We recognize the product of three macroscopic loop operators
\begin{equation}
    \average{ \Tr(e^{-\beta_1 H}\mo e^{-\beta_2 H}\mo e^{-\beta_3 H}\mo e^{-\beta_4 H}\mo)}\supset \frac{1}{L^4}\Tr(\mo\mo)^2\average{Z(\beta_1)Z(\beta_2+\beta_4)Z(\beta_3)}\,.\label{27}
\end{equation}

Consider now $4$ point functions where the operators are just two by two identical. There are $3$ topologically inequivalent options. The $s$ and $t$ channel $4$ point functions are morally time ordered whilst the $u$ channel $4$ point function is morally out of time ordered. Each of the $9$ terms in \eqref{215} contributes to precisely one of the channels. To see this we furthermore choose our initial data such that there is no connected contribution to the time ordered four point function $\Tr(\mo_1\mo_1\mo_2\mo_2)=0$ which is again inspired by the gravity calculations. With this assumption we indeed find just $3$ contributions to the $s$ channel $4$ point function
\begin{align}
    \nonumber&\sum_{i\,j\,k\,l=1}^L\sum_{I\,J\,K\,L=1}^L\mo^1_{I i}\mo^1_{J j}\mo^2_{K k}\mo^2_{L l}\,\average{U_{i b}U_{j c}U_{k d}U_{l a}U_{I a}^*U_{J b}^*U_{K c}^*U_{L d}^*}\\&\qquad=\frac{1}{L^4}\Tr(\mo_1\mo_1)\Tr(\mo_2\mo_2)(\delta_{a c})+\frac{1}{L^4}\Tr(\mo_1\mo_2\mo_1\mo_2)(\delta_{a b d}+ \delta_{b c d})\,.\label{28}
\end{align}
Similarly for the out of time ordered or $u$ channel $4$ point function we find
\begin{align}
    \nonumber&\sum_{i\,j\,k\,l=1}^L\sum_{I\,J\,K\,L=1}^L\mo^1_{I i}\mo^2_{J j}\mo^1_{K k}\mo^2_{L l}\,\average{U_{i b}U_{j c}U_{k d}U_{l a}U_{I a}^*U_{J b}^*U_{K c}^*U_{L d}^*}\\&\qquad=\frac{1}{L^4}\Tr(\mo_1\mo_1)\Tr(\mo_2\mo_2)(\delta_{a b c d})+\frac{1}{L^4}\Tr(\mo_1\mo_2\mo_1\mo_2)(1+ \delta_{a c}\delta_{b d})\,.\label{29}
\end{align}
As mentioned in the introduction correlation functions of realistic quantum chaotic systems at early times $t\ll e^{S_0}$ are self averaging. This means that the value of the correlator for a single realization of the system is identical to the ensemble averaged correlator. 

The question is what is the relevant ensemble. It is clear \cite{haake,mehta} that we should average over the energy levels as if they were the eigenvalues of a large random matrix. However it is not obvious how to average over the eigenvector components. What is clear from the current discussion is that certainly the answer is not to consider $L$ dimensional Haar random unitaries. For example both the two point function \eqref{2.10} as the $u$ channel $4$ point function \eqref{29} depend only on $\average{Z(\beta_1)\dots Z(\beta_n)}$ and $\Tr(\mo_1\dots \mo_n)$. This data is not rich enough to capture the dynamics of a realistic quantum chaotic system. For example it is insufficient to capture Lyapunov growth. 
\\~\\
\textbf{\emph{Realistic systems and local typicality}}
\\~\\
We can improve on this following \cite{ethotoc}. The assumption that the correlators of some system at early times are well described by the Haar random ensemble is identical to the assumption that $\mo$ and $U\mo U^{-1}$ have the same early time correlators for any unitary $U$. This is too strong a constraint. Realistic quantum chaotic systems are accurately described by the weaker assumption of local typicality. This assumes that $\mo$ and $U\mo U^{-1}$ have the same early time correlators but with $U$ constrained in the Hamiltonian basis to be a blockdiagonal unitary matrix $\text{diag}\,(U_1\dots U_n)$. Each of the blocks $U_1\dots U_n$ is a unitary of dimension $e^{S_0}$ whose role is to mix the $e^{S_0}$ states in an energy bin $\delta E$ around a given level.
\\~\\
The result of replacing the assumption of typicality with that of local typicality is that the original data $\Tr(\mo_1\dots \mo_n)$ is replaced by functions $\mo_1\dots \mo_n(E_1\dots E_n)$ which are smooth on energy scales of order $e^{-S_0}$. Consider the two point function for $a$ en $b$ in different bins
\begin{align}
    \average{\mo_{ab}\mo_{b a}}&=\int d U_1\,d U_2\,\bra{a} U_1\mo U_2^{-1} \ket{b}\bra{b}U_2\mo U_1^{-1} \ket{a}\nonumber\\&=\sum_{c\,f\in\delta E_1}\sum_{d\,e\in \delta E_2}\mo_{cd}\mo_{ef}\,\big\langle {U_1}_{a c}{U_1}^*_{a f}\big\rangle\big\langle{U_2}_{b e}{U_2}^*_{b d}\big\rangle \nonumber\\&=e^{-2S_0}\sum_{c\in \delta E_1}\sum_{d\in \delta E_2}\mo_{cd}\mo_{dc}=e^{-2S_0}\,\mo\mo(E_a,E_b)\,.
\end{align}
Indeed by assumption we may replace the individual correlator by its ensemble average over blockdiagonal unitary transformations.

From the gravitational side of the duality by summing over geometries we find no indications of subleading corrections due to the Weingarten functions $\text{Wg}\,(\sigma,e^{S_0})$. This suggests that the suitable ensemble averaged description of correlators in JT gravity is to consider an ensemble where the blockdiagonal matrices $U_1\dots U_n$ of eigenvector components are not Haar random unitaries of dimension $e^{S_0}$ but rather all matrix elements of the blocks are independent Gaussian random complex variables with variance $e^{S_0}$. A second example is the two point function for $a$ en $b$ are in the same energy bin
\begin{align}
\displaybreak[0]
    \average{\mo_{ab}\mo_{b a}}&=\int d U_1\,\bra{a} U_1\mo U_1^{-1} \ket{b}\bra{b}U_1\mo U_1^{-1} \ket{a}\nonumber\\&=\sum_{c\,d\,e\,f\subset \delta E_1}\mo_{cd}\mo_{ef}\,\big\langle {U_1}_{a c}{U_1}^*_{a f}{U_1}_{b e}{U_1}^*_{b d}\big\rangle \nonumber\\&=e^{-2S_0}\,\mo\mo(E_a,E_b)+e^{-2S_0}\,\mo(E_a)\mo(E_b)\,\delta_{a b}\,.
\end{align}
These calculations extend trivially to higher point functions. For example for the four point function $\average{\mo_{a b}\mo_{bc}\mo_ {c d}\mo_{d a}}$ we recover all $9$ contributions in \eqref{215} but with $L$ replaced by $e^{S_0}$ and the matrix elements replaced by smooth functions if we furthermore assume $\mo(E)=0$. Of course if certain of the labels $a,b,c$ and $d$ are not considered to be in the same bin then correspondingly certain deltas in \eqref{215} are manifestly zero.
\\~\\
We aim to describe in particular the ensemble averaged description of correlators with a dual JT gravity interpretation. The smooth functions $\mo_1\dots \mo_n(E_1\dots E_n)$ can in that case be read of from the genus zero correlation functions of JT gravity.

First of all from the bare disk amplitude in JT gravity one deduces that the potential of the random eigenvalues is tuned to result in a genus zero spectral density that matches the JT gravity spectral density close to the spectral edge \cite{sss2}
\begin{equation}
    \rho_0(E)=\frac{e^{S_0}}{2\pi}\sinh 2\pi \sqrt{E}\,\quad,\quad E\ll \Lambda\,.
\end{equation}
See the discussion around equation \eqref{curve} for more explanation. The one point function on the disk in JT gravity vanishes from which we conclude that $\mo(E)=0$. Likewise the two point function on the disk \eqref{1.7} is only nonzero if both operators are identical from which we find $\mo_1\mo_2(E_1,E_2)=0$. Similarly in the genus zero $s$ channel four point function with pairwise identical operators \eqref{49} there is no connected contribution in the fixed energy basis \cite{schwarzian,paper5} and so $\mo_1\mo_1\mo_2\mo_2(E_1,E_2,E_3,E_4)=0$.

From the explicit answer of the disk two point function \eqref{1.7} we obtain \cite{schwarzian,paper3}
\begin{equation}
    \mo\mo(E_1,E_2)=e^{S_0}\,\frac{\Gamma(\ell\pm i \sqrt{E_1}\pm i \sqrt{E_2})}{\Gamma(2\ell)}\,.\label{2.21}
\end{equation}
This same expression features in the $s$ channel genus zero four point function \eqref{2.39}. The smooth $4$ point kernel can be deduced from the genus zero $u$ channel or out of time ordered $4$ point function \cite{schwarzian,paper3,paper4}
\begin{equation}
   \average{\Tr(e^{-\beta_1 H}\mo_1e^{-\beta_2 H}\mo_2e^{-\beta_3 H}\mo_1 e^{-\beta_4 H}\mo_2)}\supset\quad\raisebox{-10mm}{\includegraphics[width=50mm]{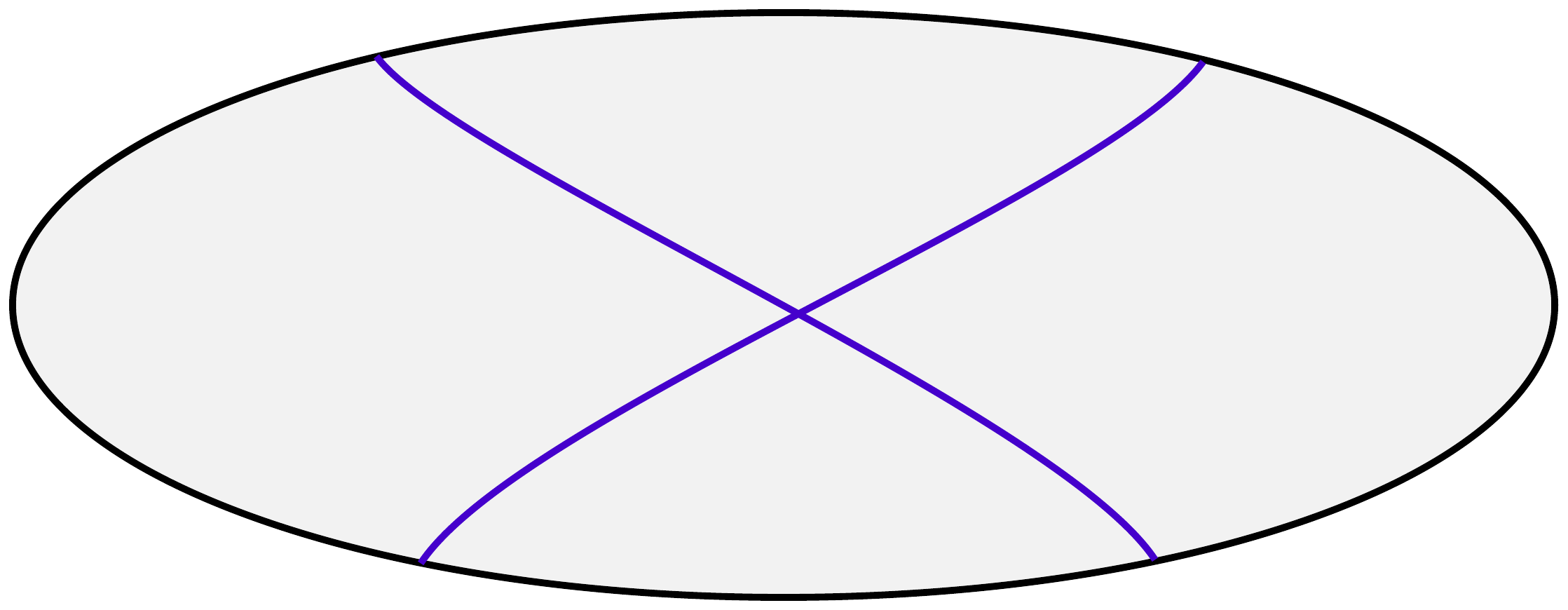}}\quad.
\end{equation}
By taking multiple inverse Laplace transforms of the exact answer one finds \cite{schwarzian,paper3,paper4}
\begin{align}
\displaybreak[0]
    &\mo_1\mo_2\mo_1\mo_2(E_1,E_2,E_3,E_4)\nonumber \\&\qquad=e^{S_0}\,\frac{\Gamma(\ell_1\pm i \sqrt{E_1}\pm i \sqrt{E_2})^{1/2}}{\Gamma(2\ell_1)^{1/2}}\frac{\Gamma(\ell_2\pm i \sqrt{E_2}\pm i \sqrt{E_3})^{1/2}}{\Gamma(2\ell_2)^{1/2}}\nonumber\\
    &\qquad\qquad\qquad\quad\frac{\Gamma(\ell_1\pm i \sqrt{E_3}\pm i \sqrt{E_4})^{1/2}}{\Gamma(2\ell_1)^{1/2}}\frac{\Gamma(\ell_2\pm i \sqrt{E_4}\pm i \sqrt{E_1})^{1/2}}{\Gamma(2\ell_2)^{1/2}}\sj{E_1}{\ell}{E_2}{E_3}{\ell}{E_4}\,.\label{25}
\end{align}
We recognize a gravitational 6j symbol whose explicit expression in terms of generalized hypergeometric functions can be found in \cite{paper3,paper4,schwarzian}.

Local typicality is a refinement of ETH which predicts higher point correlations like the one in \eqref{25}. This is key to capturing one of the hallmarks of chaotic systems, Lyapunov growth \cite{ethotoc}. The gravitational 6j symbol makes explicit the relation to Lyapunov growth as it essentially represents the gravitational shockwave S matrix in the fixed energy basis \cite{dray,schwarzian,schwarzianshocks,jackson}. We expect this 6j symbol to be a universal feature of the $4$ point function for systems with a gravitational dual. It would be interesting to see if it could be derived from the properties of fast scramblers without relying on gravity \cite{sekinosusskind,bound}.
\\~\\
With the assumption of local typicality to compute ensemble averages over eigenvector components we find for the two point function in random matrix theory \cite{phil}
\begin{equation}
    \average{\Tr(e^{-\beta_1 H}\,\mo\,e^{-\beta_2 H}\,\mo)}=e^{-2S_0}\int_\cas d E_1\,e^{-\beta_1 E_1}\int_\cas d E_2\,e^{-\beta_2 E_2}\,\mo\mo(E_1,E_2)\,\rho(E_1,E_2)\,.\label{34}
\end{equation}
Here the two level spectral density in random matrix theory is \cite{mehta,sss2,paper6}
\begin{equation}
    \rho(E_1,E_2)=\rho(E_1)\rho(E_2)+\rho(E_1)\rho(E_2)\sinc^2\,\pi\rho(E_1)(E_1-E_2)+\delta(E_1-E_2)\rho(E_1).\label{224}
\end{equation}
For the out of time ordered or $u$ channel $4$ point function we can modify the result \eqref{29} to account for local typicality. Computing the ensemble average over eigenvalues we find
\begin{align}
    \label{225}&\average{\Tr(e^{-\beta_1 H}\mo_1e^{-\beta_2 H}\mo_2 e^{-\beta_3 H}\mo_1e^{-\beta_4 H}\mo_2)}\\\nonumber &=e^{-4S_0}\int_\cas d E_1\,e^{-(\beta_1+\beta_2+\beta_3+\beta_4)}\,\rho(E_1)\,\mo_1\mo_1(E_1,E_1)\,\mo_2\mo_2(E_1,E_1)\\&\quad\nonumber +e^{-4 S_0}\int_\cas d E_1\,e^{-\beta_1 E_1}\int_\cas d E_2\,e^{-\beta_2 E_2}\int_\cas d E_3\,e^{-\beta_3 E_3}\int_\cas d E_4\,e^{-\beta_4 E_4}\\&\nonumber \qquad\qquad\qquad\qquad\qquad\qquad\qquad\qquad\quad\rho(E_1,E_2,E_3,E_4)\,\mo_1\mo_2\mo_1\mo_2(E_1,E_2,E_3,E_4)\\
    &\quad+e^{-4S_0}\int_\cas d E_1\,e^{-(\beta_1+\beta_3)E_1}\int_\cas d E_2\,e^{-(\beta_2+\beta_4)E_2}\,\rho(E_1,E_2)\, \mo_1\mo_2\mo_1\mo_2(E_1,E_2,E_1,E_2)\,.\nonumber
\end{align}
In the first term we only get the single level spectral density $\rho(E_1)$ due to the deltas in \eqref{29}. As there is no delta in the second term of \eqref{29} the ensemble averaging over eigenvalues computes a $4$ level spectral density $\rho(E_1,E_2,E_3,E_4)$. Multi level spectral densities in these Hermitian matrix integrals are known universally \cite{mehta,paper6}.\footnote{Consider for example $\rho(E_1,E_2,E_3)$ and define connected correlators as
\begin{align}
    \nonumber \rho(E_1,E_2,E_3)=&\,\rho(E_1)\rho(E_2)\rho(E_3)+\rho(E_1,E_2)_\text{conn}\, \rho(E_3)+\rho(E_2,E_3)_\text{conn}\\&+ \rho(E_1)+\rho(E_3,E_1)_\text{conn}\, \rho(E_2)+\rho(E_1,E_2,E_3)_\text{conn}.\label{66}
\end{align}
The connected correlator is the GUE cluster function  $T(E_1,E_2,E_3)$ plus contact terms
\begin{align}
    \nonumber\rho(E_1,E_2,E_3)_\text{conn}=&\,T(E_1,E_2,E_3)-\delta(E_1-E_2)T(E_1,E_3)-\delta(E_1-E_3)T(E_1,E_2)\\&-\delta(E_2-E_3)T(E_1,E_3)+\delta(E_1-E_2)\delta(E_1-E_3)\rho(E_1).\label{67}
\end{align}
The clusters have universal answers in terms of the sine kernel \cite{mehta,paper6}
\begin{align}
    T(E_1,E_2,E_3)&=2\,S(E_1,E_2)\,S(E_2,E_3)\,S(E_3,E_1)\quad,\quad T(E_1,E_2)=S(E_1,E_2)^2\,.\label{2.31}
\end{align}
The generalization to $\rho(E_1\dots E_n)$ is straightforward \cite{paper6}.} 

Following similar steps one immediately deduces a prediction for the $s$ channel $4$ point function from \eqref{28}
\begin{align}
    \label{42}&\average{\Tr(e^{-\beta_1 H}\mo_1e^{-\beta_2 H}\mo_1 e^{-\beta_3 H}\mo_2e^{-\beta_4 H}\mo_2)}\\\nonumber &=e^{-4S_0}\int_\cas d E_1\,e^{-(\beta_1+\beta_3) E_1}\int_\cas d E_2\,e^{-\beta_2 E_2}\int_\cas d E_4\,e^{-\beta_4 E_4}\\\nonumber&\qquad\qquad\qquad\qquad\qquad\qquad\qquad\qquad\quad\,\,\,\, \rho(E_1,E_2,E_4)\,\mo_1\mo_1(E_1,E_2)\mo_2\mo_2(E_1,E_4)\\\nonumber &\quad + e^{-4S_0}\int_\cas d E_1\,e^{-(\beta_1+\beta_2+\beta_4)E_1}\int_\cas d E_3\,e^{-\beta_3 E_3}\,\rho(E_1,E_3)\,\mo_1\mo_1\mo_2\mo_2(E_1,E_1,E_3,E_1)\\\nonumber &\quad +e^{-4S_0} \int_\cas d E_1\,e^{-\beta_1 E_1}\int_\cas d E_2\,e^{-(\beta_2+\beta_3+\beta_4)E_2}\,\rho(E_1,E_2)\,\mo_1\mo_1\mo_2\mo_2(E_1,E_2,E_2,E_2)\,.
\end{align}

The first term in this expression features in formula (2.31) of \cite{phil}. We note as a technical aside that there is no contribution to the kernel $\rho(E_1,E_2,E_4)$ of the type $\rho(E_1,E_2)\rho(E_1,E_4)$. The half pipe in the out of time ordered four point function therefore does not follow from the contribution identified in formula (2.31) of \cite{phil}. Nevertheless another contribution saves the day. The half pipe actually arises due to a contribution $\rho(E_1,E_2)\rho(E_1,E_4)$ in the kernel in the second term of the $u$ channel $4$ point function \eqref{225}. Note that this second term in \eqref{225} is the same that is responsible for Lyapunov growth.
%%%%%%%%%%%%%%%%%%%%%%%%%%%%%%%%%
\subsection{Taxonomy}
In the remainder of this section we consider similar boundary correlators from the bulk JT gravity point of view. We sum over all topologically different trajectories of probe quantum particles propagating through the Euclidean bulk and precisely recover the above predictions from random matrix theory. For the two point function such a study was initiated in \cite{paper5} and largely established in \cite{phil}.

For clarity of presentation we focus first on the example of the $s$ channel or time ordered $4$ point function. The correlator decomposes into quite a complicated sum
\begin{align}
    &\average{\Tr(e^{-\beta_1 H}\mo_1e^{-\beta_2 H}\mo_1e^{-\beta_3 H}\mo_2 e^{-\beta_4 H}\mo_2)}\nonumber\\ &\qquad\qquad\qquad\qquad=\sum_\Sigma \sum_{\gamma_1\,\gamma_2}\average{\Tr(e^{-\beta_1 H}\mo_1e^{-\beta_2 H}\mo_1e^{-\beta_3 H}\mo_2 e^{-\beta_4 H}\mo_2)}_{\Sigma\,\gamma_1\,\gamma_2}\,.\label{230}
\end{align}
The contributions are labeled by the bulk geometry $\Sigma$ and all the topologically inequivalent geodesics $\gamma_1$ and $\gamma_2$ on $\Sigma$ which connect the two $\mo_1$ operators and the two $\mo_2$ operators. This excludes closed matter loops around handles. We aim to order the zoo of contributions in a way that facilitates resummation. Imagine thereto slightly thickening the union of $\gamma_1,\gamma_2$ and the circular boundary of $\Sigma$. 
\\~\\
If $\gamma_1$ and $\gamma_2$ are not intersecting and not self-intersecting this defines a $4$ holed sphere $\Gamma$ as subset of $\Sigma$. One of the holes is the boundary of $\Sigma$. The other $3$ are homologous to unique circular geodesics $\kappa_1,\kappa_2$ and $\kappa_3$ in $\Gamma$. We will not consider separately the case where any of the boundaries of $\Gamma$ is contractable in $\Sigma$. It is elementary to modify the discussion to these special cases. We now cut $\Sigma$ on $\kappa_1,\kappa_2$ and $\kappa_3$. For example
\begin{equation}
   \average{\Tr(e^{-\beta_1 H}\mo_1e^{-\beta_2 H}\mo_1e^{-\beta_3 H}\mo_2 e^{-\beta_4 H}\mo_2)}\supset\quad\raisebox{-10mm}{\includegraphics[width=50mm]{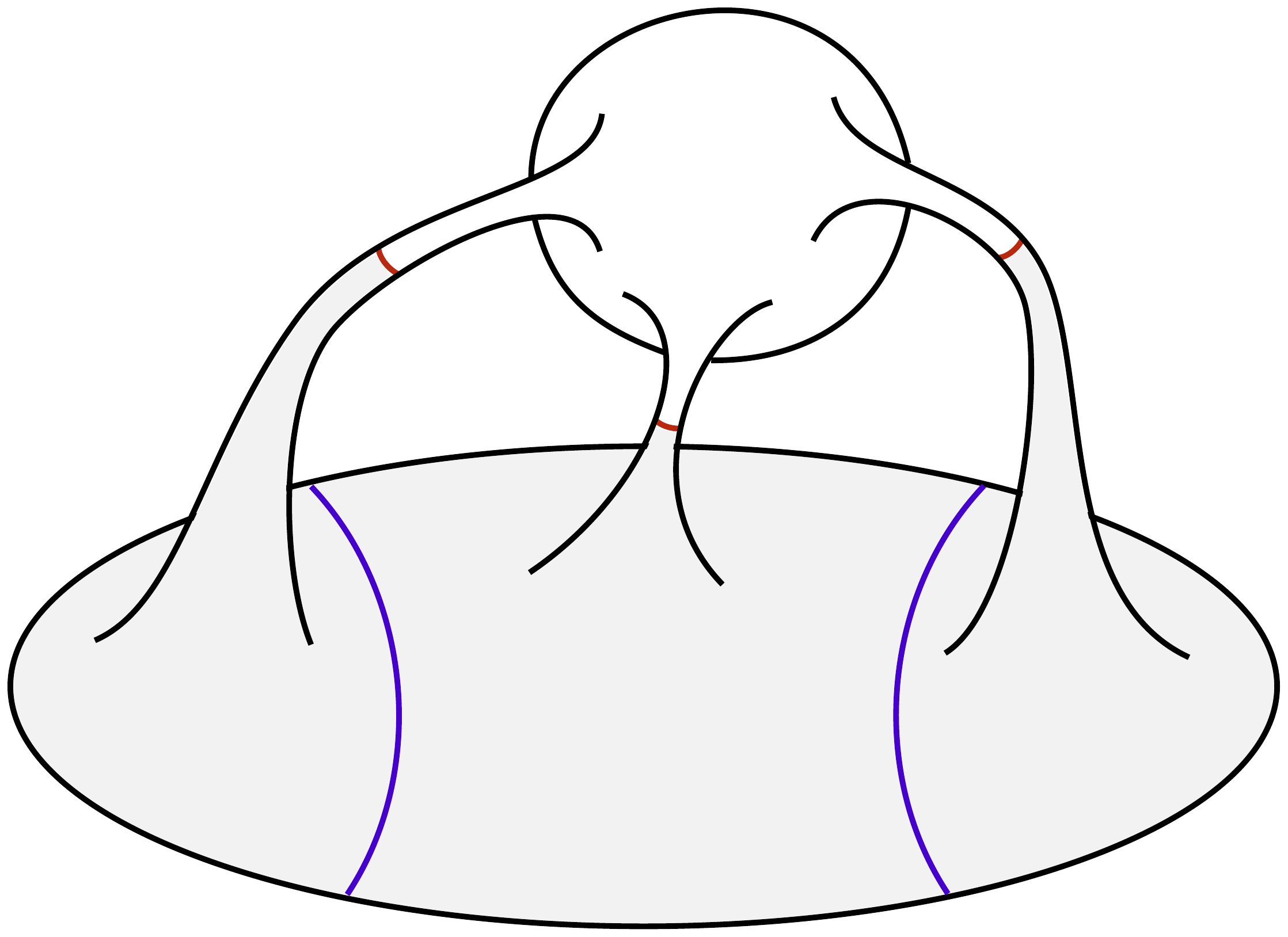}}\quad.\label{172}
\end{equation}
Here the surface K (white) is a $3$ holed sphere obtained by cutting on the $3$ geodesics $\kappa_1,\kappa_2$ and $\kappa_3$ (red) and $\Gamma$ is a $4$ holed sphere with two Wilson lines on it (gray). We have $\Sigma=\Gamma\cup\text{K}$. 

Note that there is in an infinite number of topologically inequivalent choices of $\gamma_1$ and $\gamma_2$ that give rise to topologically identical surfaces $\Gamma$ and K. The skeptical reader is advised to draw some pictures where $\gamma_1$ and or $\gamma_2$ wind around some of the handles. One tractable example is the two point function on a disk with a handle \cite{phil}. 

We denote inequivalent variants of $\Gamma$ by $\Gamma_i$ and inequivalent variants of K by K$_j$. For example a contribution with the same $\Gamma_i$ as \eqref{172} but with different K$_j$ is
\begin{equation}
   \average{\Tr(e^{-\beta_1 H}\mo_1e^{-\beta_2 H}\mo_1e^{-\beta_3 H}\mo_2 e^{-\beta_4 H}\mo_2)}\supset\quad\raisebox{-10mm}{\includegraphics[width=50mm]{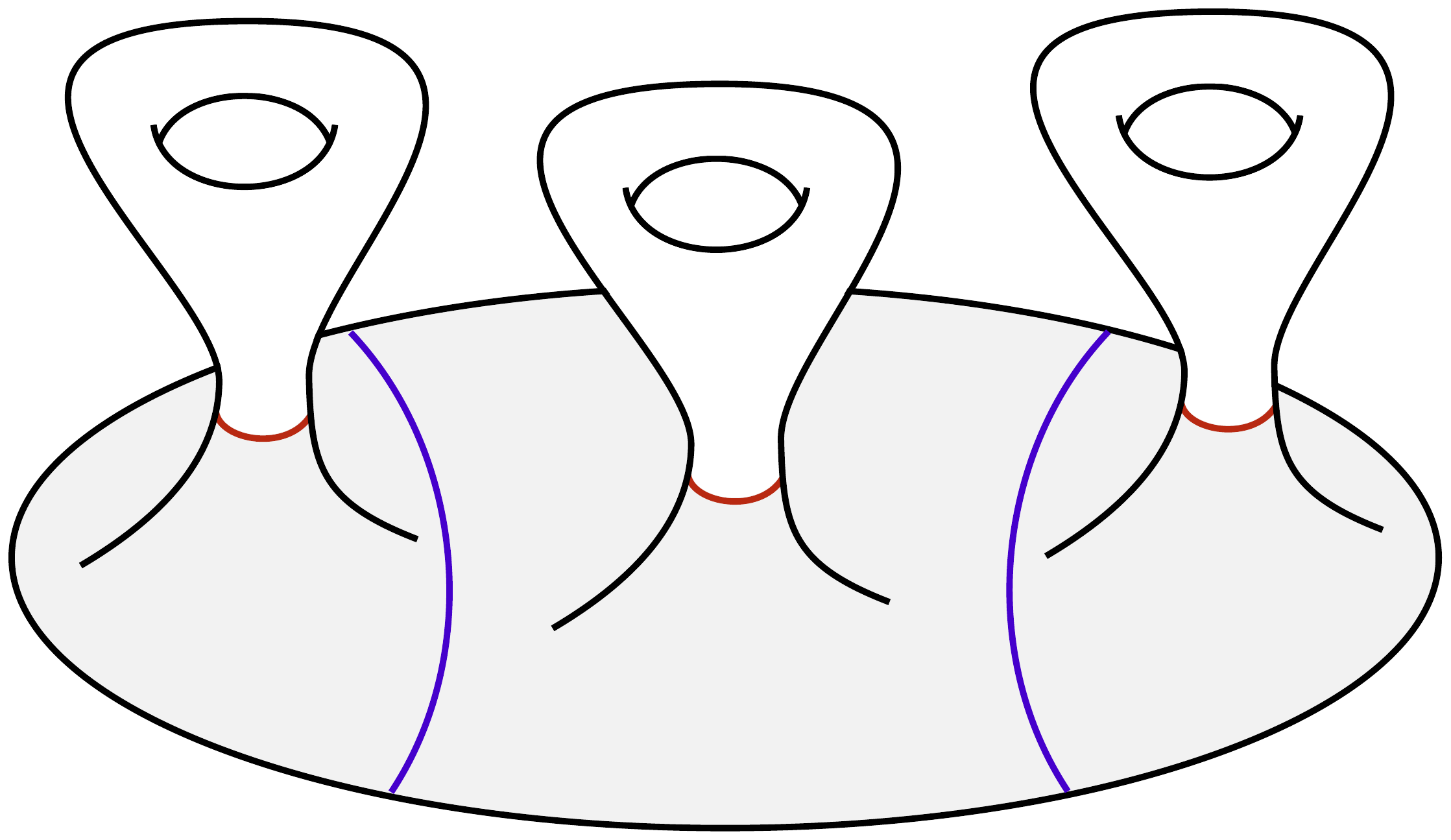}}\quad.\label{55}
\end{equation}
We define two configurations specified by geodesics $\gamma_1$ and $\gamma_2$ to be in the same ``family'' if this thickening and cutting construction associates to them topologically identical surfaces $\Gamma_i$ and K$_j$. To check if two configurations are in the same family we can imagine cutting it on the geodesics $\kappa_1,\kappa_2$ and $\kappa_3$ or equivalently on $\gamma_1$ and $\gamma_2$. We pick as representative of each family the configuration which looks like a higher genus Riemann surface with no Wilson lines on it attached to a simple amplitude with some boundary anchored Wilson lines. Examples are \eqref{172} and \eqref{55} though this choice is not important.
\\~\\
We define two configurations specified by geodesics $\gamma_1$ and $\gamma_2$ to be in the same ``class'' if this thickening and cutting construction associates to them topologically identical $\Gamma_i$. So we allow for different K$_j$ within each class. The contributions \eqref{172} and \eqref{55} are hence in the same class but not in the same family. Diagrams where some of the holes of $\Gamma_i$ are contractable in $\Sigma$ are also considered to be in the class $\Gamma_i$. Examples in the same class of \eqref{172} and $\eqref{55}$ include the $s$ channel genus zero $4$ point function \eqref{49} and
\begin{equation}
   \average{\Tr(e^{-\beta_1 H}\mo_1e^{-\beta_2 H}\mo_1e^{-\beta_3 H}\mo_2 e^{-\beta_4 H}\mo_2)}\supset\quad\raisebox{-10mm}{\includegraphics[width=50mm]{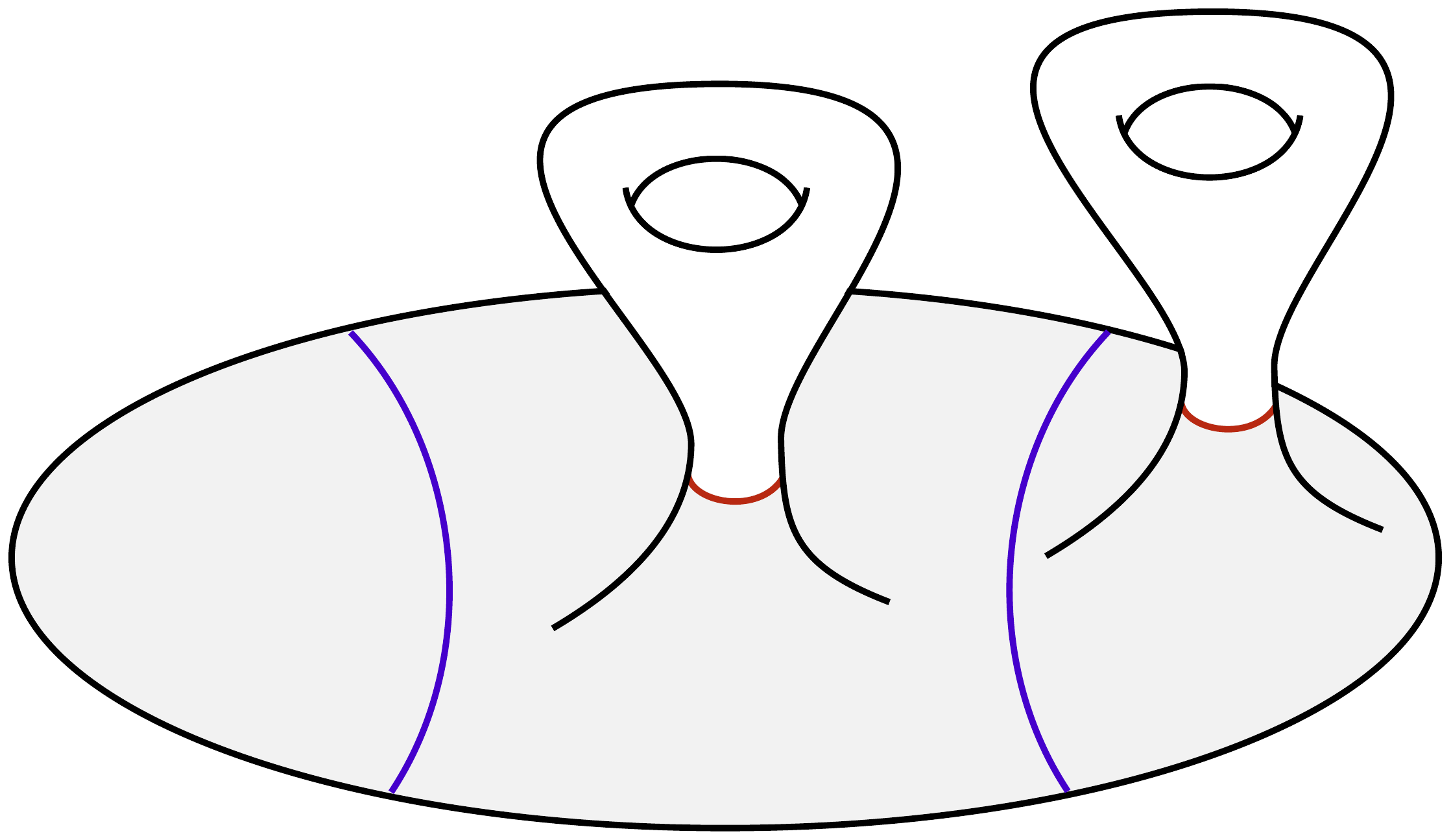}}\quad.\label{233}
\end{equation}

Different classes are obtained for example by considering curves $\gamma_1$ and $\gamma_2$ that intersect and or self-intersect. For example
\begin{equation}
   \average{\Tr(e^{-\beta_1 H}\mo_1e^{-\beta_2 H}\mo_1e^{-\beta_3 H}\mo_2 e^{-\beta_4 H}\mo_2)}\supset\quad\raisebox{-10mm}{\includegraphics[width=50mm]{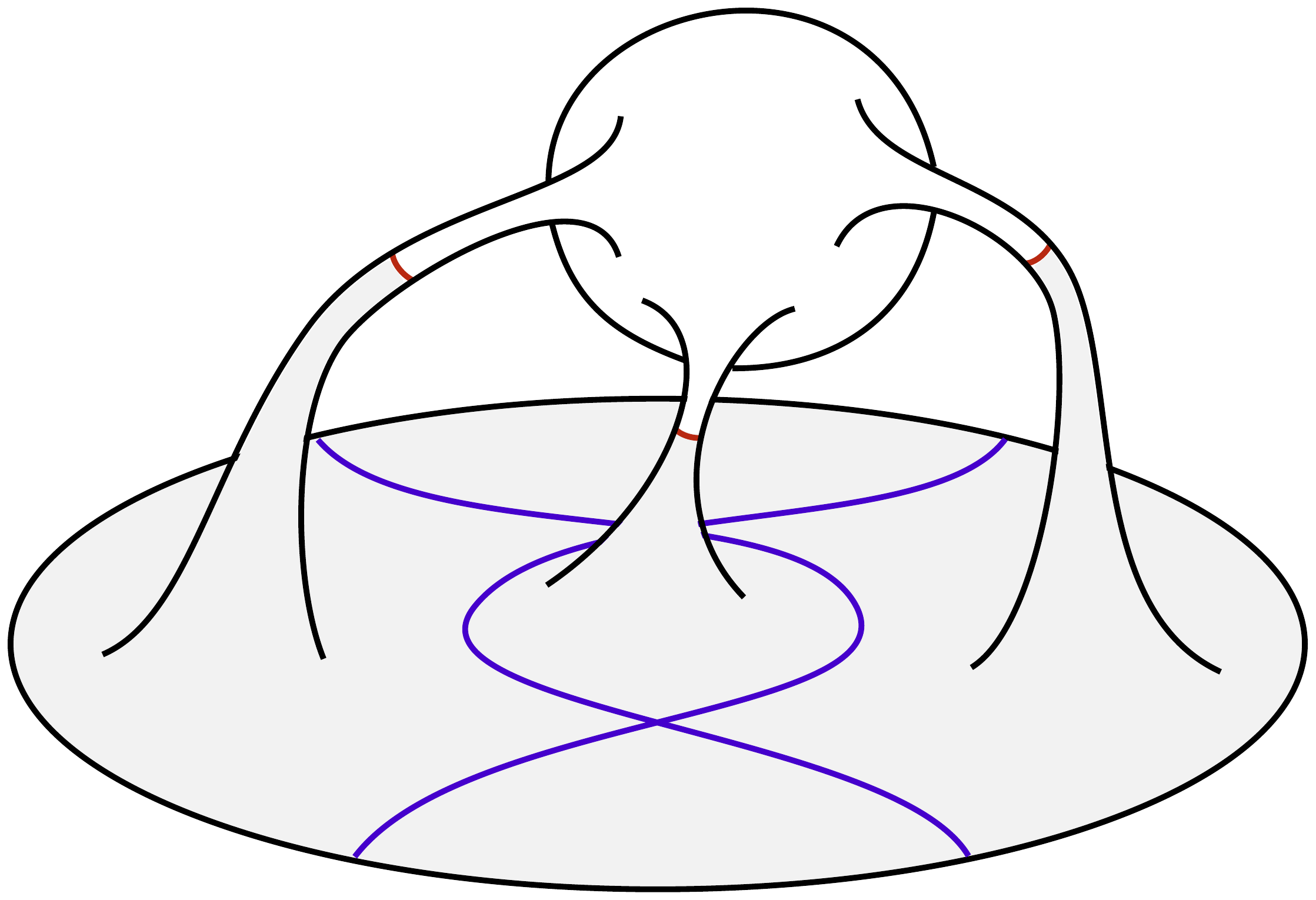}}\quad.\label{234}
\end{equation}
This has identical K$_j$ as compared to \eqref{55} however it has a different $\Gamma_i$ and is therefore to be considered in a different class. In this example we see that thickening $\gamma_1,\gamma_2$ and the boundary of $\Sigma$ defines a $6$ holed sphere. So there are contributions of geometries K$_j$ with $5$ geodesic boundaries ending on this $\Gamma_i$. For example
\begin{equation}
   \average{\Tr(e^{-\beta_1 H}\mo_1e^{-\beta_2 H}\mo_1e^{-\beta_3 H}\mo_2 e^{-\beta_4 H}\mo_2)}\supset\quad\raisebox{-10mm}{\includegraphics[width=50mm]{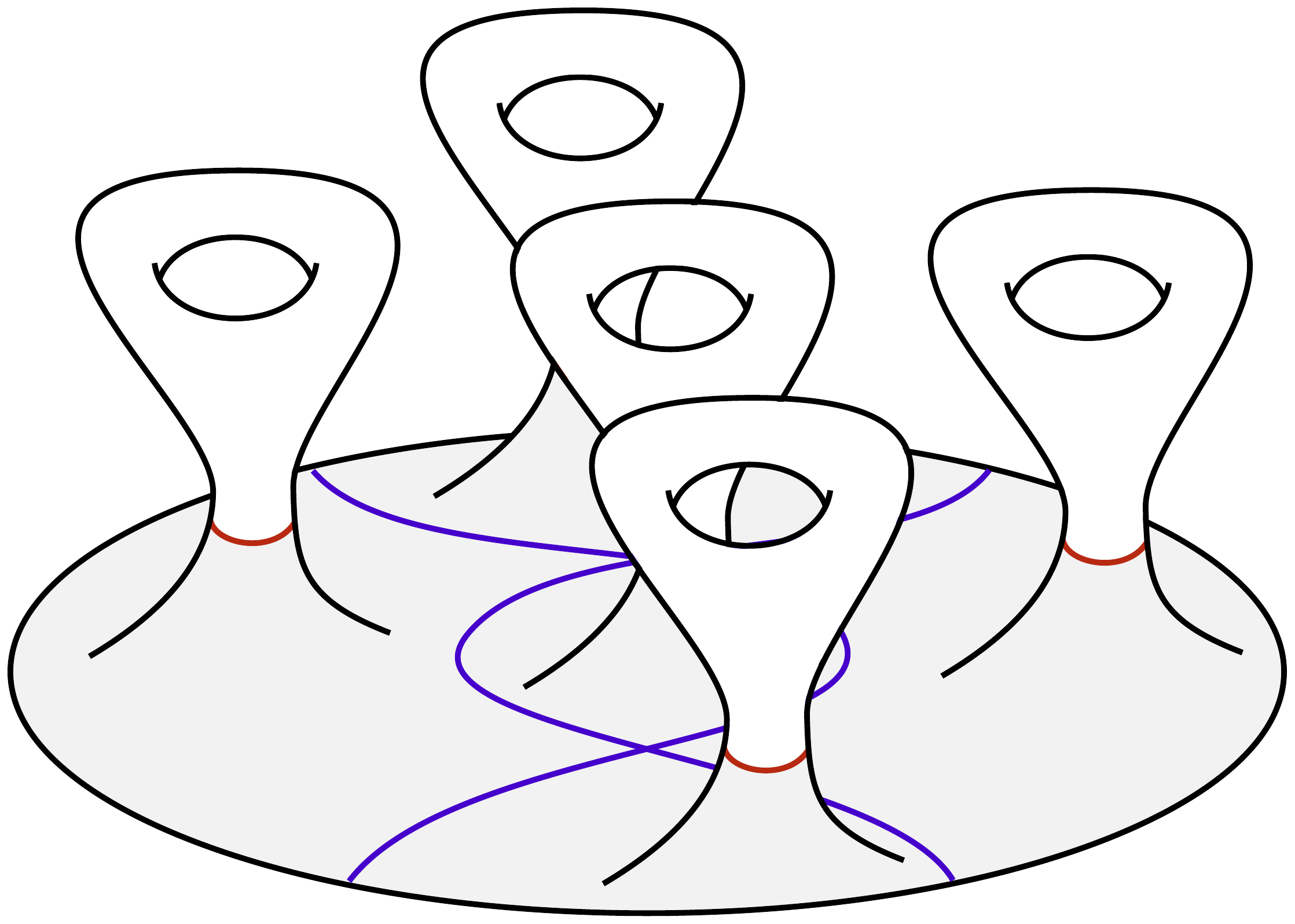}}\quad.\label{58}
\end{equation}
A class $\Gamma_i$ consists of all possible geometries K$_j$ ending on $\Gamma_i$ if we take $\Gamma_i$ to be the graph defined by the union of $\gamma_1,\gamma_2$ and the boundary of $\Sigma$. Topologically K$_j$ is identical to cutting $\Sigma$ on this graph.

Another example of a class involves just a $3$ holed sphere $\Gamma_i$
\begin{equation}
   \average{\Tr(e^{-\beta_1 H}\mo_1e^{-\beta_2 H}\mo_1e^{-\beta_3 H}\mo_2 e^{-\beta_4 H}\mo_2)}\supset\quad\raisebox{-10mm}{\includegraphics[width=50mm]{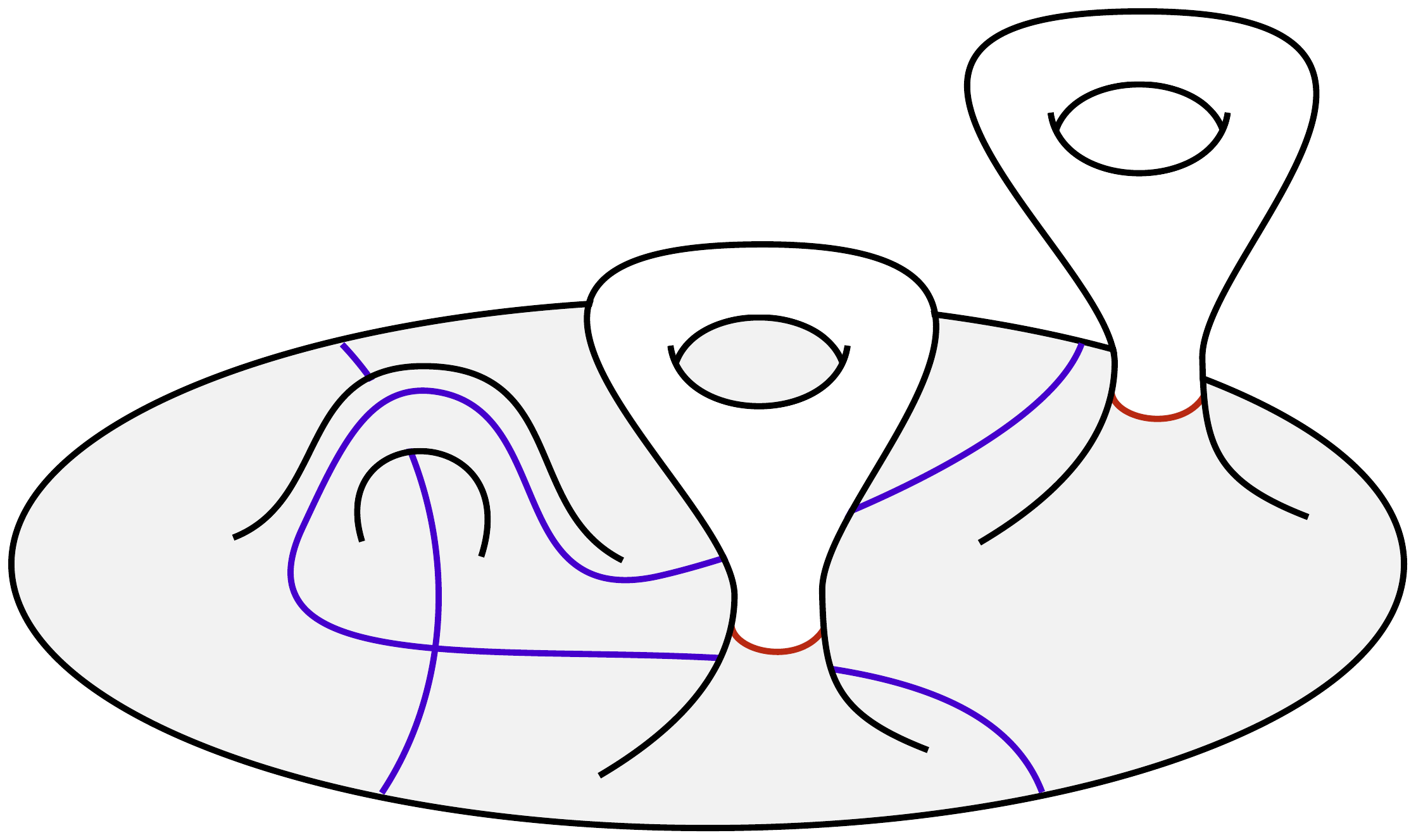}}\quad.\label{2.49}
\end{equation}
The thickening of $\gamma_1,\gamma_2$ and the boundary of $\Sigma$ indeed defines a $3$ holed sphere. Alternatively cutting $\Sigma$ on $\gamma_1$ and $\gamma_2$ one ends up with a surface with two circular boundaries. This class consist of all geometries K$_j$ ending on those two circular boundaries.
\\~\\
We pick one representative in each class as the simplest possible surface obtained as the special case where all boundaries of $\Gamma_i$ are contractable in $\Sigma$. For the class that includes \eqref{55} and \eqref{233} this is the genus zero $s$ channel $4$ point function \eqref{49}. For the class that includes \eqref{2.49} the representative is
\begin{equation}
   \average{\Tr(e^{-\beta_1 H}\mo_1e^{-\beta_2 H}\mo_1e^{-\beta_3 H}\mo_2 e^{-\beta_4 H}\mo_2)}\supset\quad\raisebox{-10mm}{\includegraphics[width=50mm]{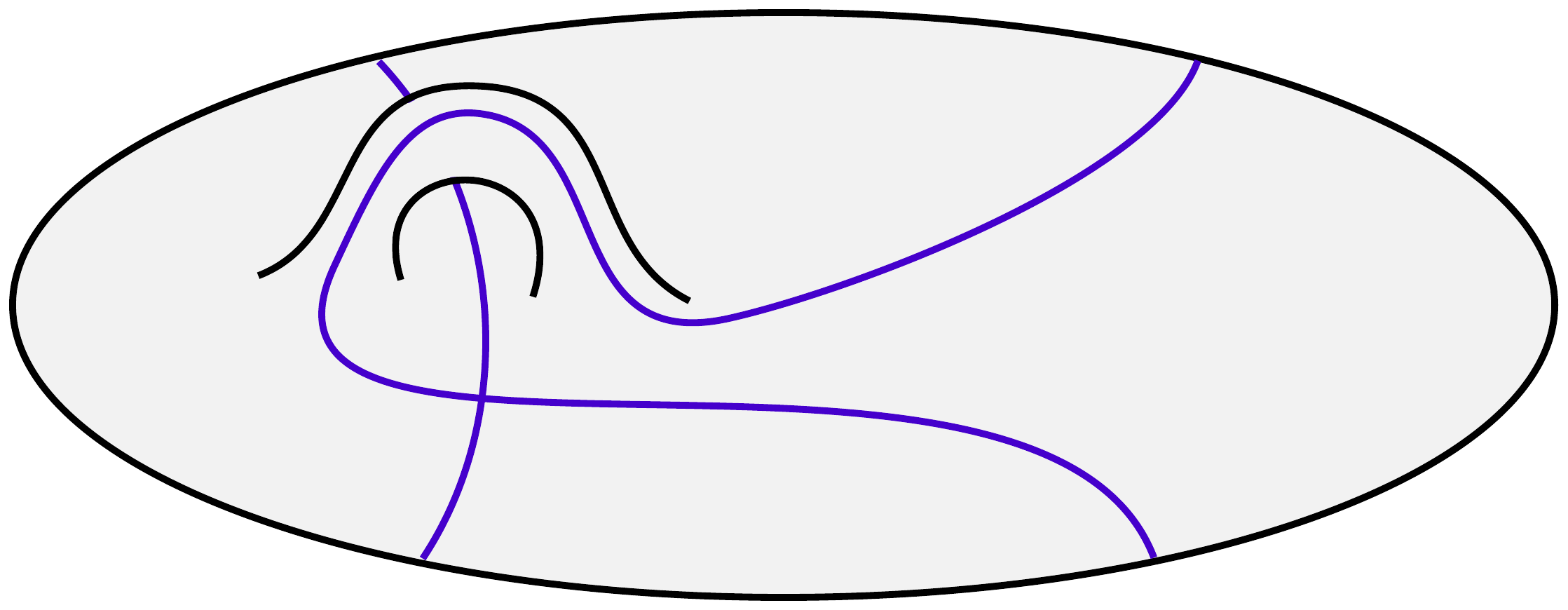}}\quad.\label{237}
\end{equation}
The class that includes \eqref{234} and \eqref{58} is represented by
\begin{equation}
   \average{\Tr(e^{-\beta_1 H}\mo_1e^{-\beta_2 H}\mo_1e^{-\beta_3 H}\mo_2 e^{-\beta_4 H}\mo_2)}\supset\quad\raisebox{-10mm}{\includegraphics[width=50mm]{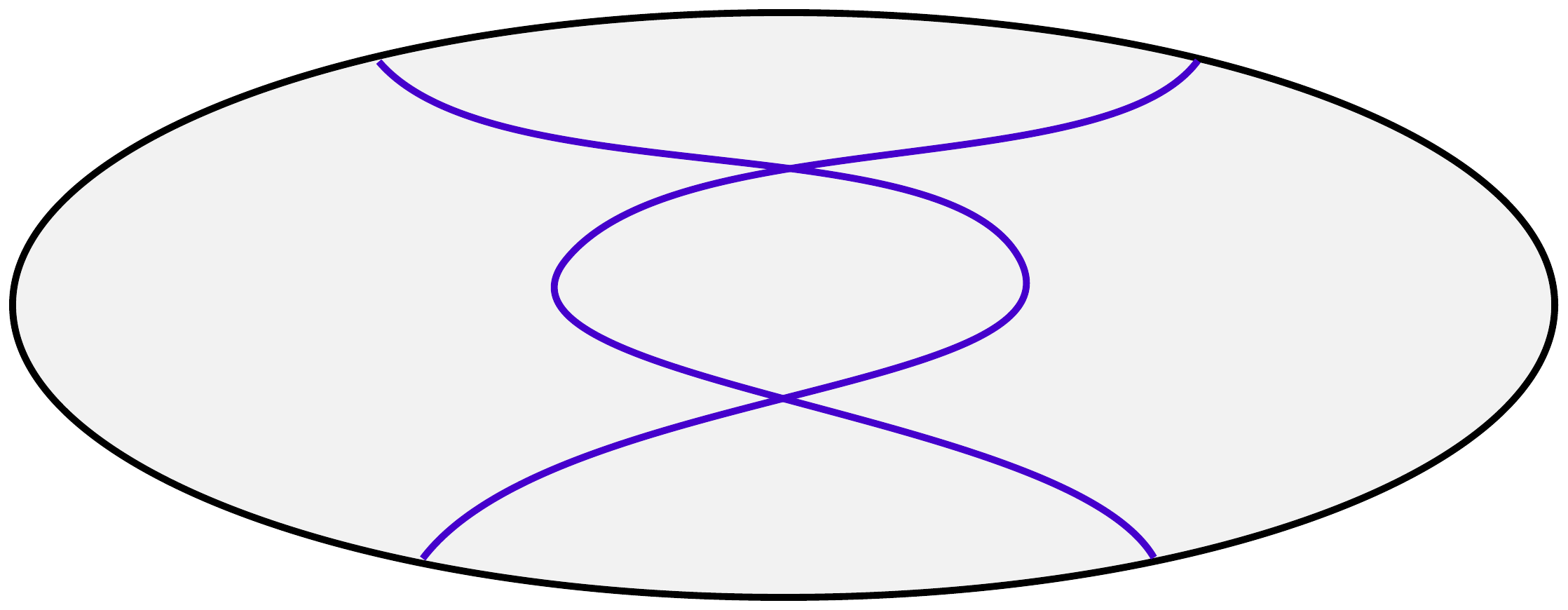}}\quad.\label{238}
\end{equation}
Note that this particular representative \eqref{238} is not separately to be counted in the sum over all diagrams in \eqref{230}. Gravitational Wilson lines may be continuously deformed through one another at no cost \cite{paper2,paper3}. So the amplitude \eqref{238} is actually equivalent to the bare $s$ channel disk amplitude \eqref{49} and we should be careful not to count it twice. The same is true for representatives with self-crossings
\begin{equation}
   \average{\Tr(e^{-\beta_1 H}\mo_1e^{-\beta_2 H}\mo_1e^{-\beta_3 H}\mo_2 e^{-\beta_4 H}\mo_2)}\supset\quad\raisebox{-10mm}{\includegraphics[width=50mm]{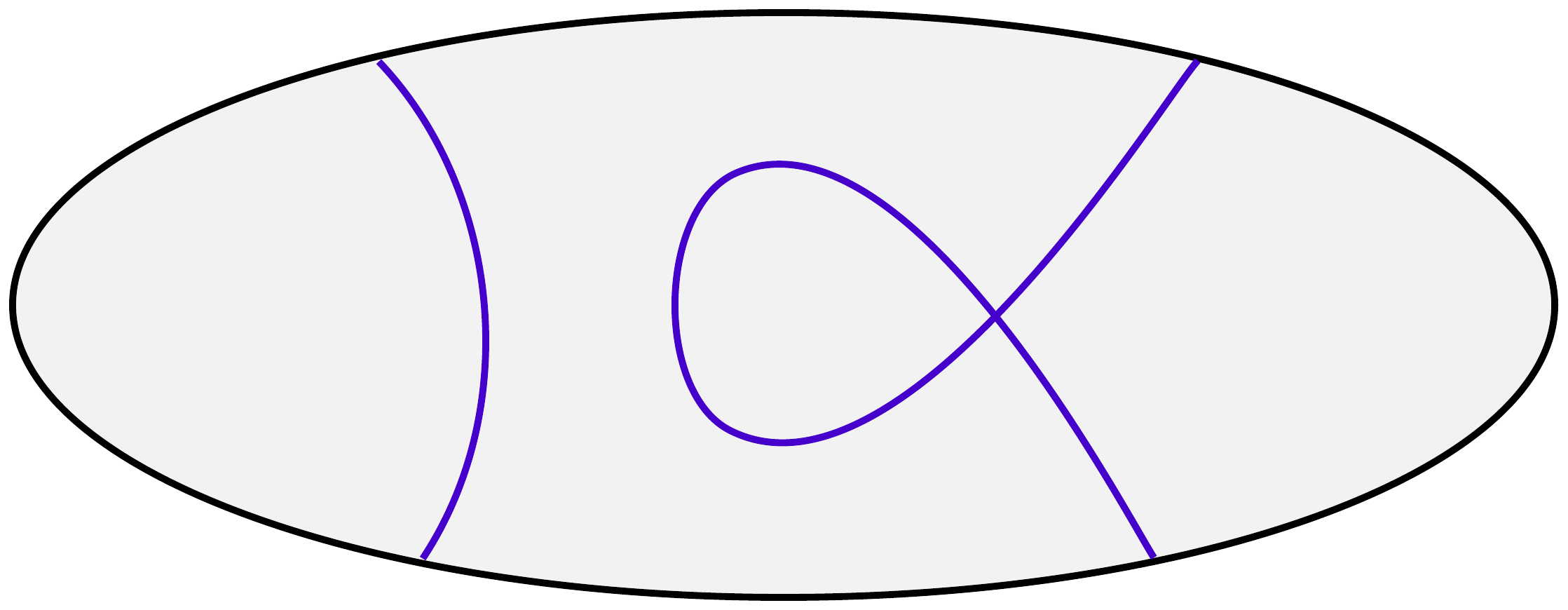}}\quad.\label{2.62}
\end{equation}
It is clear that within the set of all representatives there is a notion of simplicity. The genus zero time ordered $4$ point function \eqref{49} is simpler than \eqref{238} and \eqref{2.62} because it has no voluntary Wilson line crossings. Furthermore \eqref{49} cannot be simplified anymore. We refer to classes with the simplest possible representatives such as \eqref{49} and \eqref{237} as ``simple'' classes and classes whose representatives have voluntary crossings as ``complex'' classes. We will see that contributions of complex classes to observables are suppressed by powers of $e^{S_0}$ at all time scales hence they can be neglected.
\\~\\
The sum in \eqref{53} can be ordered into a sum over all classes $\Gamma_i$ associated with a given correlator. Furthermore within each class $\Gamma_i$ we sum over the families K$_j$ and within each family we sum over different geodesics $\gamma_1,\gamma_2$. For the complex classes we should avoid overcounting families that already contribute to simple classes.

This generalizes to other observables. What changes is the set of simple classes. For the two point function there is just one simple class with representative \eqref{1.7}. For the out of time ordered $4$ point function \eqref{42} there are $3$ simple classes whereas for the $4$ point function with all operators identical \eqref{23} there are $9$. It will become clear that there is an isomorphism between simple classes in bulk geometry and Wick contractions in random matrix theory.

To calculate the sum of all diagrams that contribute to a given correlator it is convenient to first take the sum of all contributions in a family and then sum over all families in each class.
%%%%%%%%%%%%%%%%%%%%%%%%%%%%%%%%%
\subsection{Families and the mapping class group}
Here we sum over all contributions in a family. This is the most technical part of this section. For reasons of presentation we focus on the class represented by the genus zero $s$ channel $4$ point function \eqref{49}. We want to sum over the amplitudes for the infinite set of geodesics $\gamma_1,\gamma_2$ within a class specified by $\Gamma_i$ and K$_j$ where K$_j$ has an arbitrary but fixed topology with $3$ boundary components
\begin{equation}
   \average{\Tr(e^{-\beta_1 H}\mo_1e^{-\beta_2 H}\mo_1e^{-\beta_3 H}\mo_2 e^{-\beta_4 H}\mo_2)}_{i j}\supset\quad\raisebox{-10mm}{\includegraphics[width=50mm]{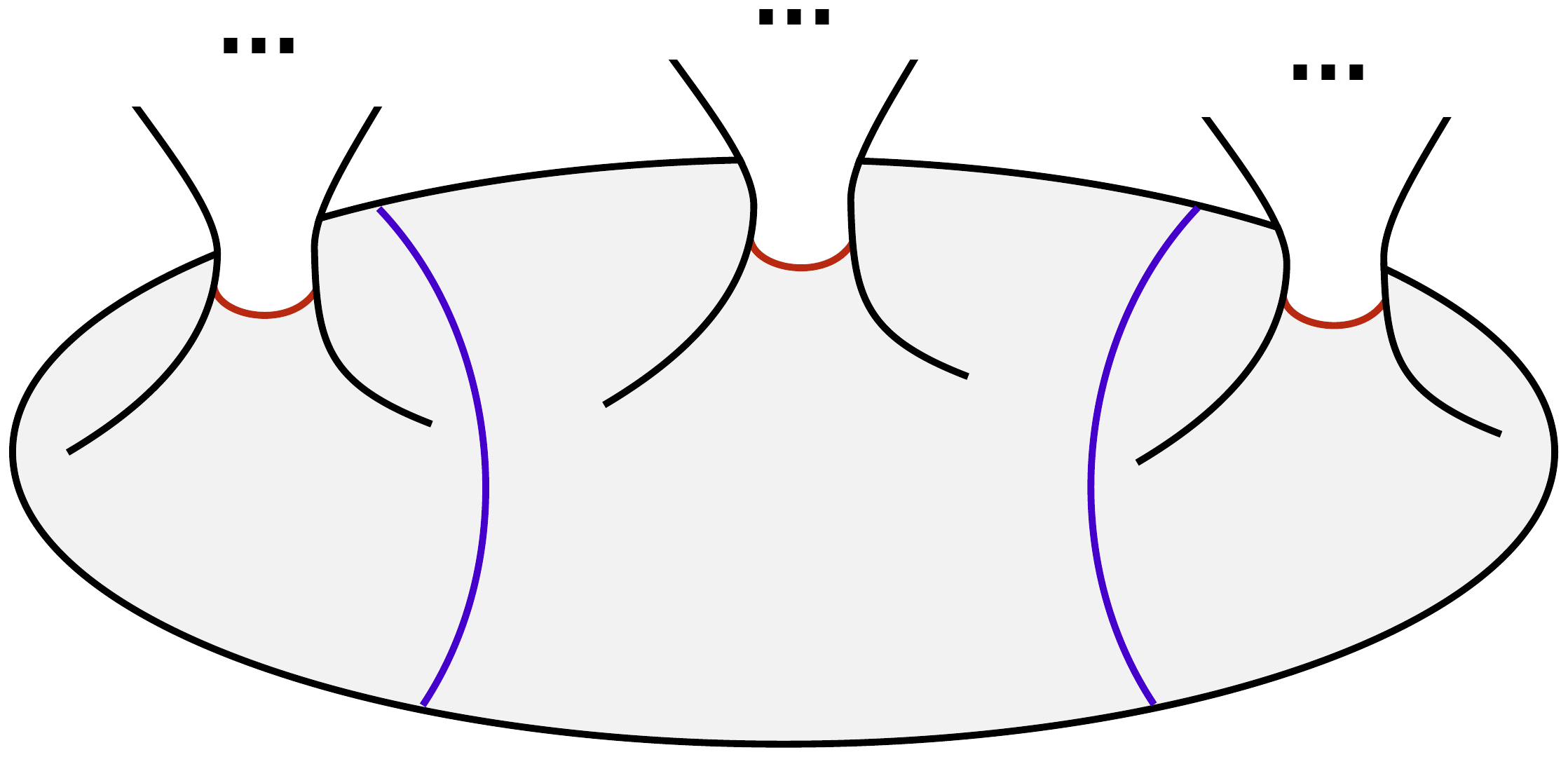}}\quad.\label{2.52}
\end{equation}

As a warm up consider the calculation of the JT gravity amplitude of topology $\Sigma$ where $\Sigma$ has a circular boundary of regularized length $\beta$. To compute the amplitude we chop $\Sigma$ into pieces. Consider a maximal set of non homologous and non intersecting geodesics $\gamma_i$ in $\Sigma$. One of these is homologous to the boundary of $\Sigma$. Denote the length of this geodesic by $b$ and the lengths of the internal geodesics by $a_i$. As mentioned in the introduction JT gravity has a first order formulation as an $\sltr$ BF theory \cite{J,paper3,paper4,lucahermanpufu}. By naively applying the rules for cutting and gluing amplitudes in BF theory the JT gravity amplitude decomposes as \cite{sss2,stanfordwitten2}
\begin{equation}
    \average{Z(\beta)}_\Sigma\overset{?}{=}\int_0^\infty db\, b\, Z(\beta,b)\,\prod_i \int_0^\infty d a_i\,a_i\,.\label{241}
\end{equation}
Here $Z(\beta,b)$ is the JT gravity amplitude of an annulus topology with a wiggly boundary of regularized length $\beta$ and a geodesic boundary of length $b$ \cite{defects,sss2}
\begin{equation}
    Z(\beta,b)=\int_0^\infty  d E\,e^{-\beta E}\int_0^\infty d b\,b\,\frac{\cos \pi b \sqrt{E}}{\sqrt{E}}=\quad \raisebox{-5mm}{\includegraphics[width=23mm]{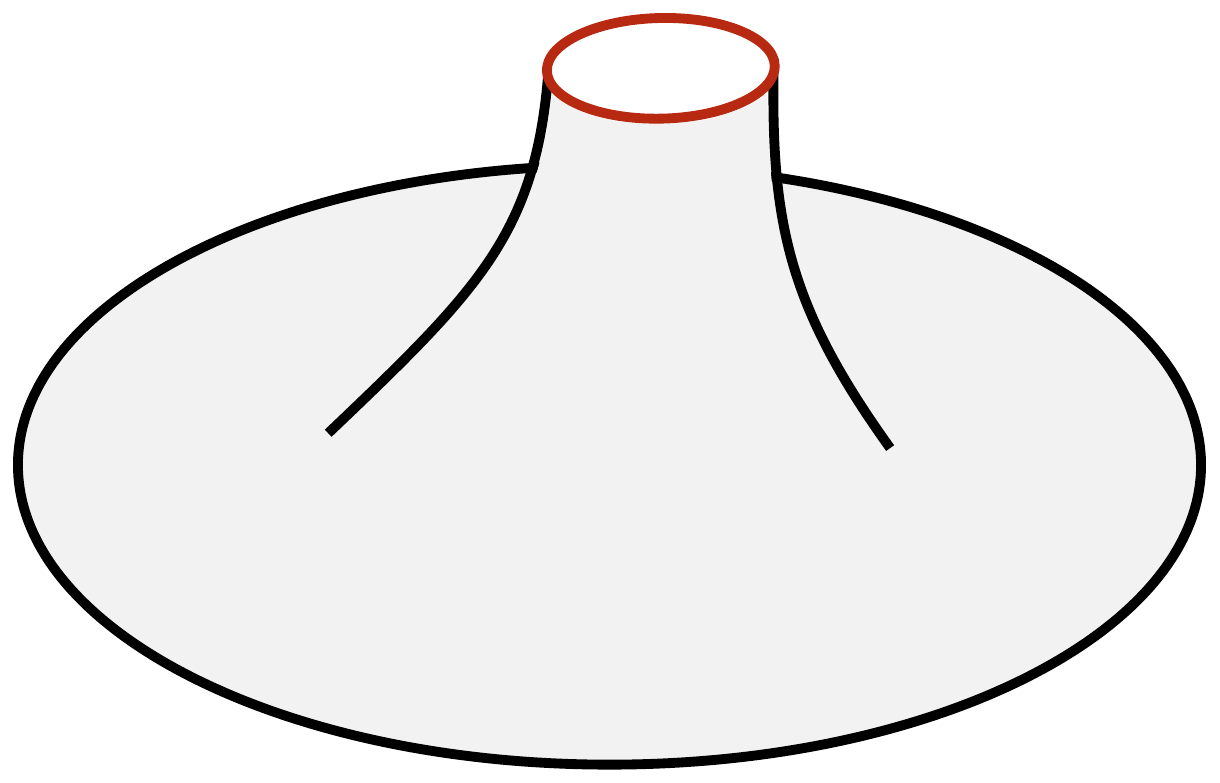}}\quad\,.\label{2.54}
\end{equation}

The right hand side of \eqref{241} is divergent so it is not quite correct. The issue is that we have integrated over the moduli space of inequivalent flat hyperbolic $\sltr$ connections. In gravity we should integrate only over inequivalent Riemann surfaces \cite{dijkgraafwitten}. The space of Riemann surface is the space of flat hyperbolic $\sltr$ connections modulo the mapping class group \cite{dijkgraafwitten}. Equivalently the space of flat hyperbolic $\sltr$ connections is the space of Riemann surfaces dressed with Moore-Seiberg graphs \cite{msclasquant}. In the current context elements of the mapping class group are identified with different choices for the complete set of geodesics on $\Sigma$. Let us denote the mapping class group of $\Sigma$ by $\text{MCG}(\Sigma)$ and the union of all inequivalent complete sets of geodesics on $\Sigma$ by $\text{M}(\gamma_i)$
\begin{equation}
    \text{MCG}(\Sigma)=\text{M}(\gamma_i).\label{159}
\end{equation}
Elements of the mapping class group map one complete set of geodesic to another such set. For example on the torus it maps any given cycle to any other cycle.

The correct formula replacing \eqref{241} is obtained by dividing out by the volume of the mapping class group. Now we are only counting inequivalent Riemann surfaces
\begin{equation}
    \average{Z(\beta)}_\Sigma=\int_0^\infty db\, b\, Z(\beta,b)\,\frac{1}{\text{MCG}(\Sigma)}\,\prod_i \int_0^\infty d a_i\,a_i\,.\label{2.53}
\end{equation}
In this expression both the nominator and denominator are infinite but their ratio is finite. Computing this ratio from first principles is difficult yet feasible \cite{mirza1,mirza2,dijkgraafwitten,stanfordwitten2}. The result is a Weil-Petersson volume that depends on the genus of $\Sigma$
\begin{equation}
    \frac{1}{\text{MCG}(\Sigma)}\,\prod_i \int_0^\infty d a_i\,a_i=V_\Sigma(b).\label{2129}
\end{equation}
Combining this with \eqref{2.54} we find \cite{sss2}
\begin{equation}
    \average{Z(\beta)}_\Sigma =\int_0^\infty d E\,e^{-\beta E}\,\rho_\Sigma(E)\quad,\quad \rho_\Sigma(E)=\int_0^\infty d b\,b\,\frac{\cos \pi b \sqrt{E}}{\sqrt{E}}\,V_\Sigma(b)\,.\label{2.58}
\end{equation}
The generalization to the $n$ level spectral density $\rho_\Sigma(E_1\dots E_n)$ is the integral of $V_\Sigma(b_1\dots b_n)$ over $b_1\dots b_n$ with a similar kernel as in \eqref{2.58} for each of the integrals. The total spectral density $\rho(E_1\dots E_n)$ is obtained by summing $\rho_\Sigma(E_1\dots E_n)$ over all topologies $\Sigma$ with $n$ fixed length boundaries.
\\~\\
Using similar cutting and gluing reasoning we can wield $\sltr$ BF techniques to compute the contribution to the amplitude \eqref{2.52} due to each of the geodesic embeddings $\gamma_1,\gamma_2$ in this family. To do so we imagine first cutting on the geodesics $\kappa_1,\kappa_2$ and $\kappa_3$. Let us denote the associated geodesic lengths by $b_1,b_2$ and $b_3$. 

For each choice of $\gamma_1,\gamma_2$ within the same family we are left with identical configurations $\Gamma_i$ and K$_j$ after these first cuts. We then further chop K$_j$ into pieces by cutting on a set of geodesics $\gamma_k$ with lengths $a_k$. The resulting formula is in spirit similar to \eqref{2.53}
\begin{align}
\label{247}
   &\average{\Tr(e^{-\beta_1 H}\mo_1e^{-\beta_2 H}\mo_1e^{-\beta_3 H}\mo_2 e^{-\beta_4 H}\mo_2)}_{i j}\\\nonumber& \qquad=\sum_{\gamma_1 \gamma_2}\prod_j \int_0^\infty d b_j\,b_j\,\average{\Tr(e^{-\beta_1 H}\mo_1e^{-\beta_2 H}\mo_1e^{-\beta_3 H}\mo_2 e^{-\beta_4 H}\mo_2)}_{i\, b_1\, b_2\, b_3}\\&\qquad\qquad\qquad\qquad\qquad\qquad\qquad\qquad\qquad\qquad\qquad\frac{1}{\text{MCG}(\Sigma)}\,\prod_k \int_0^\infty d a_k\,a_k\,.\nonumber
\end{align}
Here the first factor in the integrand denotes the JT gravity amplitude on $\Gamma_i$ with the lengths of the geodesic boundaries fixed to $b_1,b_2$ and $b_3$
\begin{equation}
  \average{\Tr(e^{-\beta_1 H}\mo_1e^{-\beta_2 H}\mo_1e^{-\beta_3 H}\mo_2 e^{-\beta_4 H}\mo_2)}_{i\, b_1\, b_2\, b_3}=\quad\raisebox{-10mm}{\includegraphics[width=50mm]{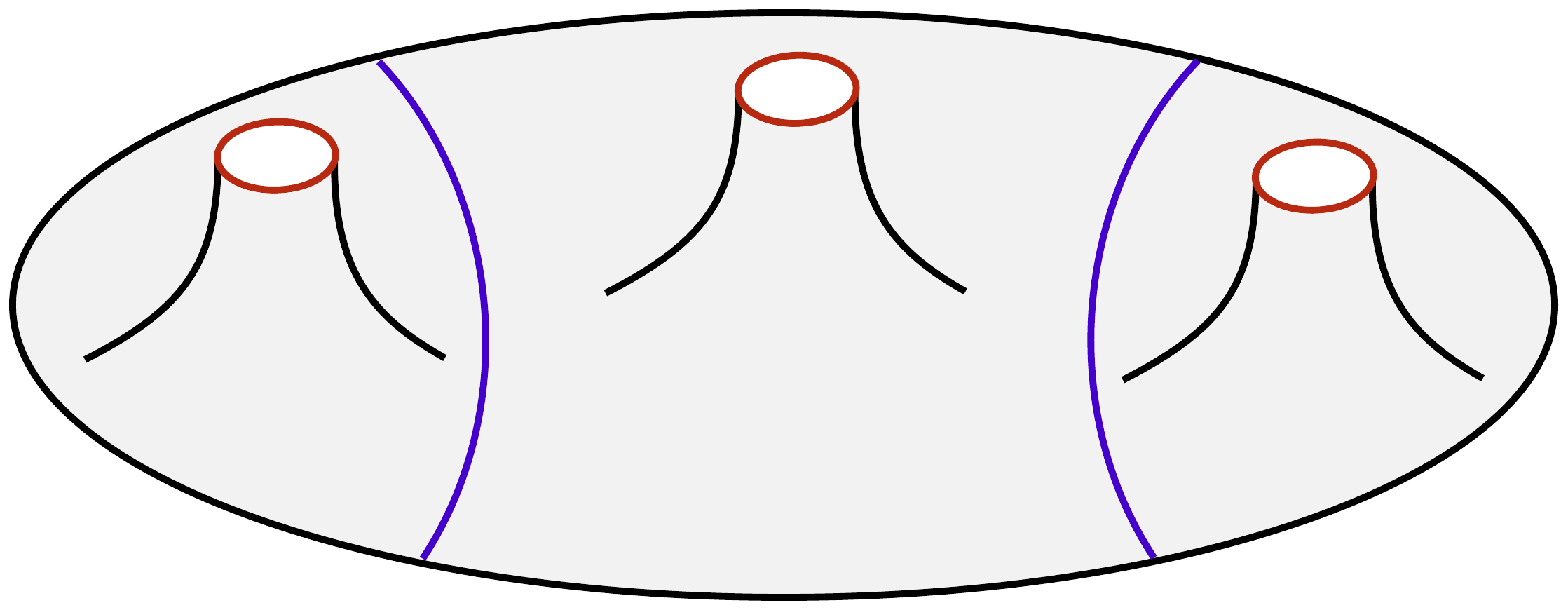}}\,\label{2.60}
\end{equation}
Note that the contributions on the second line of \eqref{247} do not combine nicely into a Weil-Petersson volume. This is because we are modding by the mapping class group of $\Sigma$ but we are only integrating over the naive moduli of K$_j$. The nontrivial kernel for the integral over $b_1,b_2$ and $b_3$ impedes us from using \eqref{2129}. But then how do we make sense of these formal infinities?

What saves the day is that the expression after the sum in \eqref{247} is independent of the choice of geodesics $\gamma_1,\gamma_2$. This is only true because we are working within one family. Let us denote union of all sets of geodesics $\gamma_1,\gamma_2$ in the family specified by $\Gamma_i$ and K$_j$ by M$(\Gamma_{i j})$. From the definition of the mapping class group it follows that for each fixed family the elements of the mapping class group decompose as \cite{stanfordwitten2}
\begin{equation}
    \text{MCG}(\Sigma)=\text{M}(\Gamma_{i j})\,\text{MCG}(K_j)\,.\label{2.61}
\end{equation}
Crucially there is no sum over $i$ nor $j$. We can use this to cancel the sum over $\gamma_1,\gamma_2$ in \eqref{247} with part of the mapping class group of $\Sigma$. We now recognize the Weil-Petersson volume of K$_j$
\begin{equation}
    \frac{1}{\text{MCG}(\text{K}_j)}\,\prod_k \int_0^\infty d a_k\,a_k=V_{\text{K}_j}(b_1,b_2,b_3).\label{2.63}
\end{equation}
We arrive at
\begin{align}
\label{2.59}
   &\average{\Tr(e^{-\beta_1 H}\mo_1e^{-\beta_2 H}\mo_1e^{-\beta_3 H}\mo_2 e^{-\beta_4 H}\mo_2)}_{i j}\\\nonumber&\qquad =\prod_j \int_0^\infty d b_j\,b_j\,\average{\Tr(e^{-\beta_1 H}\mo_1e^{-\beta_2 H}\mo_1e^{-\beta_3 H}\mo_2 e^{-\beta_4 H}\mo_2)}_{i\, b_1\, b_2\, b_3}\,V_{\text{K}_j}(b_1,b_2,b_3)\,.
\end{align}

There are no formal infinities left. This is the key identity of this subsection. It provides an alternative perspective on the whole setup. An equivalent starting point is to sum over just one representative of each family and mod just by the mapping class group of K$_j$. The Wilson lines or massive probe particles in \eqref{2.52} are partially dressing the Riemann surface with a Moore-Seiberg graph. For this configuration only the mapping class group of K$_j$ is redundant. In terms of gauge theories we have two equivalent options. Either we consider just the mapping class group of K$_j$ as redundant or we view the whole mapping class group of $\Sigma$ as redundant. In the second case the path integral instructs us to construct a gauge invariant observable by summing over gauge orbits. This sum corresponds to the sum over all $\gamma_1,\gamma_2$ within one class in \eqref{247}. The perspective where we consider the mapping class group of K$_j$ as redundant was used implicitly in \cite{paper5}. 

Note that a similar line of reasoning and a property similar to \eqref{2.61} is used to derive a recursion relation for Weil-Petersson volumes starting from the sum rule \cite{mirza1,mirza2,stanfordwitten2}. This is explained in a particularly accessible manner in appendix D of \cite{stanfordwitten2} to which we refer the skeptical reader.
\\~\\
We still need to calculate \eqref{2.60}. Because the mapping class group redundancy is already dealt with we can compute this using standard $\sltr$ BF techniques \cite{paper3,paper4,defects}. These are an application of the rules for calculating 2d Yang-Mills amplitudes \cite{witten2dgt,paper2}
\begin{align}
\displaybreak[0]
     &\nonumber \average{\Tr(e^{-\beta_1 H}\mo_1e^{-\beta_2 H}\mo_1e^{-\beta_3 H}\mo_2 e^{-\beta_4 H}\mo_2)}_{i b_1 b_2 b_3}\\\nonumber &\qquad =e^{-4S_0}\int_0^\infty d E_1\, e^{-(\beta_1+\beta_3) E_1}\int_0^\infty d E_2\,e^{-\beta_2 E_2}\int_0^\infty d E_4\,e^{-\beta_4 E_4}\\&\qquad\qquad\qquad\qquad\qquad\qquad\qquad\mo_1\mo_1(E_1,E_2)\mo_2\mo_2(E_1,E_4)\,\prod_j \frac{\cos \pi b_j \sqrt{E_j}}{\sqrt{E_j}}\,.\label{252}
\end{align}
Notice that this does not depend on K$_j$ which only enters via its Weil-Petersson volume. The integrals over $b_1,b_2$ and $b_3$ in \eqref{2.59} combine to a contribution to the $3$ level spectral density $\rho_{K_j}(E_1,E_2,E_3)$ due to the topology K$_j$
\begin{align}
\displaybreak[0]
     \nonumber&\average{\Tr(e^{-\beta_1 H}\mo_1e^{-\beta_2 H}\mo_1e^{-\beta_3 H}\mo_2 e^{-\beta_4 H}\mo_2)}_{i j}\\\nonumber&\qquad= e^{-4S_0}\int_0^\infty d E_1\,e^{-(\beta_1+\beta_3) E_1}\int_0^\infty d E_2\,e^{-\beta_2 E_2}\int_0^\infty d E_4\,e^{-\beta_4 E_4}\\&\qquad\qquad\qquad\qquad\qquad\qquad\qquad\rho_{K_j}(E_1,E_2,E_3)\,\mo_1\mo_1(E_1,E_2)\mo_2\mo_2(E_1,E_4)\,.\label{2.64}
\end{align}
This can be compared to \eqref{2.39} where K$_j$ is trivial. Note that similar reasoning can be used to calculate the amplitudes associated with the representatives of each class. For example
\begin{align}
&\raisebox{-10mm}{\includegraphics[width=50mm]{tax7.pdf}}\quad\\&\quad=e^{-4S_0}\int_\cas d E_1\,e^{-(\beta_1+\beta_2+\beta_4)E_1}\int_\cas d E_3\,e^{-\beta_3 E_3}\,\rho_0(E_1)\rho_0(E_3)\,\mo_1\mo_2\mo_1\mo_2(E_1,E_1,E_3,E_1)\,.\nonumber   
\end{align}
In this case the surface K$_j$ is trivial and so this reduces in its entirety to a standard $\sltr$ BF calculation after summing over all geodesics $\gamma_1,\gamma_2$ in the same family. Remember that there are hidden powers of $e^{S_0}$ in the integration kernels.
%%%%%%%%%%%%%%%%%%%%%%%%%%%%%%%%%
\subsection{Summing over geometries}
This generalizes immediately to generic families. Let us now sum over all families in a class. For example summing over all K$_j$ in \eqref{2.64} we recognize
\begin{equation}
    \sum_j \rho_{\text{K}_j}(E_1,E_2,E_3)=\rho(E_1,E_2,E_3).\label{2.67}
\end{equation}
Indeed this sum is identical to summing over all geometries ending on $3$ fixed energy boundaries. This genus expansion is asymptotic. We define JT gravity nonperturbatively as the double scaled matrix integral of \cite{sss2} so the explicit formula to be used for $\rho(E_1,E_2,E_3)$ is the universal matrix integral answer \eqref{66}. In summary the sum over the class $\Gamma_i$ gives
\begin{align}
\displaybreak[0]
     \nonumber&\average{\Tr(e^{-\beta_1 H}\mo_1e^{-\beta_2 H}\mo_1e^{-\beta_3 H}\mo_2 e^{-\beta_4 H}\mo_2)}_{i}\\\nonumber&\qquad= e^{-4S_0}\int_0^\infty d E_1\,e^{-(\beta_1+\beta_3) E_1}\int_0^\infty d E_2\,e^{-\beta_2 E_2}\int_0^\infty d E_4\,e^{-\beta_4 E_4}\\&\qquad\qquad\qquad\qquad\qquad\qquad\qquad\rho(E_1,E_2,E_3)\,\mo_1\mo_1(E_1,E_2)\mo_2\mo_2(E_1,E_4)\,.\label{2.68}
\end{align}
Note that this precisely matches one of the $3$ contributions in the random matrix prediction for the $s$ channel or time ordered $4$ point function \eqref{42}.

We can compare this to the genus zero $s$ channel $4$ point function \eqref{2.39} which is the representative of this class. The difference is that we have replaced the Schwarzian answer for the $3$ level spectral density $\rho_0(E_1)\rho_0(E_2)\rho_0(E_3)$ with the answer from random matrix theory $\rho(E_1,E_2,E_3)$. Given the previous discussion it should be obvious that this is true for any class contributing to any correlator. For example the $u$ channel $4$ point function has a class with representative
\begin{equation}
   \average{\Tr(e^{-\beta_1 H}\mo_1e^{-\beta_2 H}\mo_2e^{-\beta_3 H}\mo_1 e^{-\beta_4 H}\mo_2)}_i\supset\quad\raisebox{-10mm}{\includegraphics[width=50mm]{sum3.pdf}}\quad.\label{2.69}
\end{equation}
This genus zero JT gravity path integral gives \cite{paper3,paper4,paper5,altland,altland2,schwarzian,origins,kitaevsuh,zhenbin,lucahermanpufu}.
\begin{align}
\displaybreak[0]
    \label{38}&\average{\Tr(e^{-\beta_1 H}\mo_1e^{-\beta_2 H}\mo_2 e^{-\beta_3 H}\mo_1e^{-\beta_4 H}\mo_2)}_i\\&\quad\nonumber \supset e^{-4 S_0}\int_\cas d E_1\,e^{-\beta_1 E_1}\int_\cas d E_2\,e^{-\beta_2 E_2}\int_\cas d E_3\,e^{-\beta_3 E_3}\int_\cas d E_4\,e^{-\beta_4 E_4}\\&\nonumber \qquad\qquad\qquad\qquad\qquad\qquad\quad\rho_0(E_1)\rho_0(E_2)\rho_0(E_3)\rho_0(E_4)\,\mo_1\mo_2\mo_1\mo_2(E_1,E_2,E_3,E_4)\,.
\end{align}
Summing over all geometries K$_j$ ending on this topology we obtain
\begin{align}
\displaybreak[0]
    \label{38}&\average{\Tr(e^{-\beta_1 H}\mo_1e^{-\beta_2 H}\mo_2 e^{-\beta_3 H}\mo_1e^{-\beta_4 H}\mo_2)}_i\\&\quad\nonumber =e^{-4 S_0}\int_\cas d E_1\,e^{-\beta_1 E_1}\int_\cas d E_2\,e^{-\beta_2 E_2}\int_\cas d E_3\,e^{-\beta_3 E_3}\int_\cas d E_4\,e^{-\beta_4 E_4}\\&\nonumber \qquad\qquad\qquad\qquad\qquad\qquad\quad\rho(E_1,E_2,E_3,E_4)\,\mo_1\mo_2\mo_1\mo_2(E_1,E_2,E_3,E_4)\,.
\end{align}
This precisely matches one of the three contributions to the random matrix prediction for the $u$ channel $4$ point function \eqref{38}. 
\\~\\
The following comment is rather technical and can be skipped on a first reading. 

Amplitudes of complex classes come with additional dummy energy labels $M_1\dots M_n$ due to enclosed regions. Summing over all geometries we find an expression that contains a kernel $\rho(\dots M_1\dots M_n)$ and some smooth factors due to the operators. Integrals over the external energies $\dots$ are weighed by Laplace transform kernels which contributes highly oscillatory phases at late times. Higher genus and random matrix contributions to correlators are only important at late times because the late time Laplace transform probes for highly oscillatory features in $\rho(\dots M_1\dots M_n)$. This zooms in on regions in the integration domain where random matrix corrections thrive. There are no such Laplace transform kernels for the integrals over the dummy energies and so there is nothing to force these dummy integrals to zoom in on the regions of the integrand where contributions due to higher genus geometries connecting to the enclosed regions are important. This means we can replace the kernel $\rho(\dots M_1\dots M_n)$ by the genus zero kernel $\rho(\dots)\rho_0(M_1)\dots \rho_0(M_n)$ without affecting the integral up to subleading terms in $e^{S_0}$.

However the geometries which give rise to the kernel $\rho(\dots )\rho_0(M_1)\dots \rho_0(M_n)$ are not to be considered as contributing to the complex class in the first place. The enclosed regions are capped of by disks and so we can undo the voluntary crossings. These geometries are already counted in a simple class. The only contribution due to complex classes are the subleading terms in $e^{S_0}$ which we ignored previously. This argument holds on all time scales and so we may ignore complex classes for all intends and purposes. 
%%%%%%%%%%%%%%%%%%%%%%%%%%%%%%%%
\subsection{Final answer}
We can calculate a correlation function by summing the answer of all the simple classes that contribute to it. We will stick to working by example. Let us denote in this subsection the contribution of a whole class by just its representative genus zero diagram. In case of the $s$ channel or time ordered $4$ point function one identifies $3$ simple classes
\begin{align}
\displaybreak[0]
   \label{2.72}&\average{\Tr(e^{-\beta_1 H}\mo_1e^{-\beta_2 H}\mo_1e^{-\beta_3 H}\mo_2 e^{-\beta_4 H}\mo_2)}=\quad\raisebox{-10mm}{\includegraphics[width=50mm]{intro6.pdf}}\quad \\&\qquad\qquad\qquad\qquad\quad+\quad\raisebox{-10mm}{\includegraphics[width=50mm]{tax7.pdf}}\quad+\quad\raisebox{-10mm}{\includegraphics[width=50mm]{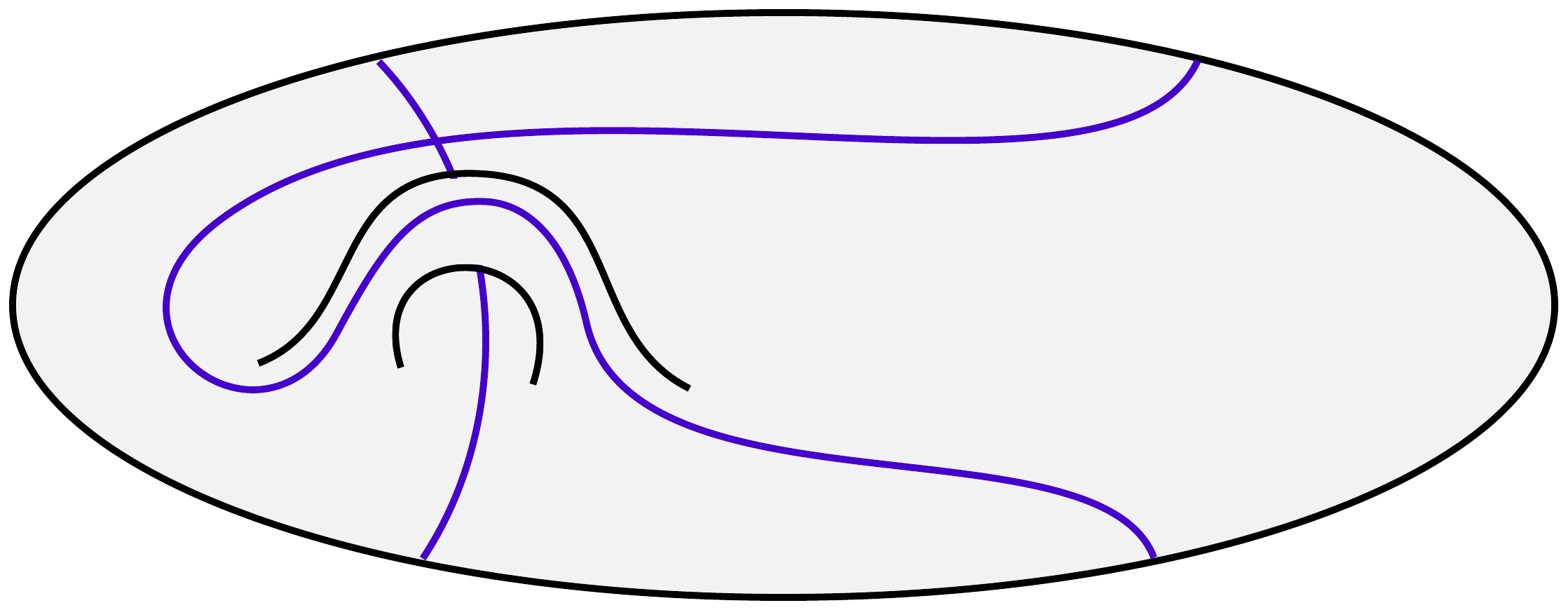}}\quad.\nonumber
\end{align}
Given the previous discussion it is elementary to check that the contributions due to each of these classes matches precisely to the $3$ contributions predicted from random matrix theory in \eqref{42}. Note that the second and third term in \eqref{42} are proportional to $e^{-S_0}$ whilst the first term in \eqref{42} is proportional to $e^{S_0}$. From the geometrical point of view this follows from the different Euler characteristic of the disk and the disk with a handle in \eqref{2.72}.

Similarly for the $s$ channel or out of time ordered $4$ point function there are $3$ simple classes
\begin{align}
\displaybreak[0]
   \label{261}&\average{\Tr(e^{-\beta_1 H}\mo_1e^{-\beta_2 H}\mo_1e^{-\beta_3 H}\mo_2 e^{-\beta_4 H}\mo_2)}=\quad\raisebox{-10mm}{\includegraphics[width=50mm]{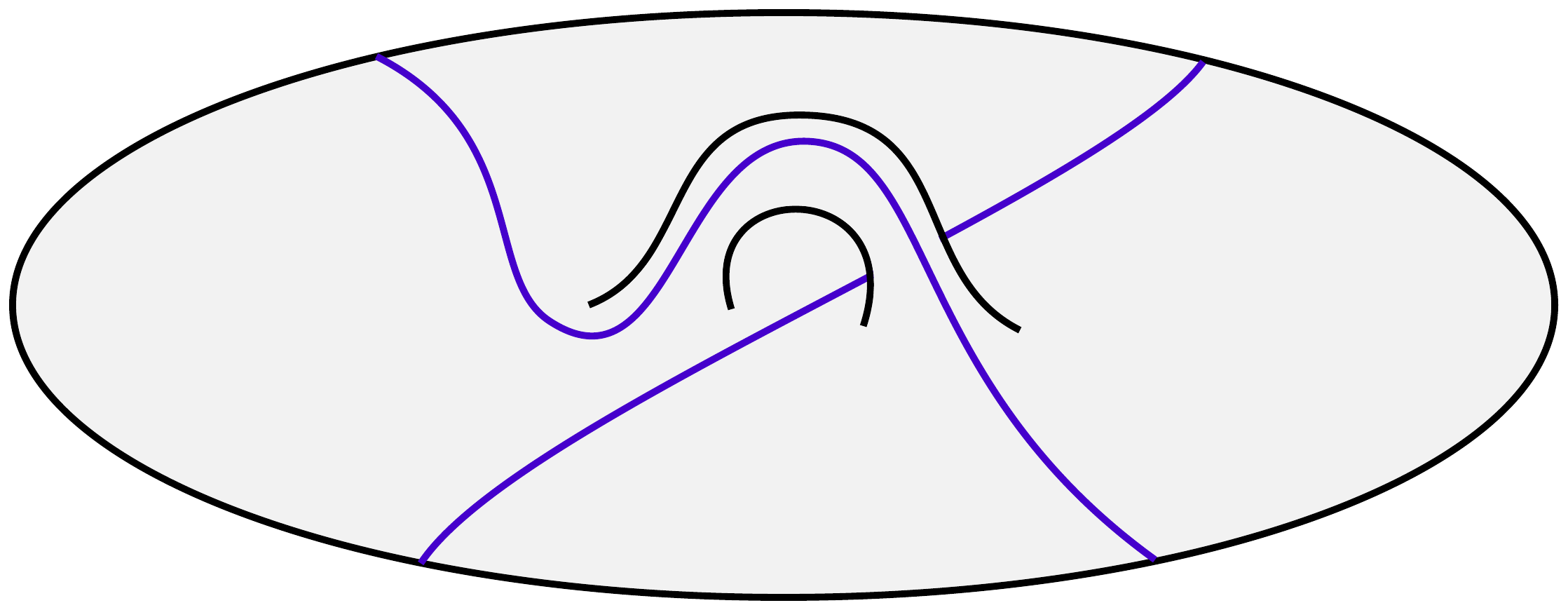}}\quad \\&\qquad\qquad\qquad\qquad\quad+\quad\raisebox{-10mm}{\includegraphics[width=50mm]{sum3.pdf}}\quad+\quad\raisebox{-10mm}{\includegraphics[width=50mm]{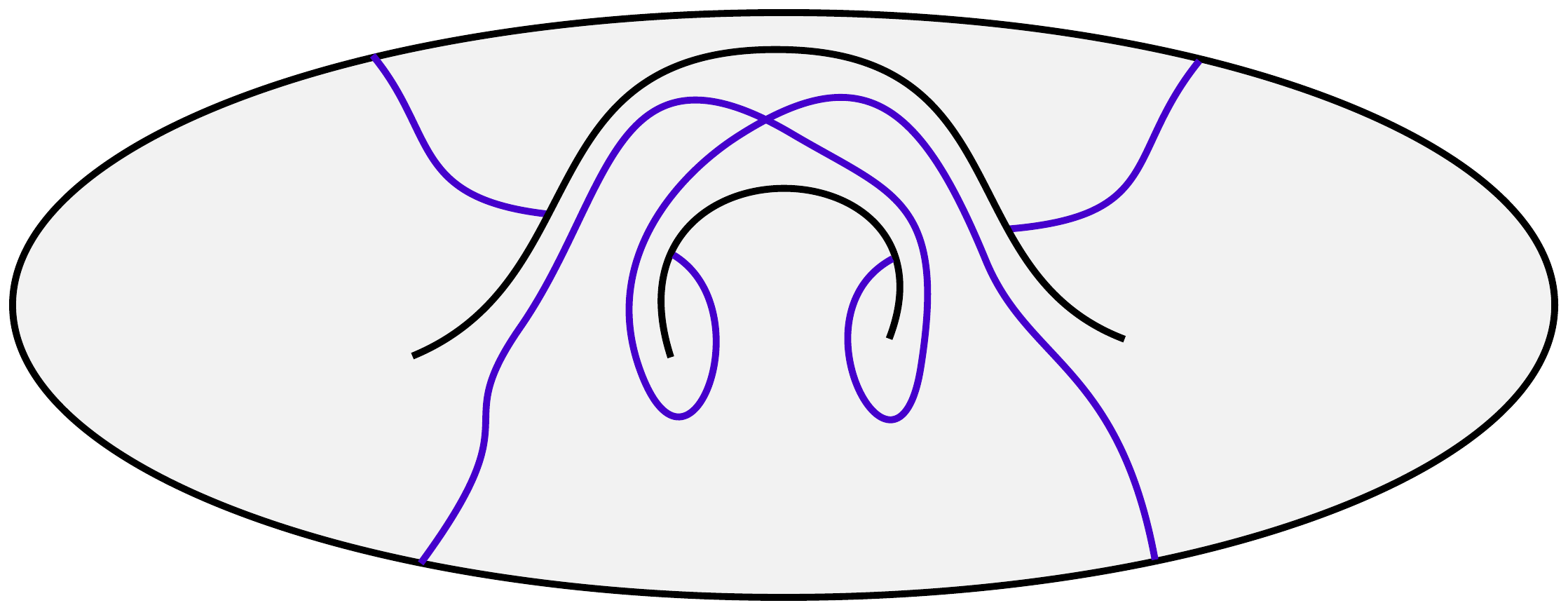}}\quad.\nonumber
\end{align}
We can sum the gravitational contributions for each of these classes. One checks that these match one by one to the $3$ terms predicted by random matrix theory in \eqref{225}. For example one finds that the second term in \eqref{225} is proportional to $e^{S_0}$ whereas the other two terms are proportional to $e^{-S_0}$. this is consistent with the geometries in \eqref{261}.

The rules to calculate boundary correlators in JT gravity are quite simple in summary. 
\begin{enumerate}
    \item Identify all simple classes that contribute to a correlator by drawing all representatives of simple classes. To check if a diagram is the representative of a simple class we cut the geometry on the particle trajectories. It is the representative of a simple class if the resulting surface is a product of disks and if there are no voluntary crossings.
    \item Compute the amplitude of each of the representatives of simple classes using standard techniques \cite{paper3,paper4}. Replace in the answer the genus zero spectral densities by a multi level spectral density in random matrix. Sum the results of all simple classes.
\end{enumerate}
In practice it is more convenient to do the calculation directly in random matrix theory where the simple classes follow immediately from Wick contractions. From the geometrical point of view it is less straightforward to determine if all simple classes have been found.

Note that at early and intermediate timescales the dominant contribution is due to simple classes of which the representative has no handles. For the $s$ channel $4$ point function this is the first class in \eqref{2.72} and for the $u$ channel $4$ point function it is the second class in \eqref{261}. This is not in general obvious at exponentially late times where overall powers of $e^{S_0}$ can change. Nevertheless one checks that even the plateau height in both the $s$ and $u$ channel $4$ point functions is dominated by these same genus zero classes.
%%%%%%%%%%%%%%%%%%%%%%%%%%%%%%%%
\subsection{Bulk reconstruction first}\label{ss:bulk}
Bulk points and bulk operators in quantum gravity are naturally defined in Lorentzian signature. One important aspect of quantum gravity is that bulk points and bulk operators are to be defined in a diff invariant manner, such that they are sensible objects in a theory with diff redundancy \cite{rec1,rec2,rec3,rec4}. Such a construction was worked out and built upon for JT gravity in \cite{paper5,thomassingle,paper201,wopjordan}. The idea is that a point in the Lorentzian bulk is uniquely defined by specifying two physical boundary times from which it is lightlike separated. We refer the interested reader to \cite{paper5,thomassingle,paper201,wopjordan} and do not repeat the construction here.

In section \ref{sect:4} we want to study bulk correlators in JT gravity at large spatial separations where higher genus and random matrix contributions are important. Unlike for boundary correlators it is not obvious how to include higher genus contributions to a certain bulk observable. The issue is that the higher genus geometries that contribute to the Euclidean path integral of JT gravity have no known sensible continuation to Lorentzian signature. 

Therefore there is no sense that we know of in which specifying a bulk point in Lorentzian signature specifies a point on each of the higher genus geometries that contribute to the Euclidean path integral. This means bulk observables are not to be computed by calculating the propagator from one bulk point to another on each of the higher genus Riemann surfaces before integrating over the moduli space of Riemann surfaces. So how should we include higher genus and random matrix corrections to bulk correlation functions?
\\~\\
One idea is to use bulk operator reconstruction. It is well known how to write Lorentzian correlators of free bulk matter in terms of boundary correlators in JT gravity \cite{paper5,paper201,wopjordan} if we ignore higher genus contributions to the Euclidean path integral. This is implementing bulk operator reconstruction in quantum gravity \cite{hkll1,hkll2,kll,kl}. 

From the previous section we furthermore know how to include higher genus and random matrix corrections to boundary correlators. We claim that bulk correlators in JT gravity can be described by implementing the same bulk operator reconstruction on fully dressed boundary correlators. Examples in section \ref{sect:4} will clarify the setup.

Though natural, this is ultimately a prescription with no obvious guaranteed success. This same caveat applies to any calculation using the gravitational path integral. It is still very much unclear what the rules of the game are. The best we can do it take our best shot at a definition and see if it leads to sensible results. 

In this sense our definition of bulk observables is doing reasonably well. As argued in the introduction we expect in a well defined theory of quantum gravity (a finite entropy system) that the cluster decomposition principle does not hold up. We expect that the exponential decay of correlation functions at large spatial separations eventually makes way for erratic oscillations around an analogue to the plateau. Bulk observables in JT gravity defined via this bulk reconstruction first principle turn out to indeed have this property. This is as an incentive that our definition is a step in the right direction. 

One further positive sign in this regard are the results of \cite{paper201}. It was argued that Hawking radiation should be expected to carry imprints of level repulsion in the chaotic level statistics of quantum black holes. Repulsion between black hole levels implies there should be barely Hawking particles at energies $\omega\ll e^{-S_0}$. This effect was then observed in an eternal setup via explicit calculations in JT gravity. On a technical level those calculations used the same prescription as we are using here for including higher genus and random matrix contributions to the bulk two point function of a massless scalar field.
%%%%%%%%%%%%%%%%%%%%%%
% SECTION %
%%%%%%%%%%%%
%%%%%%%%%%%
% SECTION %
%%%%%%%%%%%
\section{Cluster decomposition}\label{sect:4}
In this section we compute several bulk matter two point functions in JT gravity with the purpose of investigating large spatial separations. Correlators at large spatial separations behave structurally similar to boundary operators at late times \cite{bhrm}. We distinguish $4$ regimes depending on the distance $d$ between any two operators. The relevant scales are set by the Newton constant $G$ and the inverse string coupling $e^{S_0}$.
\begin{enumerate}
    \item At short spatial distances $d\ll 1/G$ the exact correlator is accurately approximated as a semiclassical matter correlators on the gravitational saddle. These semiclassical correlators decay exponentially with $d$. There is a yet shorter distance scale at which there are high energy backreaction effects. Here perturbative and nonperturbative corrections in $G$ are important. Such effects were discussed in \cite{paper5}. 
    \item At longer distances $d\gg 1/G$ we probe the low energy features of Schwarzian quantum mechanics which is dual to JT gravity if we ignore higher genus and random matrix contributions \cite{jensen,malstanyang,ads2}. In this regime we probe the square root spectral edge of $\rho_0(E_1)\dots \rho_0(E_n)$ which result is a transition from exponential decay to power law decay known as the slope \cite{altland,altland2,schwarzian,paper5}.
    \item Contributions from higher genus configurations are suppressed for $d\ll e^{S_0}$. The first higher genus contributions which become relevant are due to annuli connecting distinct regions \cite{sss,sss2,phil,paper6}. These give rise to contributions which grow with $d$. This generalization of the ramp competes with the slope for dominance and eventually wins \cite{bhrm,sss}. Other perturbative contributions in $e^{S_0}$ contribute negligibly for energies far enough from the spectral edge. Closer to the edge we can use the Airy model \cite{paper6}.
    \item For $d\gg e^{S_0}$ the correlators go to some analogue to the plateau \cite{bhrm}. This is due to nonperturbative effects in $e^{S_0}$ associated with branes. The transition from ramp to plateau is due to the sine kernels in random matrix correlators such as \eqref{224} and \eqref{2.31}. The plateau itself is due to contact terms in $\rho(E_1\dots E_n)$ \cite{sss2,paper6}.
\end{enumerate}
This analogue to the plateau at large distances is in violation of cluster decomposition. We focus on identifying such contributions to bulk correlators in JT gravity. 
\\~\\
Up to and including in this section we take JT gravity as the ensemble averaged description of \cite{sss2}. Large distances correlators in a more microscopic theory are characterized by erratic oscillations around an analogue to the plateau. These oscillations are recovered only if we consider more microscopic versions of JT gravity that include alpha branes \cite{maxfieldmarolf}.

These erratic oscillations are inherent to finite entropy chaotic quantum system \cite{maldainfo,susskindlindesay}. It is this fundamental discreteness which excludes cluster decomposition in quantum gravity. For some observables such as the partition function the plateau actually vanishes but the erratic oscillations are always there \cite{bhrm}. In the averaged description of gravity we can access more details of these oscillations by computing higher moments. For example the spectral form factor computes the variance of the partition function. It has a plateau so the partition function is nonzero at late times too.
%%%%%%%%%%%%%%%%%%%%%%%%%%%%%%%%%%
\subsection{Conformal matter}
Consider the two point function of a spinless weight $h$ conformal primary field with Neumann boundary conditions. The semiclassical matter correlator is
\begin{equation}
    \average{\phi_h(u_1,v_1)\phi_h(u_2,v_2)}_\text{matter}=\frac{\sqrt{M}^h}{\sinh^{2h} \sqrt{M}(u_1-u_2)}\,\frac{\sqrt{M}^h}{\sinh^{2h} \sqrt{M}(v_1-v_2)}\,.
\end{equation}
Here $M$ is the black hole mass. We consider the microcanonical ensemble throughout this section. For large separations $\abs{z_1-z_2}\gg 1$ this classical correlator decays exponentially. Ignoring higher genus contributions this observable can be coupled to JT gravity by evaluating it a set of reparameterized Lorentzian metrics \cite{paper5,thomassingle,paper201,wopjordan}
\begin{equation}
    ds^2=-\frac{f'(u)f'(v)}{(f(u)-f(v))^2}\, du\,dv.\label{60}
\end{equation}
Here we denote by $f(t)$ the reparameterization $T=f(t)$ with the saddle $f(t)=\tanh \sqrt{M} t$. This field is weighted by a Schwarzian action if we ignore higher genus contributions to the gravitational path integral \cite{jensen,malstanyang,ads2}. So at genus zero the bulk two point function is
\begin{equation}
    \int[\dpi f]\, \frac{f'(u_1)^h f'(u_2)^h}{(f(u_1)-f(u_2))^{2h}}\,\frac{f'(v_1)^h f'(v_2)^h}{(f(v_1)-f(v_2))^{2h}}\,e^{-S[f]}= \quad\raisebox{-10mm}{\includegraphics[width=50mm]{intro6.pdf}}\quad\,.\label{63}
\end{equation}

Indeed the Schwarzian $4$ point function is identical to a genus zero JT gravity calculation with two boundary anchored Wilson lines. This is the leading early time contribution to the $s$ channel boundary $4$ point function \eqref{2.72} at $t\ll e^{S_0}$ but with slightly different labels. Let us take two operators on the same time slice to isolate large distance physics. Computing \eqref{63} in the microcanonical ensemble we recover a variant of \eqref{2.39}
\begin{align}
    \average{\phi_h(0,z_1)\phi_h(0,z_2)}\supset  e^{-2S_0}\int_0^\infty d E_1\int_0^\infty d E_2\,e^{i(z_1-z_2)(E_1-E_2)}\,\rho_0(E_1)\rho_0(E_2)\,\rvert \mo^h_{M E_1}\rvert^2\rvert \mo^h_{M E_2}\rvert^2.\nonumber
\end{align}
Here $\rvert \mo^h_{E_1 E_2}\rvert^2=e^{-S_0}\,\mo_h\mo_h(E_1,E_2)$ is the order one smooth function in \eqref{2.21}. The results of the previous section explain how to include higher genus contributions to the $s$ channel $4$ point function in \eqref{63}. By summing over geometries we find a microcanonical variant of \eqref{42}. We take this sum over geometries in \eqref{2.72} as a sensible definition of a fully dressed conformal primary bulk two point function in JT gravity for reasons explained in section \ref{ss:bulk}.
\\~\\
\textbf{\emph{Large distance behavior}}
\\~\\
The correlator \eqref{2.72} receives contributions from $3$ simple classes. We focus on the class of which the representative has no handles and which dominates at early and intermediate time and distance scales. Summing over all geometries which end on \eqref{63} and furthermore including random matrix corrections one finds
\begin{align}
    &\average{\phi_h(0,z_1)\phi_h(0,z_2)}\nonumber \\
    &\qquad \qquad \supset e^{-2S_0}\int_\cas d E_1\int_\cas d E_2\,e^{i(z_1-z_2)(E_1-E_2)}\,\frac{\rho(M,E_1,E_2)}{\rho(M)}\,\rvert \mo^h_{M E_1}\rvert^2\rvert \mo^h_{M E_2}\rvert^2.\label{65}
\end{align}
This is similar to the first term in \eqref{42}.
We are interested in the behavior of this expression for $\abs{z_1-z_2}\gg e^{S_0}$. The dominant contribution at these distances is due to geometries which connect the two probe regions
\begin{equation}
    \frac{\rho(M,E_1,E_2)}{\rho(M)}\supset \rho(E_1,E_2)_\text{conn} = \delta(E_1-E_2)\,\rho(E_1)-\rho(E_1)\rho(E_2)\,\sinc^2\,\pi\rho(E_1)(E_1-E_2).\label{77}
\end{equation}
The perturbative series of this connected contribution starts with an annulus
\begin{equation}
    \rho(E_1,E_2)_\text{conn}\,\rvert \mo^h_{M E_1}\rvert^2\rvert \mo^h_{M E_2}\rvert^2\supset \quad\raisebox{-10mm}{\includegraphics[width=50mm]{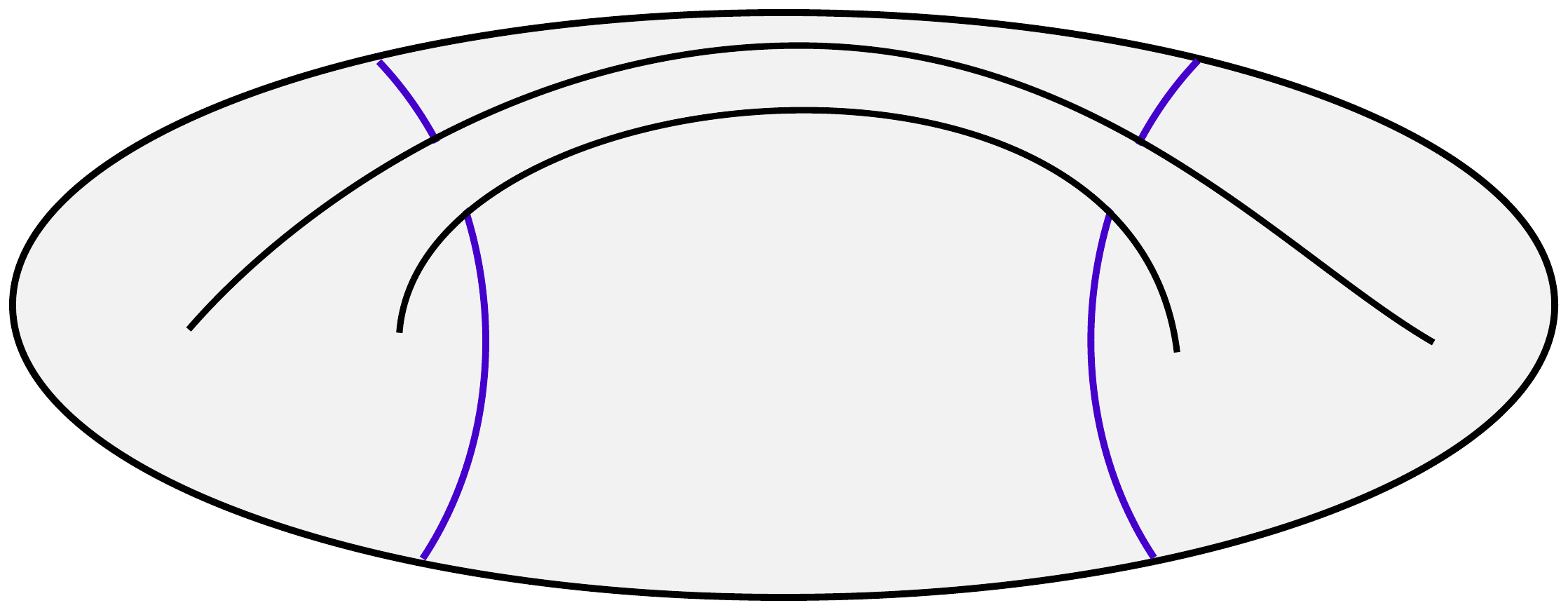}}\quad. \label{70}
\end{equation}
The contribution to the bulk two point function due to these connected geometries is the Fourier transform of \eqref{77} at exponentially large distances where the Fourier transform probes for quickly oscillating features of the integrand. These fast features are due to the sine kernel and the contact term in \eqref{77} which have significant support only for $\abs{E_1-E_2}\ll 1$. By the stationary phase approximation we can Taylor expand the kernel due to the operators to leading order around $E_1=E_2$. We then recognize the familiar ramp and plateau \cite{bhrm,sss,sss2}
\begin{align}
    \average{\phi_h(0,z_1)\phi_h(0,z_2)}\supset\,&\nonumber e^{-2S_0}\int_\cas d E\,\rho(E)\,\rvert \mo^h_{M E}\rvert^4\\&\qquad+e^{-2S_0}\int_\cas d E\,\rho(E)\,\text{Ramp}\left(\frac{z_1-z_2}{2\pi\rho(E)}\right)\,\rvert \mo^h_{M E}\rvert^4\,.\label{3.8}
\end{align}
The contribution due to the ramp term decays to zero for exponentially large distances yet it does not stop abruptly. In the end only the contribution due to the contact term in \eqref{77} survives for very large conformal distances $\abs{z_1-z_2}\gg e^{S_0}$
\begin{equation}
    \average{\phi_h(0,z_1)\phi_h(0,z_2)}\supset e^{-2S_0}\int_\cas d E\,\rho(E)\,\rvert \mo^h_{M E}\rvert^4\,.\label{3.9}
\end{equation}
One checks that all other contributions to \eqref{65} either decay to zero at exponentially large distances or are suppressed by powers of $e^{S_0}$. In fact we see that the Fourier transform at asymptotically large distances can only get contributions due to terms in $\rho(M,E_1,E_2)$ which are non analytic in $\rvert E_1-E_2\rvert $. The smooth terms have a maximal frequency of order $e^{S_0}$ and so their Fourier transform decays to zero at order $e^{S_0}$ distances. There are just two such non analytic terms
\begin{equation}
    \frac{\rho(M,E_1,E_2)}{\rho(M)}\supset\delta(E_1-E_2)\rho(E_1)+\delta(E_1-M)\delta(E_2-M).\label{84}
\end{equation}
The second contribution is due to a contact term in the series of fully connected geometries which is explicit in \eqref{67}. This contributes a term of order $e^{-2S_0}$ which can be neglected as compared to the order $e^{-S_0}$ plateau in \eqref{3.9}.

Each of the simple classes contributes to the plateau but one checks that in this case the two other classes give constant contributions to the bulk two point function of order $e^{-2S_0}$ which can be neglected for all distances.
\\~\\
\textbf{\emph{Dirichlet boundary conditions}}
\\~\\
Consider the contribution due to the same genus zero class to the primary bulk two point function with Dirichlet boundary conditions. Ignoring higher genus we have
\begin{equation}
    \nonumber \int [\dpi f]\,\frac{f'(u_1)^h f'(u_2)^h}{(f(u_1)-f(u_2))^{2h}}\,\frac{f'(v_1)^h f'(v_2)^h}{(f(v_1)-f(v_2))^{2h}} -\frac{f'(u_1)^h f'(v_2)^h}{(f(u_1)-f(v_2))^{2h}}\,\frac{f'(v_1)^h f'(u_2)^h}{(f(v_1)-f(u_2))^{2h}}\,e^{-S[f]}\,.\label{85}
\end{equation}
The second term is new as compared to \eqref{63} and can be thought of as due to an imagine charge. Going through the gears one finds eventually that the plateau vanishes because the second term gives the same answer as the first but with $\abs{z_1-z_2}$ replaced by $\abs{z_1+z_2}$. The plateau is a constant and so the two contributions cancel due to the relative minus sign.

This does not mean that the microscopic version of this correlator vanishes at large distances. It means that is oscillates erratically around zero. To confirm this we can compute the variance at large distances
\begin{equation}
    \average{(\phi_h(0,z_1)\phi_h(0,z_2))(\phi_h(0,z_1)\phi_h(0,z_2))}.
\end{equation}
From the geometric point of view we are led to consider two copies of the configuration in \eqref{63} and are instructed to summing over all topologies which end on these two copies. The result is
\begin{align}
   \nonumber &4\,e^{-4S_0} \int_\cas d E_1\int_\cas d E_2\int_\cas d E_3\int_\cas d E_4\,\sin z_1(E_1-E_2)\sin z_1(E_3-E_4)\\&\nonumber \qquad\qquad \qquad \qquad \sin z_2(E_1-E_2)\sin z_2(E_3-E_4)\,\frac{\rho(M_1,E_1,E_2,M_2,E_3,E_4)}{\rho(M_1)\rho(M_2)}\, \\&\qquad \qquad \qquad \qquad \qquad \qquad \qquad \qquad \qquad \qquad \rvert \mo^h_{M_1 E_1}\rvert^2\rvert \mo^h_{M_1 E_2}\rvert^2\rvert \mo^h_{M_2 E_3}\rvert^2\rvert \mo^h_{M_2 E_4}\rvert^2. \label{89} 
\end{align}
We distinguish independent values of $M_1$ and $M_2$ but consider the same microcanonical ensemble for both. The variance is nonzero due to geometries which connect the two copies. The dominant contribution to the large distance variance is due to contact terms
\begin{align}
    &\nonumber\frac{\rho(M_1,E_1,E_2,M_2,E_3,E_4)}{\rho(M_1)\rho(M_2)}\supset \rho(E_1,E_3)_\text{conn}\,\rho(E_2,E_4)_\text{conn}+\rho(E_1,E_4)_\text{conn}\,\rho(E_2,E_3)_\text{conn}\\&\qquad\supset \delta(E_1-E_3)\delta(E_2-E_4)\rho(E_1)\rho(E_2)+\delta(E_1-E_4)\delta(E_2-E_3)\rho(E_1)\rho(E_2).
\end{align}
The leading geometry that contributes to this connected correlator has two annuli connecting the copies
\begin{align}
  \nonumber \rho(E_1,E_3)_\text{conn}\,\rho(E_2,E_4)_\text{conn}\,\rvert \mo^h_{M_1 E_1}\rvert^2\rvert \mo^h_{M_1 E_2}\rvert^2&\rvert \mo^h_{M_2 E_3}\rvert^2\rvert \mo^h_{M_2 E_4}\rvert^2 \supset\\&\qquad \quad\raisebox{-10mm}{\includegraphics[width=50mm]{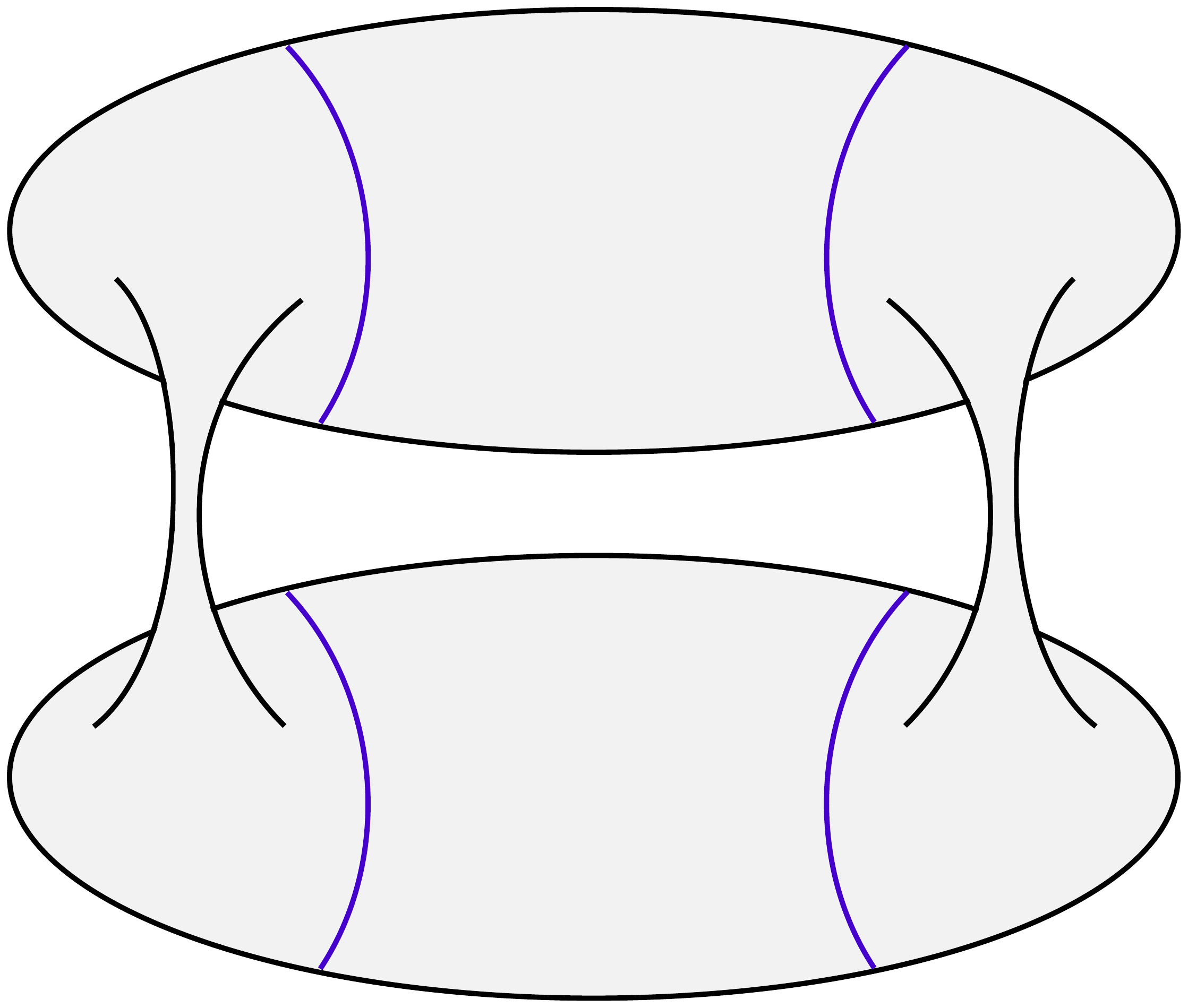}}\quad.
\end{align}
In the end we indeed find a constant plateau value at large distances $z_1,z_2\gg e^{S_0}$
\begin{align}
    \nonumber &\average{(\phi_h(0,z_1)\phi_h(0,z_2))(\phi_h(0,z_1)\phi_h(0,z_2))}\\&\qquad\qquad\qquad\qquad\qquad\supset 8\,e^{-4S_0} \int_\cas d E_1\, \rho(E_1)\,\rvert \mo^h_{M E_1}\rvert^4\int_\cas d E_2\, \rho(E_2)\,\rvert \mo^h_{M E_2}\rvert^4\,.
\end{align}
Other contributions can be checked to either decay or be suppressed by powers of $e^{S_0}$. By comparing powers of $e^{S_0}$ we see that the erratic oscillations in the large distance bulk two point function with Dirichlet boundary conditions are just as prominent as in the Neumann case. The relative suppression of $e^{S_0}$ between the short distance and the large distance bulk two point functions is the same as the relative size of the plateau of the boundary two point function \cite{bhrm}.
%%%%%%%%%%%%%%%%%%%%%%%%%%%%%%%%%%%%%%%%%%
\subsection{Extremal massive fields}
Consider the massive scalar two point function with Dirichlet boundary conditions on an extremal black hole with semiclassical geometry
\begin{equation}
    ds^2=\frac{dz^2-dt^2}{z^2}\,.
\end{equation}
In the gravitational quantum theory we study a microcanonical ensemble near the vacuum. The massive scalar Wightman two point function on the gravitational saddle is \cite{spradlin}
\begin{equation}
    \average{\phi_{m^2}(t_1,z_1)\phi_{m^2}(t_2,z_2)}_\text{matter}=x^\ell\, _2F_1(\ell,\ell,2\ell,x)\quad,\quad x= \frac{4z_1z_2}{(t_1-t_2)^2-(z_1-z_2)^2}\,.\label{101}
\end{equation}
On one time slice and at large distances for heavy fields this semiclassical two point function decays exponentially with proper distance
\begin{equation}
    \average{\phi_{m^2}(0,z_1)\phi_{m^2}(0,z_2)}_\text{matter}=\exp(-m\, d(z_1,z_2))\,.
\end{equation}
This decays to zero for asymptotic values of $d(z_1,z_2)$ as required by cluster decomposition.

For finite temperature black holes the spatial separation $d(z_1,z_2)$ is bound from above if we regularize the divergences due to the asymptotic boundary. Asymptotic divergences are important to understand the holographic dictionary in finite entropy dS quantum gravity \cite{susskindlindesay}. In AdS quantum gravity these are not probing long time boundary physics but instead zoom in on ultra short times for which JT gravity is inaccurate.

Here instead we want to probe for large distances $d(z_1,z_2)$ by placing one operator deep into the bulk. The distance to the semiclassical horizon at finite temperature is finite so this setup does not allow to probe long distance physics. The proper distance to the horizon of an extremal black hole does diverge due to the infinite throat so we focus on the extremal setup.
\\~\\
\textbf{\emph{Coupling to Schwarzian modes}}
\\~\\
We first ignore higher genus contributions in which case the JT gravity path integral reduces to a Schwarzian. The two point function is coupled to the Schwarzian reparameterization mode by evaluating it in each of the metrics \eqref{60} and then taking the Schwarzian path integral \cite{paper5,wopjordan}. We rewrite the crossratio as
\begin{equation}
    x=\frac{(u_1-v_1)(u_2-v_2)}{(u_1-u_2)(v_1-v_2)}\,.\label{3.23}
\end{equation}
The two point function in the reparameterized metrics \eqref{60} differs only by the definition of the crossratio
\begin{align}
    \nonumber &\average{\phi_{m^2}(t_1,z_1)\phi_{m^2}(t_2,z_2)}_\text{matter}=x^\ell\, _2F_1(\ell,\ell,2\ell,x)\quad,\quad x=\frac{(f(u_1)-f(v_1))(f(u_2)-f(v_2))}{(f(u_1)-f(u_2))(f(v_1)-f(v_2))}\,.
\end{align}
To compute the Schwarzian path integral of this it is technically convenient to rewrite the bulk two point function in terms of boundary operators by using bulk operator reconstruction \cite{paper5,wopjordan,hkll1,hkll2,kll,kl}. On the gravitational saddle we have up to normalization
\begin{equation}
    x^\ell\, _2F_1(\ell,\ell,2\ell,x) = \int_{v_1}^{u_1} d x_1\, K_\ell(x_1\rvert u_1,v_1)\int_{v_2}^{u_2} d x_2\, K_\ell(x_2\rvert u_2,v_2)\, \average{\mo_\ell(x_1)\mo_\ell(x_2)}\,.\label{3.24}
\end{equation}
Here the kernel and boundary two point function are
\begin{equation}
    K_\ell(x\rvert u,v)=\frac{(u-v)^{1-\ell}}{(u-x)^{1-\ell}(x-v)^{1-\ell}}\quad,\quad \average{\mo_\ell(x_1)\mo_\ell(x_2)}=\frac{1}{(x_1-x_2)^{2\ell}}\,.
\end{equation}
Equation \eqref{3.24} also reproduces the reparameterized bulk two point function by inserting the reparameterized kernel and boundary two point function \cite{paper5,wopjordan}
\begin{align}
    K_\ell(x\rvert u,v)&= f'(x)^{1-\ell} \frac{(f(u)-f(v))^{1-\ell}}{(f(u)-f(x))^{1-\ell}(f(x)-f(v))^{1-\ell}}\\\average{\mo_\ell(x_1)\mo_\ell(x_2)}&=\frac{f'(x_1)^\ell f'(x_2)^\ell}{(f(x_1)-f(x_2))^{2\ell}}\,.
\end{align}
One recognizes the kernel as a product of $3$ boundary bilocals which means we can write the reparameterized bulk two point function as a double integral over $7$ bilocals
\begin{align}
    &\average{\phi_{m^2}(t_1,z_1) \phi_{m^2}(t_2,z_2)}_\text{matter} \\\nonumber &=\int_{v_1}^{u_1} d x_1\,\average{\mo_{\ell-1}(u_1)\mo_{\ell-1}(v_1)}\average{\mo_{1-\ell}(u_1)\mo_{1-\ell}(x_1)}\average{\mo_{1-\ell}(x_1)\mo_{1-\ell}(v_1)}\\\nonumber &\qquad\int_{v_2}^{u_2} d x_2\,\average{\mo_{\ell-1}(u_2)\mo_{\ell-1}(v_2)}\average{\mo_{1-\ell}(u_2)\mo_{1-\ell}(x_2)}\average{\mo_{1-\ell}(x_2)\mo_{1-\ell}(v_2)}\,\average{\mo_\ell(x_1)\mo_\ell(x_2)}.
\end{align}
This corresponds to a genus zero JT gravity calculation with $7$ boundary anchored Wilson lines \cite{paper3,paper4}. Note that there is an ambiguity associated to how we order the Lorentzian endpoints of the bilocals along the Euclidean boundary circle when computing the path integral \cite{paper5,paper201}. Different orderings give different answers for the correlation functions.

Here we but aim to prove that the decay at large distances eventually stops and makes way for erratic oscillations around a plateau. We do not expect this behavior to structurally depend on a choice of operator ordering. One can check that the height of the plateau does numerically depend on the operator ordering but the overall power of $e^{S_0}$ does not.

We choose one ordering to be done with it and focus on the bulk to boundary correlator for purposes of presentation
\begin{equation}
    \average{\phi_{m^2}(t,z)\mo_\ell(x)}_\text{matter}=\int_v^u d y\, K_\ell(y\rvert u,v)\, \average{\mo_\ell(y)\mo_\ell(x)}\,.\label{112}
\end{equation}
We are led to compute a JT gravity disk amplitude with $4$ boundary anchored Wilson lines
\begin{equation}
    \int [\dpi f] \average{\phi_{m^2}(0,z)\mo_\ell(0)}_\text{matter}\,e^{-S[f]}=\int_{-z}^{+z} d y\quad\raisebox{-10mm}{\includegraphics[width=50mm]{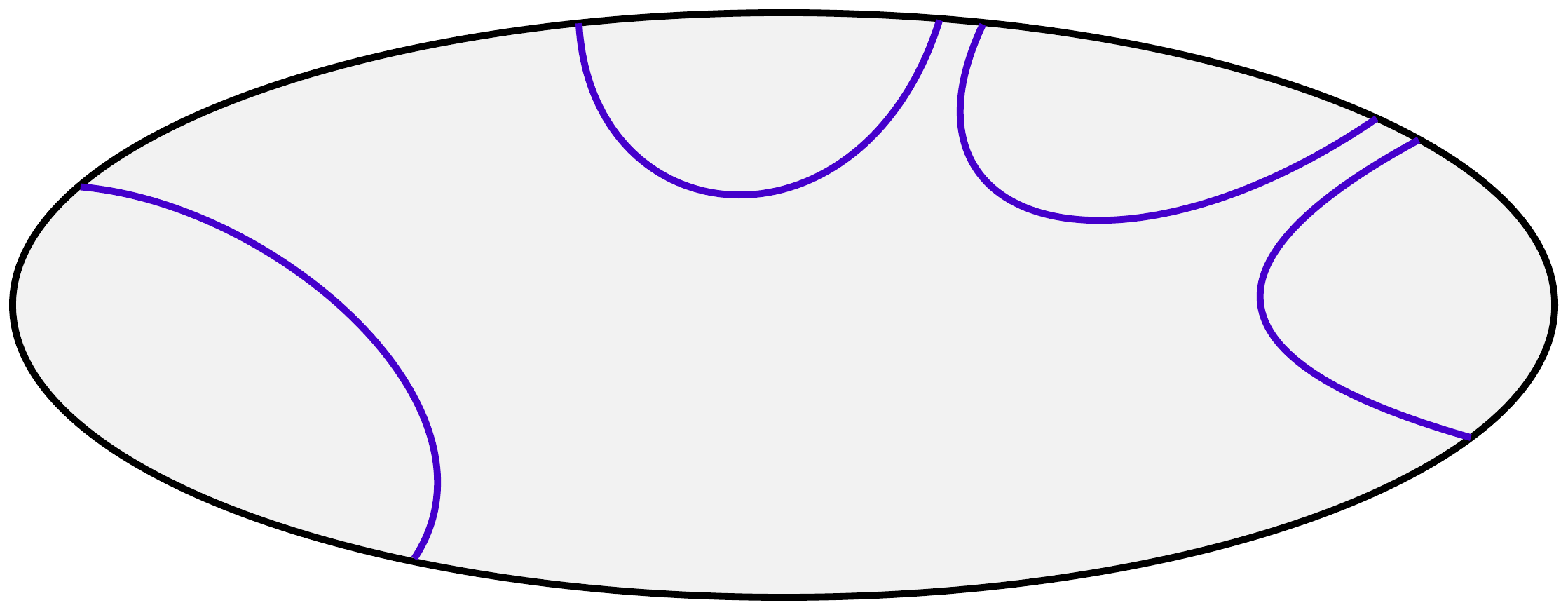}}\quad.\label{325}
\end{equation}
The energy of the central region is fixed microcanonically to the vacuum because we consider the extremal setup. The amplitude is the integral is the leading early time contribution to the boundary $8$ point function in JT gravity where we consider only pairwise identical operators so that there is no sum over channels. This amplitude evaluates to \cite{paper3,paper4,zhenbin}
\begin{align}
    &\average{\phi_{m^2}(0,z) \mo_\ell(0)}\label{114}\\
    &\supset e^{-4S_0}\nonumber \int_{-z}^{+z} d y\, \prod_{i=1}^4\int_0^\infty d E_i\,\rho_0(E_i)\,e^{iy(E_3-E_4-E_1)}\,e^{i z(E_3+E_4-2E_2)}\,\rvert\mo^\ell_{0 E_1}\rvert^2\,\rvert\mo^{\ell-1}_{0 E_2}\rvert^2\,\rvert\mo^{1-\ell}_{0 E_3}\rvert^2\,\rvert\mo^{1-\ell}_{0 E_4}\rvert^2\,.
\end{align}
The semiclassical answer is recovered by approximating $E_i\gg 1$ in the integrand. The slope region at $z\gg 1$ is due to comparable dominant contributions from the integration regions $E_1,E_2,E_3\ll 1$ and $E_1,E_2,E_4\ll 1$ which result in a $z^{-9/2}$ power law decay. Note that this depends heavily on the choice of operator ordering.
\\~\\
\textbf{\emph{Large distance behavior}}
\\~\\
We include contributions due to higher genus geometries to this bulk correlator by replacing the genus zero boundary $8$ point function amplitude in \eqref{325} by the full boundary $8$ point function in JT gravity. This includes a sum over geometries and classes and can be derived using the matrix integral techniques of section \ref{sect:2}. We focus on the contribution due to the simple class with representative \eqref{325}. Summing over all geometries in this class one finds
\begin{align}
    \nonumber \average{\phi_{m^2}(0,z)\mo_\ell(0)}\supset\,&e^{-4S_0}\prod_{i=1}^4\int_\cas d E_i\, \frac{\rho(0,E_1,E_2,E_3,E_4)}{\rho(0)}\,\int_{-z}^{+z}d y\,e^{iy(E_3-E_4-E_1)}\\&\qquad\qquad\qquad\qquad e^{i z(E_3+E_4-2E_2)}\,\rvert\mo^\ell_{0 E_1}\rvert^2\,\rvert\mo^{\ell-1}_{0 E_2}\rvert^2\,\rvert\mo^{1-\ell}_{0 E_3}\rvert^2\,\rvert\mo^{1-\ell}_{0 E_4}\rvert^2.\label{115}
\end{align}
At asymptotic distances the Hamiltonian weights are highly oscillatory and so the the contributions of all smooth terms in $\rho(0,E_1,E_2,E_3,E_4)$ vanishes. Only the non analytic contact terms which equate two or more of the energies can be relevant to the plateau. The dominant contribution is due to
\begin{equation}
    \frac{\rho(0,E_1,E_2,E_3,E_4)}{\rho(0)}\supset \frac{\rho(0,E_1)_\text{conn}}{\rho(0)}\rho(E_2,E_3,E_4)_\text{conn}\supset\delta(E_1)\delta(E_2-E_3)\delta(E_3-E_4)\rho(E_2).\label{3.34}
\end{equation}
The first term in this genus series of connected geometries is
\begin{align}
    \nonumber\frac{\rho(0,E_1)_\text{conn}}{\rho(0)}\rho(E_2,E_3,E_4)_\text{conn}\,&\rvert\mo^\ell_{0 E_1}\rvert^2\,\rvert\mo^{\ell-1}_{0 E_2}\rvert^2\,\rvert\mo^{1-\ell}_{0 E_3}\rvert^2\,\rvert\mo^{1-\ell}_{0 E_4}\rvert^2\supset\\&\qquad\qquad\qquad\qquad\raisebox{-10mm}{\includegraphics[width=55mm]{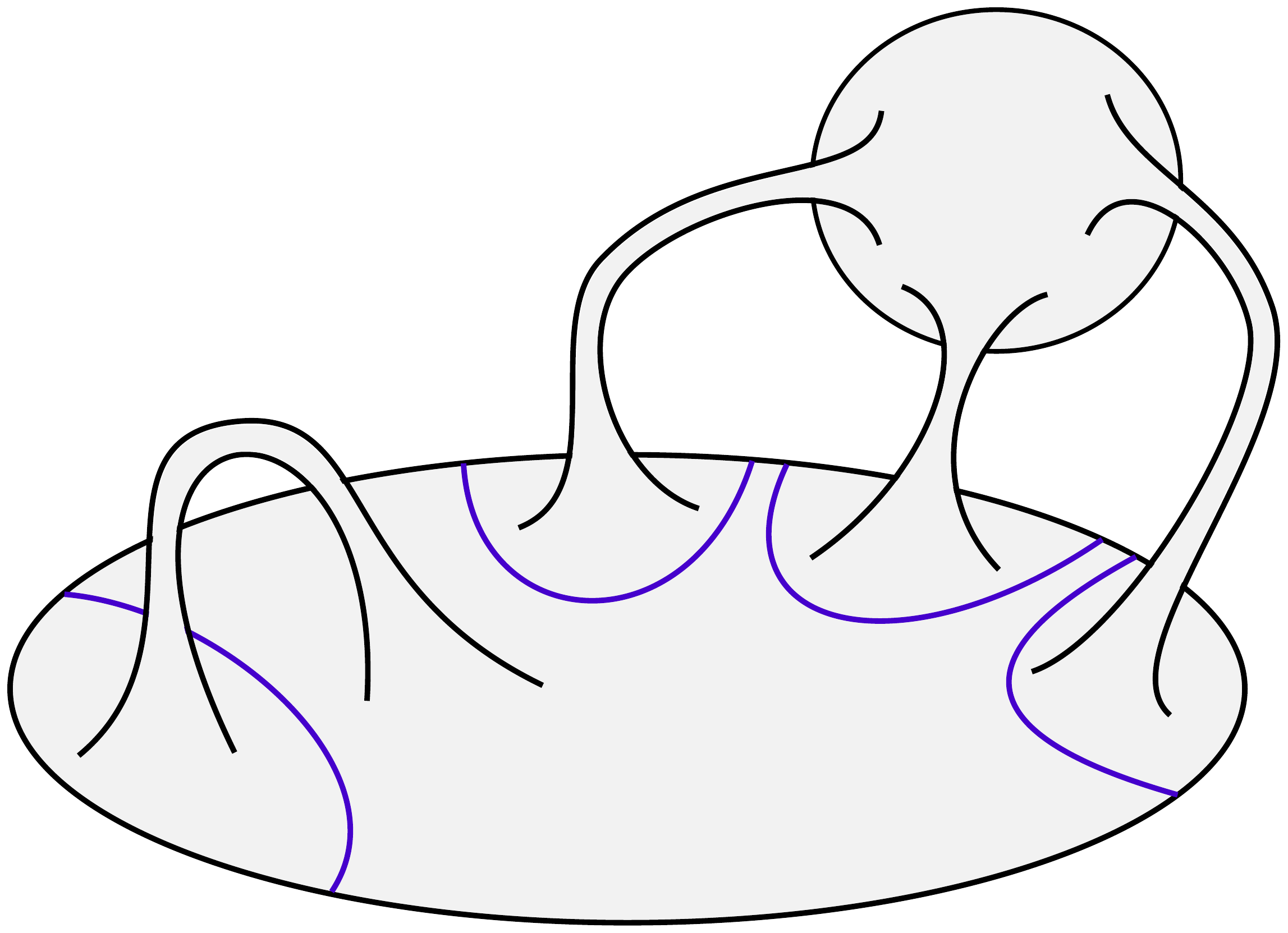}}\quad.
\end{align}
This is the first example where the details of $\rho(E_1,E_2,E_3)_\text{conn}$ are relevant to large distance physics. Sometimes we need more than just annuli and the associated genus expansion. The contact term in \eqref{3.34} gives a linearly rising contribution that persists on all distance scales
\begin{equation}
    \average{\phi_{m^2}(0,z)\mo_\ell(0)} \supset\,2z\,e^{-4S_0}\, \rvert\mo^\ell_{00}\rvert^2\int_\cas d E\,\rho(E)\,\rvert\mo^{\ell-1}_{0 E}\rvert^2\,\rvert\mo^{1-\ell}_{0 E}\rvert^4\,.\label{118}
\end{equation}

Note that this grows linearly in $z$. From the perspective of bulk operator reconstruction it is clear where this linear growth comes from. There are constant plateau contributions in all boundary correlators. We are integrating that plateau over the operator reconstruction patch in \eqref{325}. Eventually for large enough $z$ this becomes the dominant contribution.

From the bulk perspective however this is quite surprising. This might be the sign of a deeper yet to be appreciated lesson about large distance physics in gravity so it would be interesting to understand this linear growth from the bulk point of view. Alternatively this might be an indication that our definition of bulk correlators is not adequate. Could there be found more bulk intrinsic ways to define contributions due to Euclidean wormholes?

There seems to be a similar tension between the comments on the extrapolate dictionary in dS quantum gravity \cite{susskindlindesay} and our definition of bulk operators. But how could you even make sense of bulk operators in gravity without using bulk operator reconstruction?

Note that all classes contribute such positive linearly rising contributions to the asymptotic bulk two point function however these contributions are in general subdominant. Regardless the total asymptotic answer is clearly nonzero so there is no cluster decomposition in JT gravity with our definition of bulk correlators. 
%%%%%%%%%%%
% SECTION %
%%%%%%%%%%%
\section{Factorization}\label{sect:5}
Thus far we defined JT gravity as summing over all Riemann surfaces of arbitrary topology that end on probe boundaries \cite{sss2}. For example for the partition function
\begin{equation}
    \average{Z(\beta)}=\quad \raisebox{-5mm}{\includegraphics[width=23mm]{intro1.pdf}}\quad +\quad\raisebox{-5mm}{\includegraphics[width=23mm]{intro2.pdf}}\quad+\dots\label{genus}
\end{equation}
Such genus expansions are completed nonperturbatively as a particular double scaled matrix integral \cite{sss2}. In part of this section we focus on a finite dimensional matrix integral
\begin{equation}
    \mathcal{Z}=\int \prod_{a=1}^L d\lambda_a\,e^{-L V(\lambda_a)}\,\prod_{b<a}^L (\lambda_a-\lambda_b)^2\prod_{i=1}^L d U_{i a}\,d U_{i a}^* e^{- L \rvert U_{i a}\rvert^2}=\mathcal{Z}_\lambda\,\mathcal{Z}_U\,.\label{matint}
\end{equation}
The potential $V(E)$ determines the distribution of eigenvalues. The eigenvalue density is
\begin{equation}
    \rho(E)=\frac{1}{\mathcal{Z}_\lambda}\int \prod_{a=1}^L d\lambda_a\,e^{-L V(\lambda_a)}\,\prod_{b<a}^L (\lambda_a-\lambda_b)^2\sum_{c=1}^L \delta(E-\lambda_c)\,.
\end{equation}
We consider a matrix integral for which the genus zero spectral density has finite support
\begin{equation}
    \rho_0(E)= \quad \raisebox{-5mm}{\includegraphics[width=56mm]{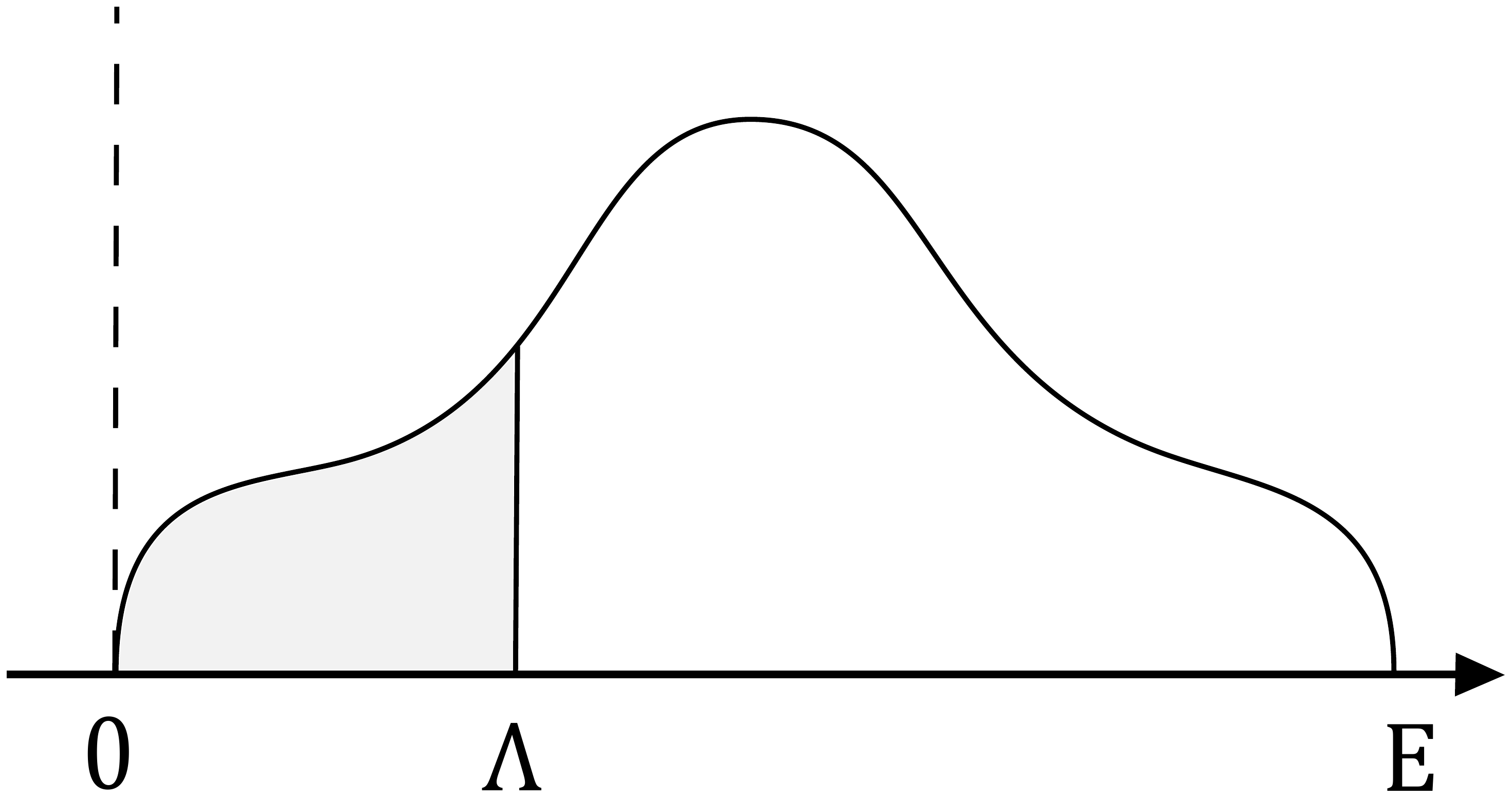}}\quad.\label{curve}
\end{equation}
We choose the potential such that there is a region close to the edge of the spectrum (gray) where we find the JT gravity spectral curve \cite{stanfordwitten,malstanyang,schwarzian}
\begin{equation}
    \rho_0(E)=\frac{e^{S_0}}{2\pi^2} \sinh 2\pi \sqrt{E}\,\theta(E)\quad,\quad E \ll \Lambda\,.\label{jtcurve}
\end{equation}
If we denote the cumulative spectral occupation as $N(E)$ then we imagine $L\gg N(\Lambda)\gg 1$ so just a tiny part of the curve near the spectral edge is described by JT gravity. 

Multi spectral densities $\rho(E_1\dots E_n)$ in the matrix ensemble correspond in perturbative JT gravity to multi boundary amplitudes. For example
\begin{equation}
    \rho(E_1,E_2)= \quad \raisebox{-5mm}{\includegraphics[width=50mm]{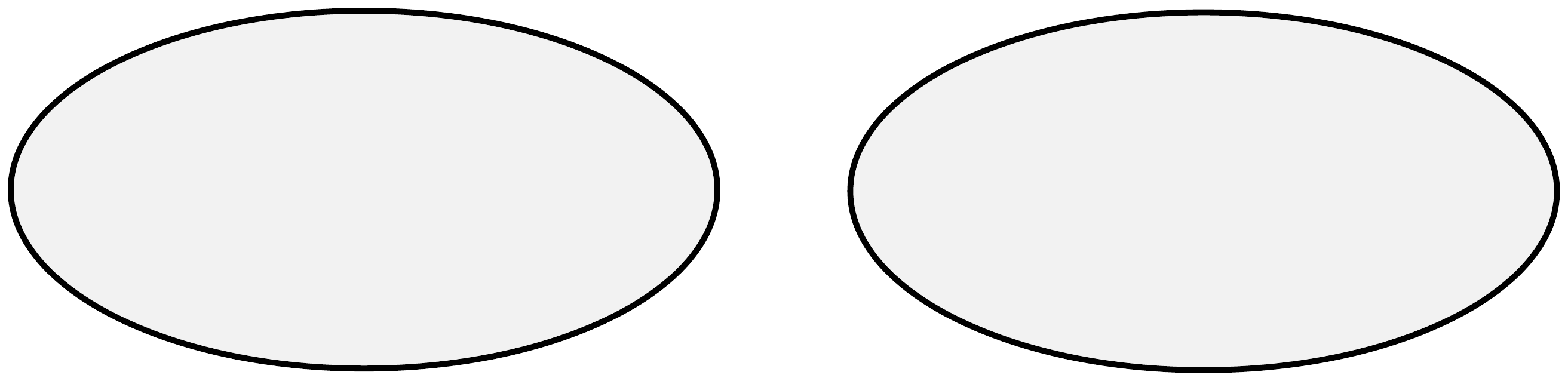}}\quad+\quad \raisebox{-5mm}{\includegraphics[width=50mm]{intro3.pdf}}\quad+\dots \label{4.6}
\end{equation}
The perturbative genus expansion is controlled by $e^{S_0}$ so perturbation theory breaks down when there are order $e^{S_0}$ boundaries. But we can make sense of such complicated gravity amplitudes with order $e^{S_0}$ boundaries via the matrix integral definition. The bulk dual to a finite dimensional matrix integral with spectral edge \eqref{jtcurve} contains not only pure gravity but also a tower of massive fields. In an $L$ dimensional matrix integral we can make sense of correlators with up to $L$ boundaries. JT gravity does not capture all the information of these complex observables but it does capture their low energy features.

In this section we construct alternative definitions of JT gravity which differ from that of \cite{sss2} by the inclusion of certain spacetime D branes in the gravitational path integral. These are much like the D branes in string theory except now pieces of spacetime attach to the branes instead of pieces of worldsheet. In particular we aim to build intuition about the role of the ensemble in gravity and the role of branes in factorization. We first work within the controlled setup of a finite dimensional matrix integral. We can interpret the results of that analysis in the double scaling limit to link with the role of branes in JT gravity. 

There is no strict notion of a complete set of states within pure JT gravity as it is just meant do describe the low energy sector of some parent theory. However it makes perfect sense to construct a basis of states in the parent theory and ask what these look like from the low energy JT gravity point of view.
%%%%%%%%%%%%%%%%%%%%%%%%%%%%%%%%%%%%%%%%%%%%%%%%%
\subsection{Hilbert space of baby universes in random matrix theory}\label{sect41}
We construct the Hilbert space of baby universes of an $L$ dimensional matrix integral which we choose to interpret as a gravitational theory of discrete surfaces. One could read section \ref{sect:eigenvalue} as the double scaling limit of this more controlled matrix integral construction.
\\~\\
\textbf{\emph{Eigenvalues}}
\\~\\
Macroscopic loops $Z(\beta)$ and the inverse Laplace transforms $\rho(E)$ introduce boundaries in the discrete geometries. We can construct a Hilbert space of baby universes by acting with these boundary creating operators on the no boundary state. Their eigenstates form a basis of alpha states \cite{maxfieldmarolf}. Consider now \cite{paper6,mehta}
\begin{align}
    \bra{\rho(E_1)}\ket{\rho(E_2)}&=\rho(E_1,E_2)\label{85}\\&=\frac{1}{\mathcal{Z}_\lambda}  \int \prod_{a=1}^L d\lambda_a\,e^{-L V(\lambda_a)}\,\prod_{b<a}^L (\lambda_a-\lambda_b)^2\sum_{c=1}^L \delta(E_1-\lambda_c) \sum_{d=1}^L \delta(E_2-\lambda_d)\nonumber \\
    &=\rho(E_1)\,\delta(E_1-E_2)+\rho(E_1)\rho(E_2)-\rho(E_1)\rho(E_2)\,\sinc^2\,\pi\rho(E_1)(E_1-E_2)\,.\nonumber
\end{align}
The contact term is due to the diagonal terms in the sum on the second line. The smooth terms vanish when the energies are close together as testimony to level repulsion in quantum chaotic systems. In some sense these states are locally orthogonal
\begin{equation}
    \bra{\rho(E_1)}\ket{\rho(E_2)}=\rho(E_1)\,\delta(E_1-E_2)+\text{smoother}\,.
\end{equation}
More in general we consider products of such fixed energy boundaries. Taking $E_1<\dots <E_n$ and $M_1<\dots <M_n$ one deduces
\begin{equation}
    \bra{\rho(E_1)\dots \rho(E_n)}\ket{\rho(M_1)\dots \rho(M_n)}=\rho(E_1\dots E_n)\prod_{i=1}^n\delta(E_i-M_i)+\text{smoother}\,.
\end{equation}
This least analytic contribution is immediately recognized as a contact term contribution in the matrix integral calculation of $\rho(E_1\dots E_n,M_1\dots M_n)$. For $n\ll L$ the smooth contributions have a universal answer in terms of the sine kernel \cite{mehta}. Here we are interested in a more extreme scenario with $L$ boundaries. One finds that such states are orthogonal
\begin{equation}
    \bra{\rho(E_1)\dots \rho(E_L)}\ket{\rho(M_1)\dots \rho(M_L)}=\rho(E_1\dots E_L) \prod_{i=1}^L \delta (E_i-M_i)\,.\label{93}
\end{equation}
This is elementary to prove by manipulating the operator insertion in the matrix integral
\begin{align}
\displaybreak[0]
    \nonumber \prod_{i=1}^L \sum_{a_i=1}^L \delta(E_i-\lambda_{a_i})\prod_{j=1}^L \sum_{b_j=1}^L \delta(M_j-\lambda_{b_j})&=\prod_{j=1}^L \delta(E_j-M_j)\sum_{\pi}\prod_{i=1}^L \delta(E_i-\lambda_{\pi(i)})\\&=\prod_{i=1}^L \delta(E_i-M_i)\prod_{i=1}^L \sum_{a_i=1}^L \delta(E_i-\lambda_{a_i}).
\end{align}
There are $L$ different energies $E_1\dots E_L$ that are each fixed to one of the $L$ eigenvalues. The only nonzero result is when each energy is fixed to a different eigenvalue. The first term on the second line factors out of the ensemble \eqref{matint} proving \eqref{93}. We can define a basis as
\begin{equation}
    \ket{E_1\dots E_L}=\frac{\ket{\rho(E_1)\dots \rho(E_L)}}{\sqrt{\rho(E_1\dots E_L) }}\,.\label{estates}
\end{equation}
Note that states with more than $L$ boundaries are null.

Using the definition of the matrix integral \eqref{matint} we can compute the inner product with the no boundary state
\begin{equation}
    \bra{\text{HH}}\ket{\rho(E_1)\dots \rho(E_L)}=\frac{L!}{\mathcal{Z}_\lambda}\prod_{a=1}^L e^{-L V(E_a)}\,\prod_{b<a}^L (E_a-E_b)^2=\rho(E_1\dots E_L).\label{96}
\end{equation}
The prefactor is due to the sum over permutations $\pi$. This set of states for $E_1<\dots <E_L$ is complete if we restrict to observables which are independent of eigenvector components in the basis $\ket{i}$. Within this subset of observables we have constructed a basis for the Hilbert space of baby universes. For example by inserting a complete set of these states one writes
\begin{equation}
    \bra{\text{HH}}\Tr(\mo)\ket{\text{HH}}=\int_{E_1<\dots <E_L} \prod_{a=1}^L d E_a\, \bra{\text{HH}}\ket{E_1\dots E_L} \bra{E_1\dots E_L}\Tr(\mo)\ket{\text{HH}}\,. \label{hhtro}
\end{equation}
One finds via a direct matrix integral calculation
\begin{equation}
    \bra{E_1\dots E_L}\Tr(\mo)\ket{\text{HH}}=\sum_{a=1}^L \mo(E_a) \bra{E_1\dots E_L}\ket{\text{HH}}\,.
\end{equation}
Using \eqref{96} and the symmetries of the integrand we can rewrite \eqref{hhtro} as
\begin{equation}
    \bra{\text{HH}}\Tr(\mo)\ket{\text{HH}}=\frac{1}{\mathcal{Z}_\lambda}\int \prod_{a=1}^L d E_a\,e^{-L V(E_a)}\,\prod_{b<a}^L (E_a-E_b)^2\sum_{c=1}^L \mo(E_c)\,=\average{\Tr(\mo)}.\label{4.27}
\end{equation}
This is a realization of the general phenomenon that the gravitational path integral naturally computes ensemble averages. Here it implements a matrix integral. Observables in an alpha state correspond to observables in one quantum system with data $\alpha$ (here the energies) and the measure of the ensemble is due to the overlap between the no boundary state and alpha states
\begin{equation}
    \average{\Tr(\mo)}=\int d\alpha\,\rvert \bra{\text{HH}}\ket{\alpha}\rvert^2\,\Tr(\mo_\alpha)\,.
\end{equation}

One checks by computing overlaps with a generic state that these alpha states diagonalize any operator which is independent of the eigenvector components. For example
\begin{equation}
    Z(\beta_1)Z(\beta_2)\ket{E_1\dots E_L}=\sum_{a=1}^L e^{-\beta E_a}\sum_{b=1}^L e^{-\beta E_b} \ket{E_1\dots E_L}.\label{100}
\end{equation}
This factorization property is quite trivial from the matrix integral point of view. However it is rather surprising from the point of view of the gravitational path integral \cite{sss2,phil,paper6,rw1}.

Before proceeding we note that these fixed energy boundaries can be interpreted as gravitational branes. For example we have for $L\gg 1$ the matrix integral identity \cite{sss2,paper6}
\begin{equation}
    \rho(E)=\frac{1}{2\pi}\langle \psi^2(E)\rangle\quad,\quad \psi(E)=\exp(\text{Disk}(E))\,.\label{4.30}
\end{equation}
Disk$(E)$ is a different type of fixed energy gravitational boundary \cite{sss2,paper6} which in Liouville gravity corresponds to a circular boundary with FZZT boundary conditions \cite{fzz,t}. The exponential reflects that any number of such indistinguishable boundaries may exist and contribute to observables. It is essentially the definition of what me mean by a brane.
\\~\\
\textbf{\emph{Eigenstates}}
\\~\\
The matrix integral \eqref{matint} is an ensemble average over eigenvalues but also over components of the eigenvector $\ket{a}$ of the Hamiltonian in the rigid basis $\ket{i}$. We want to associate alpha states with these eigenvector components too. This is ultimately important to understand how products of matrix elements such as $\bra{i}e^{-\beta H}\ket{i}$ can factorize in gravity.

The point is that eigenvector component operators $\bra{i}\ket{a}$ also define in some sense boundary conditions for the gravitational path integral as explained in more detail in section \ref{sect:eigenvalue}. At the level of the matrix integral this can be appreciated by comparing with a discussion in appendix D of \cite{rw1} which relates mass $\mu$ geodesic EOW branes in gravity to states $\ket{i}$
\begin{equation}
    \ket{i}=\sum_{a=1}^L \psi(E_a)\,U_{i a}\ket{a}\quad,\quad \psi(E_a)=2^{1-2\mu}\Gamma(\mu-1/2\pm i\sqrt{2 E_a})\,.\label{104}
\end{equation}
The $U_{i a}$ are taken as independent Gaussian random complex variables in \cite{rw1}. Eigenvectors fit into the same framework by choosing a different boundary wavefunction in the energy basis $\psi(E_a)=1$. So eigenvector boundaries are just EOW branes although they differ from the mass $\mu$ branes in \cite{rw1} by the boundary conditions in the gravitational path integral. 

We proceed with the understanding that we can construct a multi boundary gravitational Hilbert space by acting with products of matrix elements $U_{i a}$ on the no boundary state. We are looking for the alpha states that diagonalize these operators
\begin{equation}
    U_{i a}\ket{Z}=Z_{i a}\ket{Z}\,.\label{106}
\end{equation}
The idea is to make such states by acting with a delta operator on the no boundary state
\begin{equation}
    \ket{\psi(Z_{i a})}=\delta(Z_{i a}-U_{i a})\ket{\text{HH}}=\int d P_{i a}\, e^{-i P_{i a} Z_{i a}}\,\ket{\psi(P_{i a})}\quad,\quad \ket{\psi(P_{i a})}=e^{i P_{i a}\,U_{i a}}\ket{\text{HH}}\,.\label{4.33}
\end{equation}
The operator $\psi(P_{i a})$ is a brane of eigenvector boundaries $U_{i a}$ with coupling $P_{i a}$ which makes $\psi(Z_{i a})$ a Fourier transformed brane. Similar branes were identified as alpha states in the model of \cite{maxfieldmarolf}. Via elementary Gaussian integration using the definition of the matrix integral \eqref{matint} one finds
\begin{equation}
    \bra{\psi(P_{i a})\psi(P^*_{i a})}\ket{\psi(W_{i a})\psi(W_{i a}^*)}=e^{-\rvert P_{i a}+K_{i a}\rvert^2/L}\,.
\end{equation}
By taking multiple Fourier transforms one obtains an orthogonal set
\begin{equation}
    \bra{\psi(Z_{i a})\psi(Z^*_{i a})}\ket{\psi(W_{i a})\psi(W_{i a}^*)}=\delta(Z_{i a}-W_{i a})\,\delta(Z_{i a}^*-W_{i a}^*)\,\average{\psi(Z_{i a})\psi(Z^*_{i a})}\,.
\end{equation}
Via similar steps one computes
\begin{equation}
    \bra{\text{HH}}\ket{\psi(Z_{i a})\psi(Z^*_{i a})}=\average{\psi(Z_{i a})\psi(Z^*_{i a})}=\frac{1}{\mathcal{Z}_{Z_{i a}}} e^{- L \rvert Z_{i a}\rvert^2}\,.\label{109}
\end{equation}
We obtain a basis by normalizing and taking direct products
\begin{equation}
    \ket{Z}=\bigotimes_{a=1}^L \bigotimes_{i=1}^L\frac{\ket{\psi(Z_{i a})\psi(Z_{i a}^*)}}{\sqrt{\average{\psi(Z_{i a})\psi(Z_{i a}^*)}}}\,.\label{zstates}
\end{equation}
By inserting a complete set of these states we write
\begin{equation}
    \bra{\text{HH}}U_{j b}\, U_{k c}^*\ket{\text{HH}}=\int \prod_{a=1}^L \prod_{i=1}^L d Z_{i a}\,d Z_{i a}^*\,\bra{\text{HH}} \ket{Z  }\bra{Z}U_{j b}\, U_{k c}^*\ket{\text{HH}}\,.\label{113}
\end{equation}
Because the branes $\psi(P_{i a})$ are plane waves we have
\begin{align}
    \bra{\text{HH}}U_{i a}\ket{\psi(P_{i a})\psi(P_{i a}^*)}&=-i\partial_{P_{i a}} \bra{\text{HH}}\ket{\psi(P_{i a})\psi(P_{i a}^*)}\,.
\end{align}
From this we derive that the Fourier transformed branes are eigenstates
\begin{align}
    \bra{\text{HH}}U_{i a}\ket{Z}&=Z_{i a} \bra{\text{HH}}\ket{Z}\,.\label{4.40}
\end{align}
Combining this with \eqref{109} we can rewrite \eqref{113} as
\begin{equation}
     \bra{\text{HH}}U_{j b}\, U_{k c}^*\ket{\text{HH}}=\frac{1}{\mathcal{Z}_U}\int \prod_{a=1}^L \prod_{i=1}^L d Z_{i a}\,d Z_{i a}^*\,e^{- L \rvert Z_{i a}\rvert^2 }\, Z_{j b}\,Z_{k c}^*\,=\average{Z_{j b}\,Z_{k c}^*}.
\end{equation}
We see that the ensemble average over eigenvector components emerges from a gravitational path integral in the no boundary state. As anticipated in the introduction the alpha states of the matrix ensemble are a direct product of eigenvalue alpha states and eigenvector alpha states. These direct products diagonalize more complex boundary creating operators like
\begin{equation}
    \frac{\bra{\text{HH}}\bra{i} e^{-\beta_1 H}\ket {i}\bra{i}e^{-\beta_2 H}\ket{i}\,\ket{Z\,E_1\dots E_L}}{\bra{\text{HH}}\ket{Z\,E_1\dots E_L}}=\sum_{a=1}^L e^{-\beta_1 E_a}\,\rvert Z_{i a}\rvert^2 \sum_{b=1}^L e^{-\beta_2 E_b}\,\rvert Z_{i b}\rvert^2\,.\label{4.42}
\end{equation}
In other words the ratio of gravitational path integrals on the left spits out a result that is identical to computing $\bra{i} e^{-\beta_1 H}\ket {i}\bra{i}e^{-\beta_2 H}\ket{i}$ in a quantum mechanical system with fixed Hamiltonian matrix.
%%%%%%%%%%%%%%%%%%%%%%%%%%%%%%%%%%%%%%%%%%
\subsection{Branes and factorization}\label{sect:eigenvalue}
The more interesting part of this exercise is to understand what these alpha states look like from a geometrical point of view. For example we would like to reproduce this answer for the expectation value of $\bra{i} e^{-\beta_1 H}\ket {i}\bra{i}e^{-\beta_2 H}\ket{i}$ from the gravitational path integral. Here we include the branes which define the alpha states in the gravitational path integral and sum over geometries.
\\~\\
\textbf{\emph{Eigenvalues}}
\\~\\
The role of eigenvalues in JT gravity can be understood by defining the gravitational path integral to include $L$ fixed energy boundaries $\rho(E_1\dots E_L)$ which were called eigenbranes in \cite{paper6}. The correlator $\rho(E_1\dots E_L)$ corresponds in perturbative JT gravity to summing over all geometries which end on these fixed energy boundaries
\begin{equation}
    \rho(E_1\dots E_L)\supset\quad \raisebox{-7mm}{\includegraphics[width=42mm]{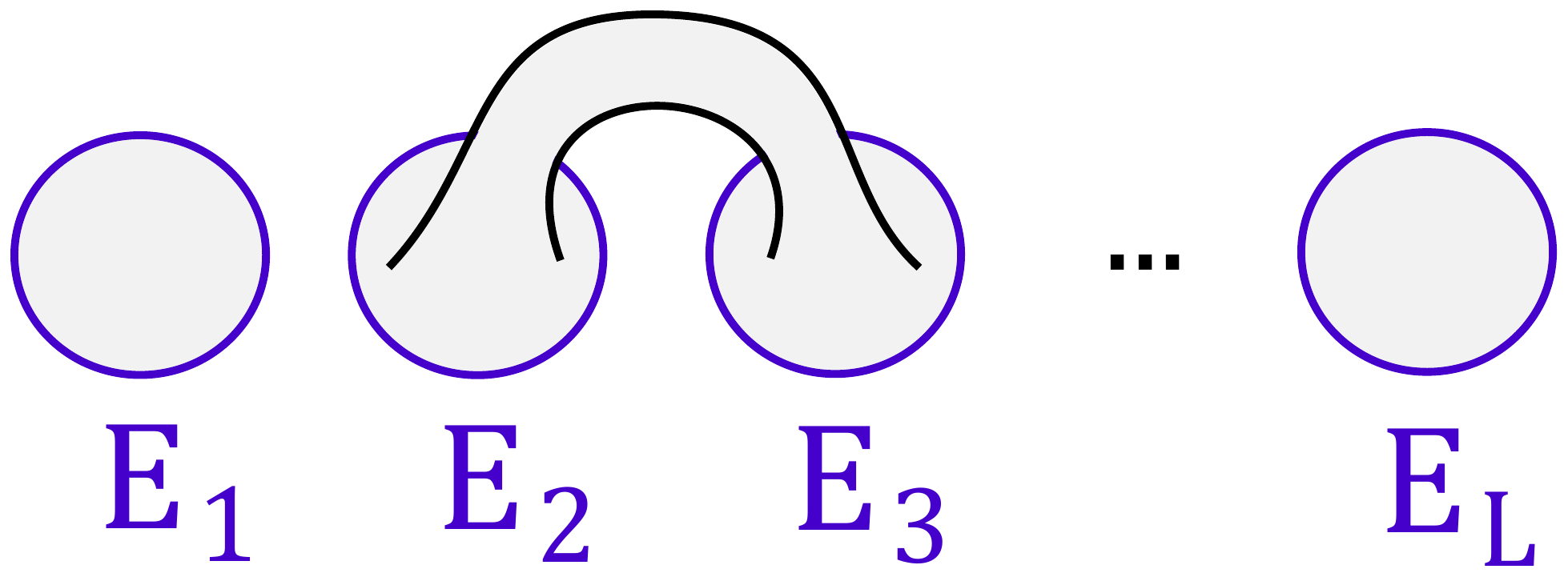}}\quad .\label{123}
\end{equation}
This eigenbrane partition function is defined here to be the bare JT gravity path integral. Within JT gravity we can only genuinely makes sense of the low energy boundaries $E \ll \Lambda$. The other branes get their meaning by the embedding in an $L$ dimensional matrix integral.

We can compute observables in this gravitational theory with branes \cite{paper6}. For example inserting a probe fixed energy boundary $\rho(M)$ we are led sum over all geometries which end on the union of this probe boundary and the eigenbranes
\begin{equation}
    \rho(M,E_1\dots E_L)\supset \quad \raisebox{-7mm}{\includegraphics[width=65mm]{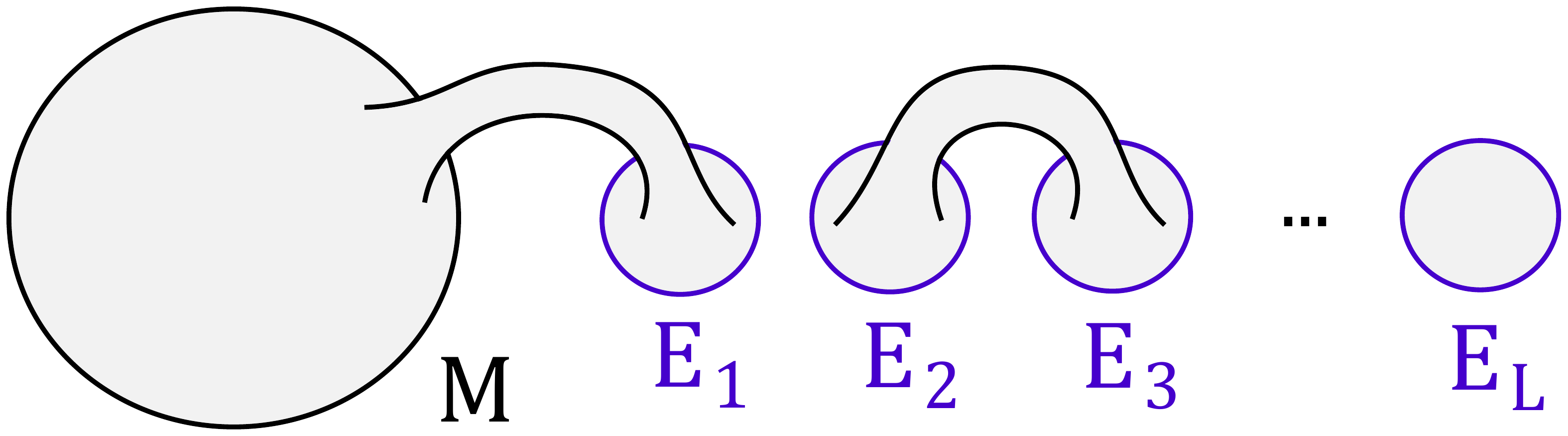}}\quad.\label{124}
\end{equation}
Probe boundaries (black) are distinguished from eigenbranes (blue). There are also contact terms due to contributions where several of the fixed energy boundaries have merged \cite{paper6,mergers}. For example
\begin{equation}
    \rho(M,E_1\dots E_L)\supset\quad \raisebox{-7mm}{\includegraphics[width=65mm]{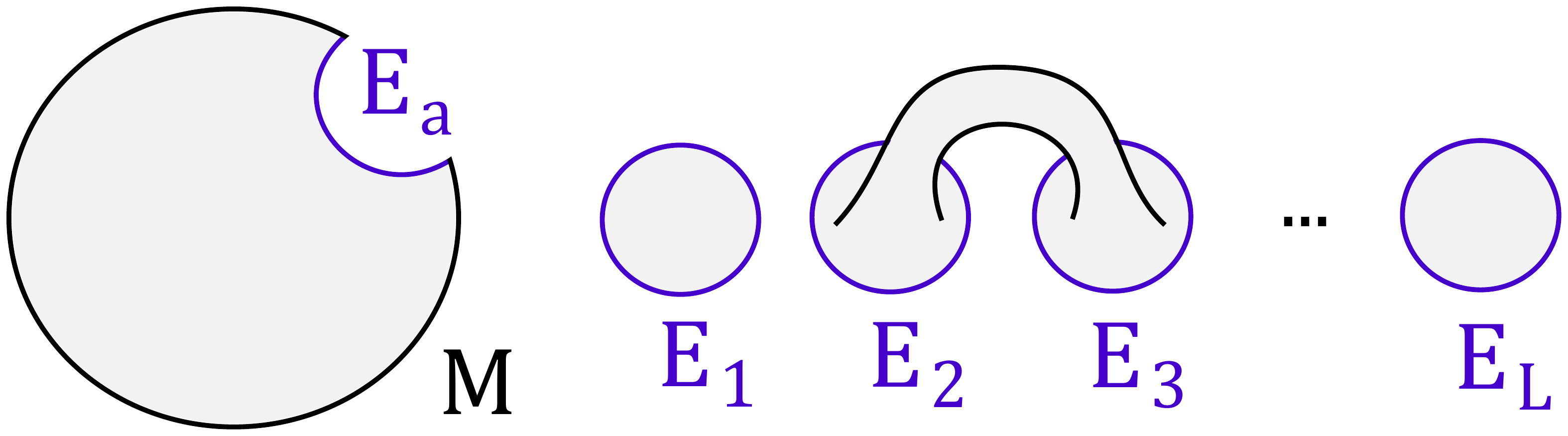}}\quad.\label{125}
\end{equation}
Summing over such mergers one indeed recovers the contact terms in the matrix integral language \cite{paper6}
\begin{equation}
    \rho(M,E_1\dots E_L)\supset \sum_{a=1}^L \delta(M-E_a)\,\rho(E_1\dots E_L)\,.\label{125}
\end{equation}
In taking the sum note that there are no contributions to the gravitational path integral from geometries connecting the merged probe boundary to the eigenbranes. Furthermore eigenbranes cannot merge with each other as their energies are all different \cite{paper6}. 

Normalizing by the eigenbrane partition function \eqref{123} we find that the expectation value of a probe boundary is a conditional probability in the original matrix ensemble
\begin{equation}
    \rho(M)_{E_1\dots E_L}=\frac{\rho(M, E_1\dots E_L)}{\rho(E_1\dots E_L)}=\sum_{a=1}^L \delta(M-E_a)\,.\label{127}
\end{equation}
Only the contact term contributions remain so the sum of all smooth contributions vanishes. This can be appreciated using the gravitational path integral by calculating for $n\ll e^{S_0}$ \cite{paper6}
\begin{equation}
    \rho(M)_{E_1\dots E_n}=\frac{\rho(M, E_1\dots E_n)}{\rho(E_1\dots E_n)}=\sum_{a=1}^n \delta(M-E_a)+\text{smooth}\,.
\end{equation}
The smooth term is due to geometries with no mergers. Via explicit calculations using the gravitational path integral one finds that the smooth term tends to vanish if $M$ is close to the fixed eigenvalues. This is the gravitational translation of level repulsion in random matrix theory. To see this it is important on a technical level to include appropriate nonperturbative corrections. These are captured by working with the gravitational brane pairs introduced in \eqref{4.30}. For details see \cite{sss2,paper6,shih}. For more eigenbranes the gravitational calculations become inaccurate. Fortunately we can make sense of them via the matrix integral definition. By Laplace transforming we get the answer for a fixed length probe boundary in JT gravity
\begin{equation}
    \average{Z(\beta)}_{E_1\dots E_L}=\sum_{a=1}^L e^{-\beta E_a}=\sum_{a=1}^L \quad \raisebox{-7mm}{\includegraphics[width=21mm]{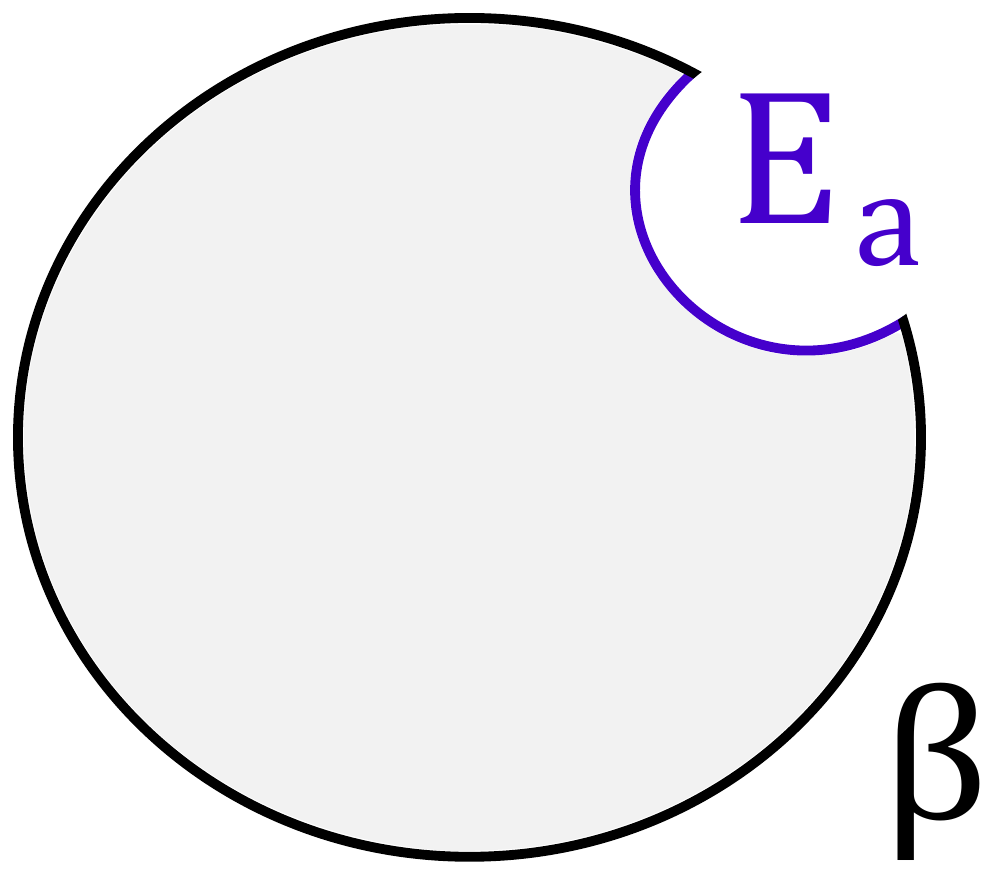}}\quad.\label{128}
\end{equation}

We would now like to understand factorization of $Z(\beta_1)Z(\beta_2)$ in this version of JT gravity with eigenbranes. Consider first the inverse Laplace transform as was discussed in \cite{paper6}. We sum over geometries that end on the two probe boundaries and the eigenbranes 
\begin{equation}
    \rho(M_1,M_2,E_1\dots E_L)\supset \quad \raisebox{-7mm}{\includegraphics[width=88mm]{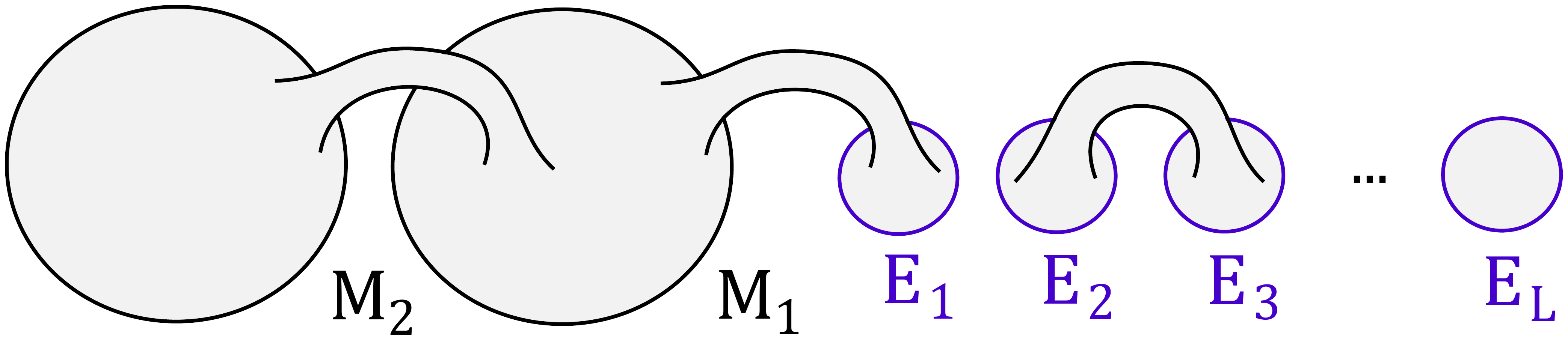}}\quad. \label{129}
\end{equation}
The contact terms arise due to mergers. For example
\begin{equation}
    \rho(M_1,M_2,E_1\dots E_L)\supset \quad \raisebox{-7mm}{\includegraphics[width=88mm]{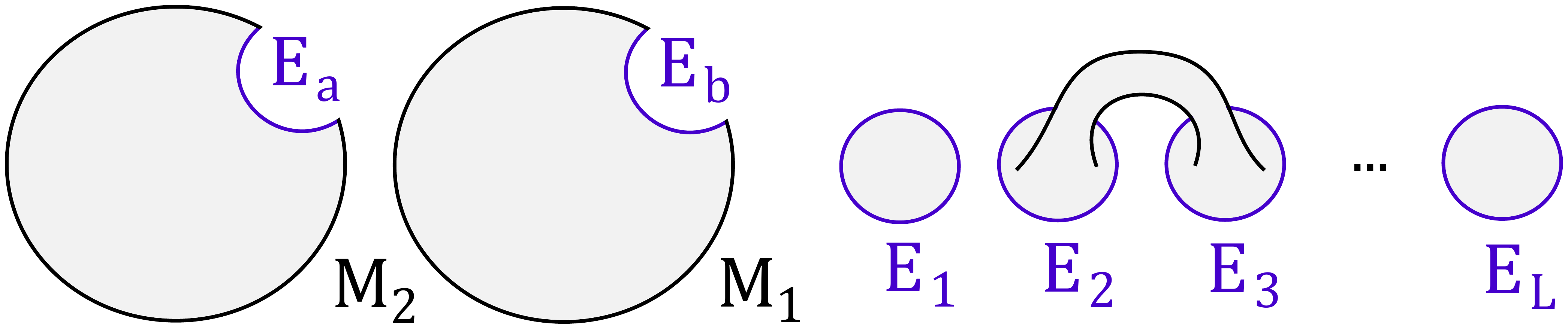}}\quad.
\end{equation}
This includes no manifestly diagonal terms as there is only one of each eigenbrane. However we must account for the possibility that the two probe boundaries merge and eigenbranes can merge into this merged probe boundary hence providing diagonal contact terms. In the end one finds for the expectation value \cite{paper6}
\begin{align}
    \rho(M_1,M_2)_{E_1\dots E_L}=\frac{\rho(M_1,M_2, E_1\dots E_L)}{\rho(E_1\dots E_L)}&= \sum_{a=1}^L\delta(M_1-E_a)\sum_{b=1}^L \delta(M_2-E_b)\,%\nonumber\\&=\rho(M_1)_{E_1\dots E_L}\,\rho(M_2)_{E_1\dots E_L}\,.
    \label{131}
\end{align}
Again only the contact terms survive and the smooth contributions vanish. This is motivated from the gravitational path integral by repeating this calculation in a theory with $n\ll e^{S_0}$ eigenbranes \cite{paper6}. This conclusion generalizes to any number of probe boundaries. Laplace transforming to the fixed length basis the surviving contributions due to mergers are
\begin{align}
    &\average{Z(\beta_1)Z(\beta_2)}_{E_1\dots E_L}=\sum_{a=1}^L\quad \raisebox{-7mm}{\includegraphics[width=22mm]{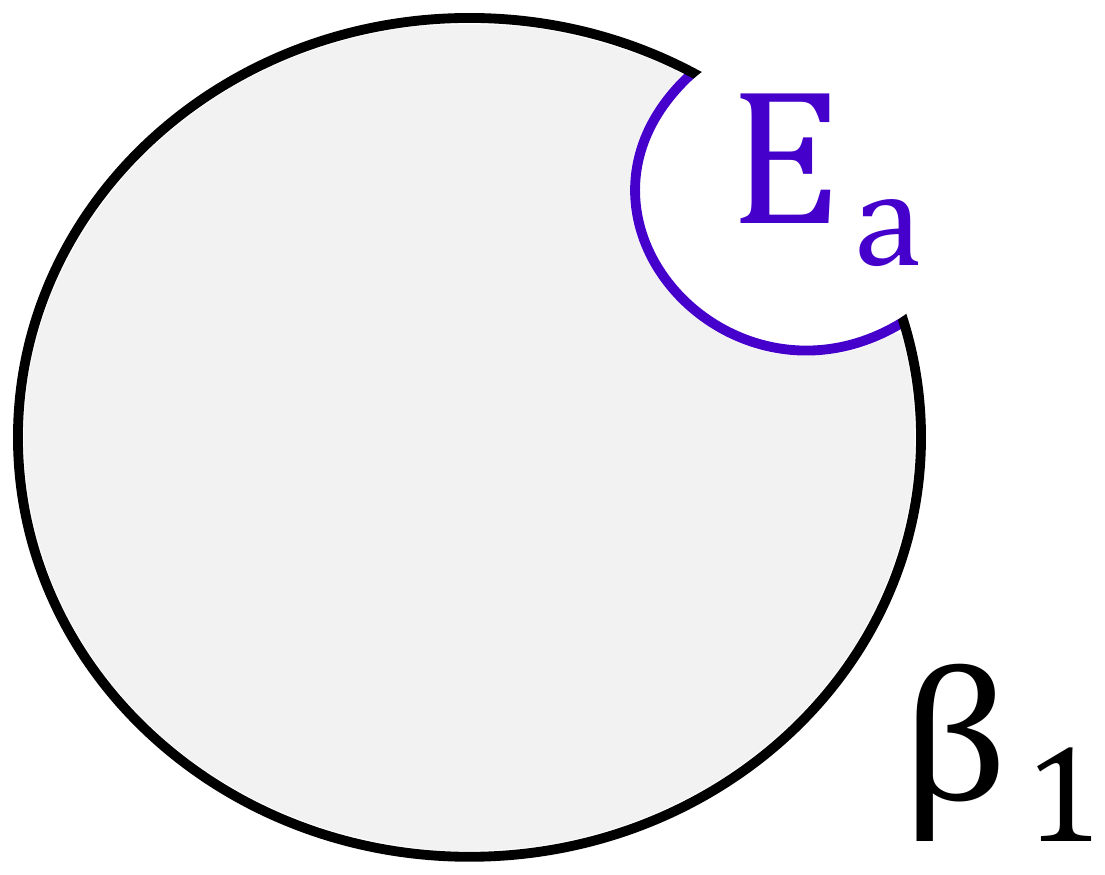}}\quad \sum_{b\neq a}^L \quad \raisebox{-7mm}{\includegraphics[width=22mm]{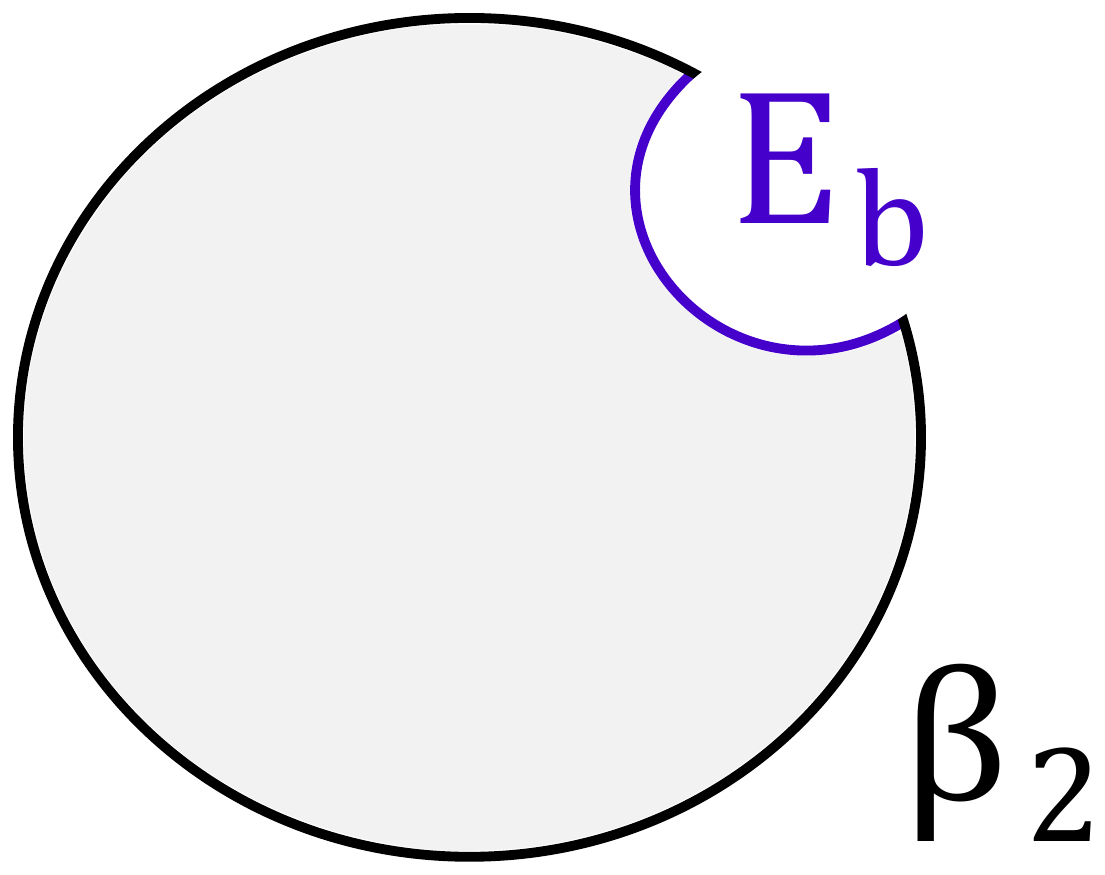}}\quad \label{132}\\
    &\qquad\qquad\qquad\qquad\qquad\quad\,\,\,\, +\sum_{a=1}^L\quad \raisebox{-7mm}{\includegraphics[width=42mm]{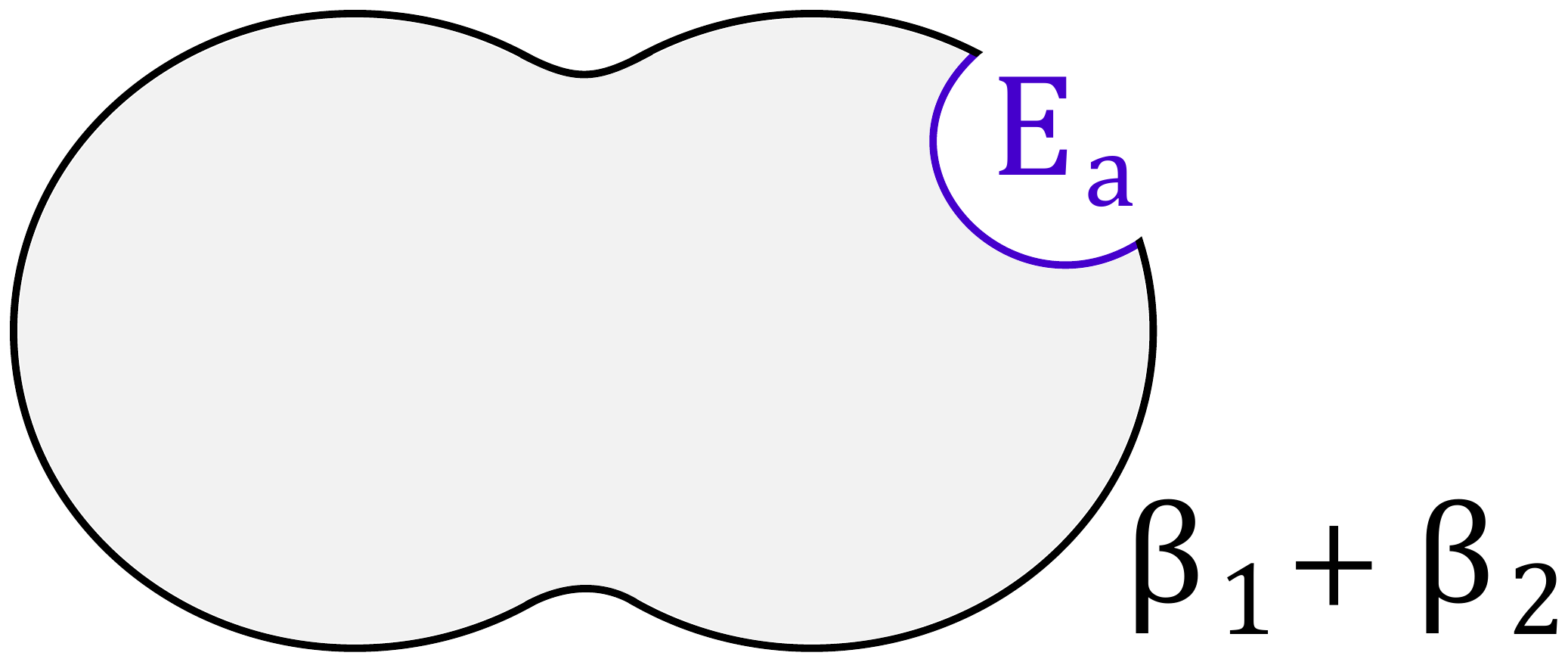}} =\average{Z(\beta_1)}_{E_1\dots E_L}\,\average{Z(\beta_2)}_{E_1\dots E_L}\nonumber \,.
\end{align}
The connected contribution is numerically identical to the missing diagonal terms and so this observable factorizes.

Let us pause and contemplate the geometric implications of this factorization. Drawing all geometries that contribute to the left and right of \eqref{132} we see that most terms are manifestly identical. Scratching them only connected terms (in a broad sense) remain. For example on the left there is an annulus connecting the probe boundaries and on the right there are diagonal terms where the probe boundaries connect to matching eigenbranes \cite{paper6}\footnote{In the first sum the probe boundaries can only connect to the top row of eigenbranes. In the second sum the left probe boundary only connects to the top row and the right boundary only connects to the bottom row. Furthermore different rows of eigenbranes are not allowed to connect to each other. }
\begin{align}
\label{connect1}
\displaybreak[0]
		  \nonumber &\sum_{\text{connected}} \,  \raisebox{-11mm}{\includegraphics[width=85mm]{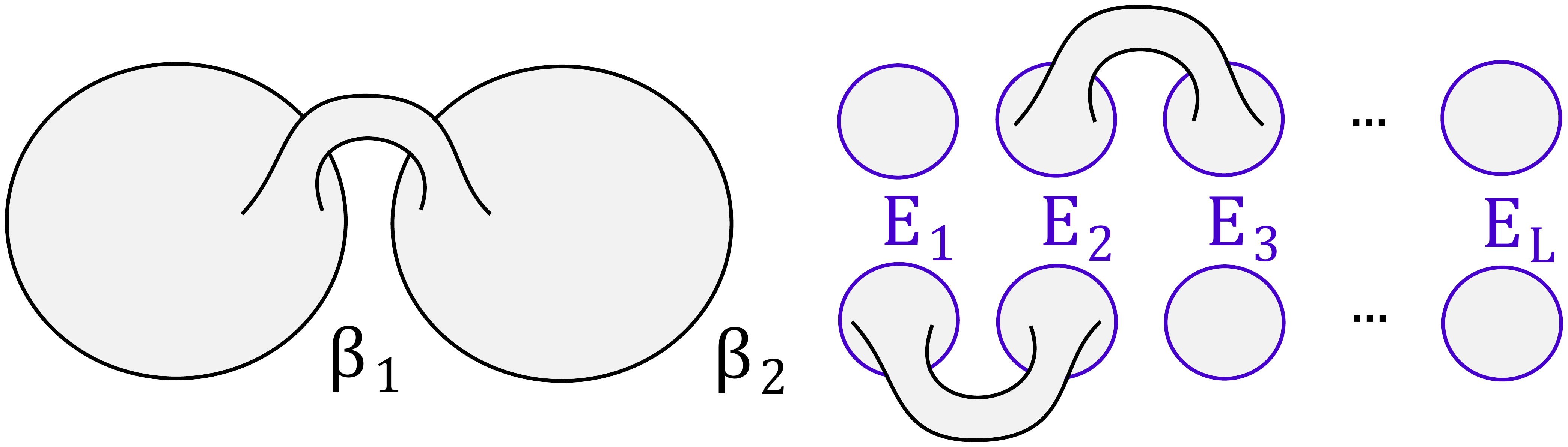}}\, \\&\qquad\qquad\qquad=\,\sum_{\text{connected}} \, \raisebox{-12mm}{\includegraphics[width=90mm]{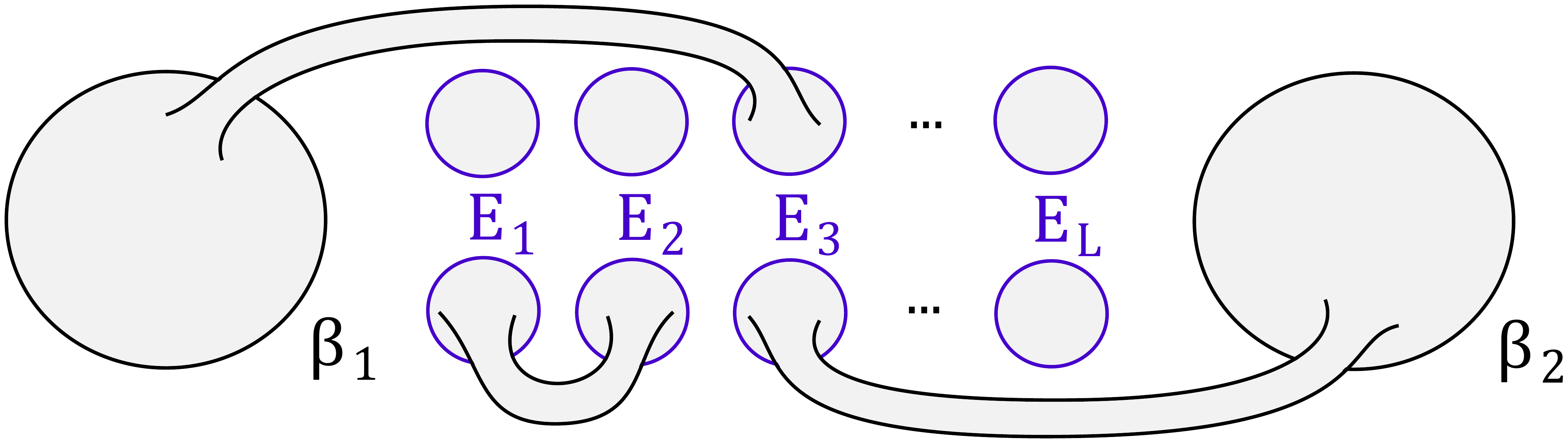}}\quad.
\end{align}
From a gravitational point of view it is surprising that these sums are numerically identical and nonzero. Factorization of multi boundary observables in gravity is not a consequence of the connected or diagonal contributions vanishing, they do not. Rather observables factorize because there are additional off diagonal terms due to the probe boundaries connecting to eigenbranes \cite{phil,rw1,malmaoz}.
\\~\\
\textbf{\emph{Eigenstates}}
\\~\\
We consider the effect of including eigenstate branes in the definition of the gravitational path integral. Such branes explain from the bulk point of view the factorization of observables that depend on eigenstate components. Consider first such observables in the ensemble averaged gravitational theory with no branes \cite{sss2}. We define a rigid basis of states $\ket{i}$ as
\begin{equation}
    \ket{i}=\sum_{a=1}^L U_{i a}\ket{a}\,.
\end{equation}
The matrix elements of $U$ are independent Gaussian random complex variables \eqref{18}. We see that these are structurally identical to the EOW brane states \eqref{104} considered in \cite{rw1}. The ensemble averaging over $U$ generates all Wick contractions between bras and kets in a certain collection of matrix elements. This sum was formulated as a rule for EOW branes in gravity in \cite{rw1}.

As a first example consider matrix element $\bra{i}e^{-\beta H}\ket{i}$
\begin{align}
    \average{\bra{i}e^{-\beta H}\ket{i}}&=
    \frac{1}{L}\,\average{Z(\beta)} =\quad \raisebox{-7mm}{\includegraphics[width=22mm]{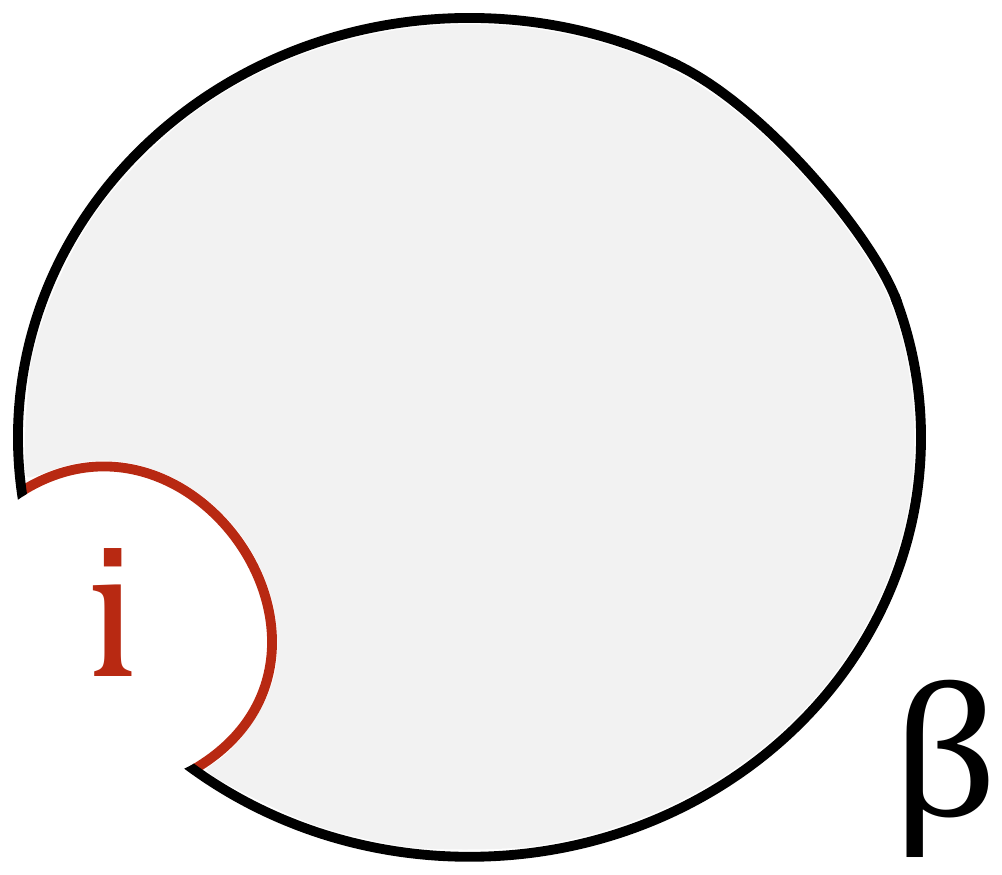}}\quad+\quad \raisebox{-7mm}{\includegraphics[width=22mm]{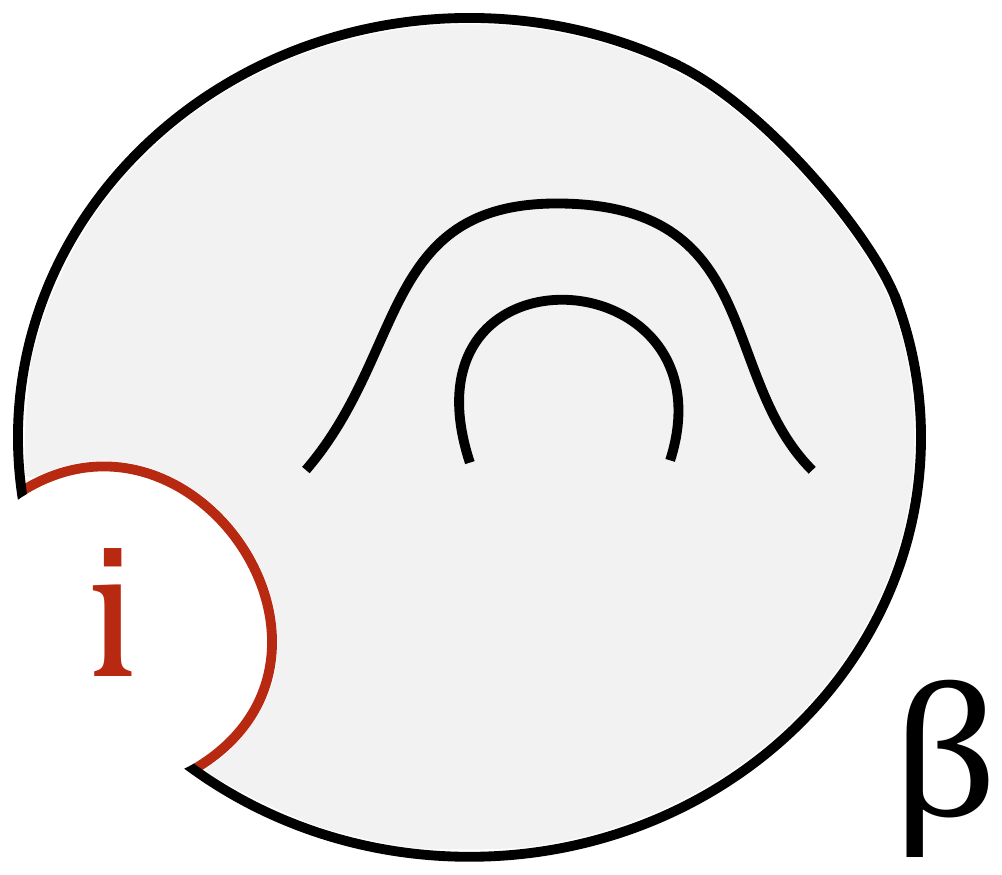}}\quad + \dots \label{133}
\end{align}
The only difference to the mass $\mu$ geodesic EOW branes of \cite{rw1} is the wavefunction $\psi(E)$ associated to the piece of eigenvector boundary (red) in the energy basis. The most generic EOW branes are characterized by a wavefunction $\psi(E)$
\begin{equation}
    \average{\bra{i}e^{-\beta H}\ket{i}}= \int_C d E\,e^{-\beta E}\,\rho(E)\,\frac{\psi(E)}{L}.\label{4.57}
\end{equation}
Different wavefunctions $\psi(E)$ imply different boundary conditions in the path integral at the vector boundary segment. The mass $\mu$ geodesic EOW branes of \cite{rw1} correspond to choosing $\psi(E)$ as in \eqref{104} whilst for the eigenvectors we choose $\psi(E)$ constant.

Another important example is the product of two matrix elements. Summing over Wick contractions between bras and kets one obtains
\begin{align}
\displaybreak[0]
    &\nonumber \average{\bra{i}e^{-\beta_1 H}\ket{i}\bra{i}e^{-\beta_2 H}\ket{i}}
    =\frac{1}{L^2}\,\average{Z(\beta_1+\beta_2)}+\frac{1}{L^2}\,\average{Z(\beta_1)Z(\beta_2)}\nonumber \\&=\quad \raisebox{-7mm}{\includegraphics[width=42mm]{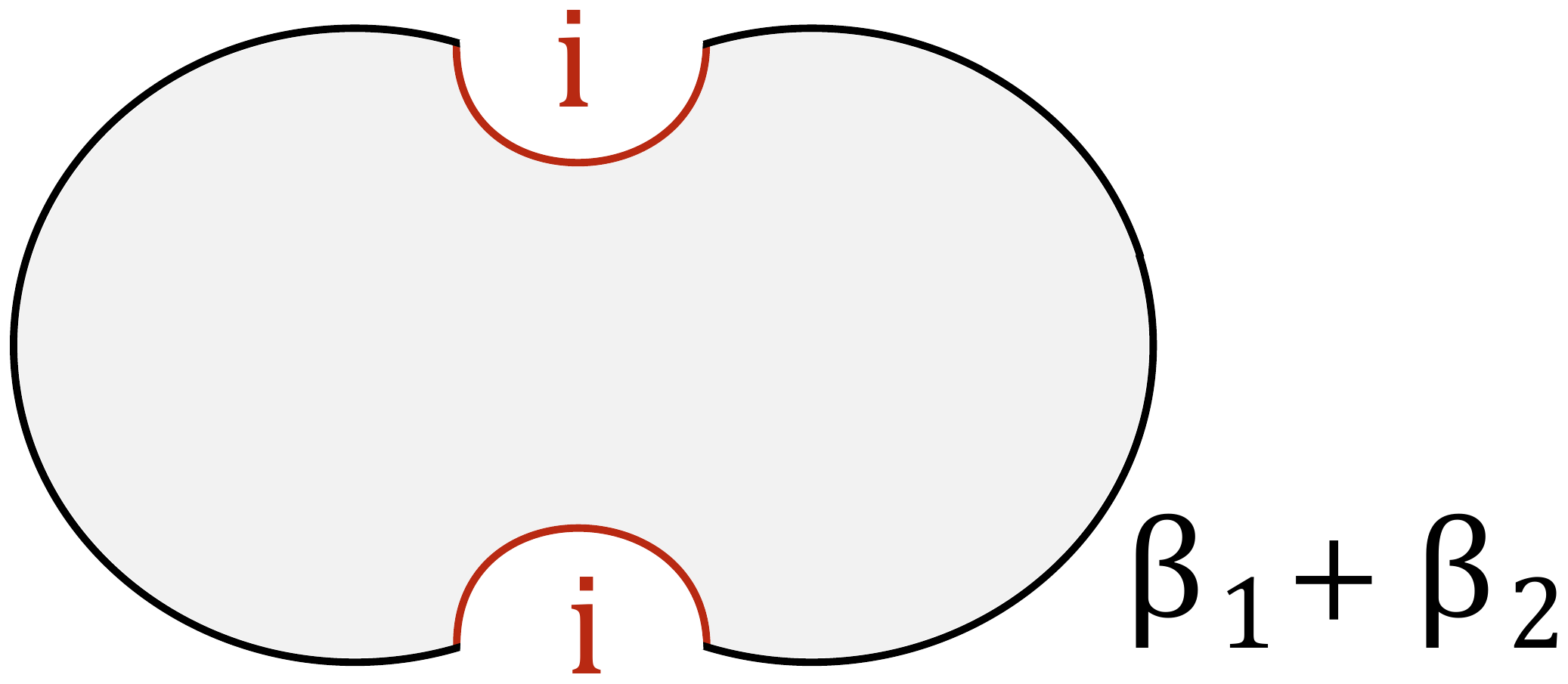}} +\quad \raisebox{-7mm}{\includegraphics[width=42mm]{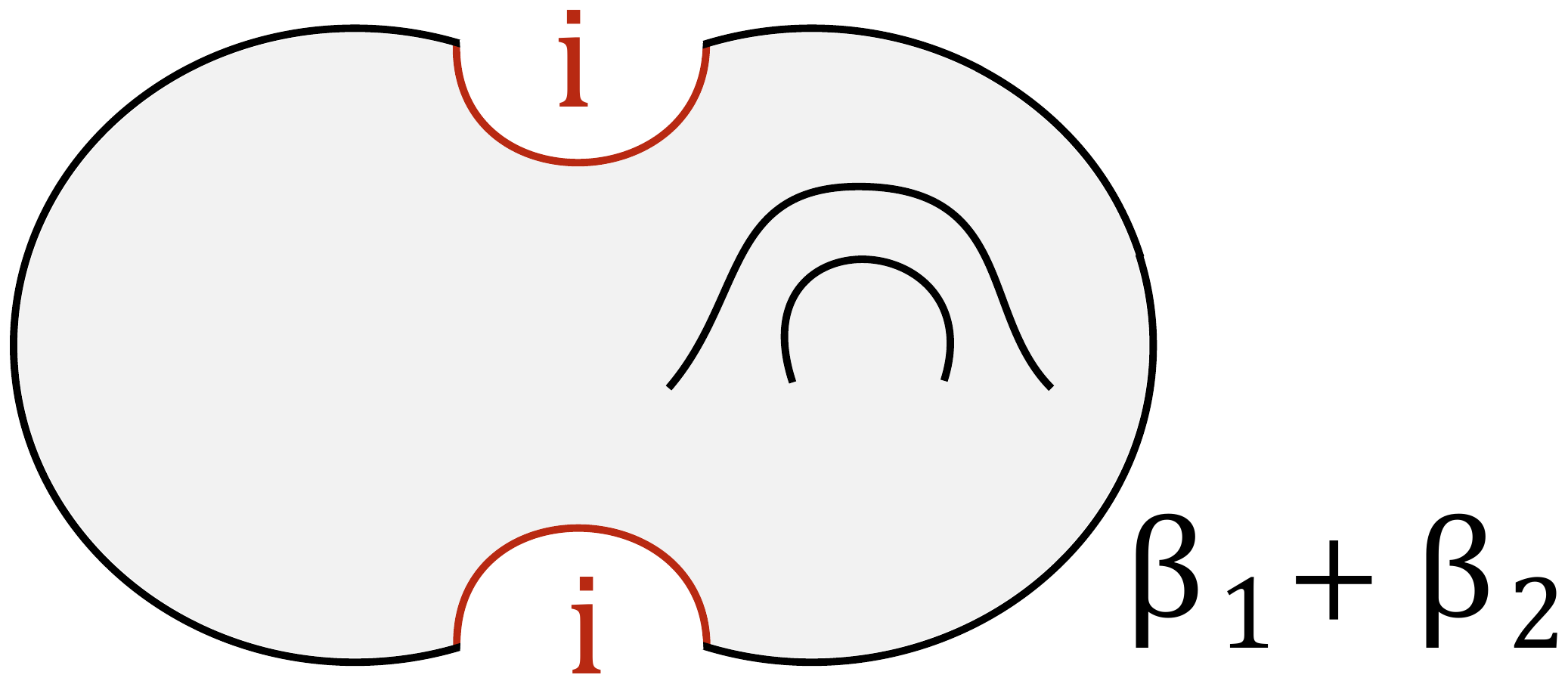}}+\dots\nonumber \\&\qquad\qquad + \quad \raisebox{-7mm}{\includegraphics[width=44mm]{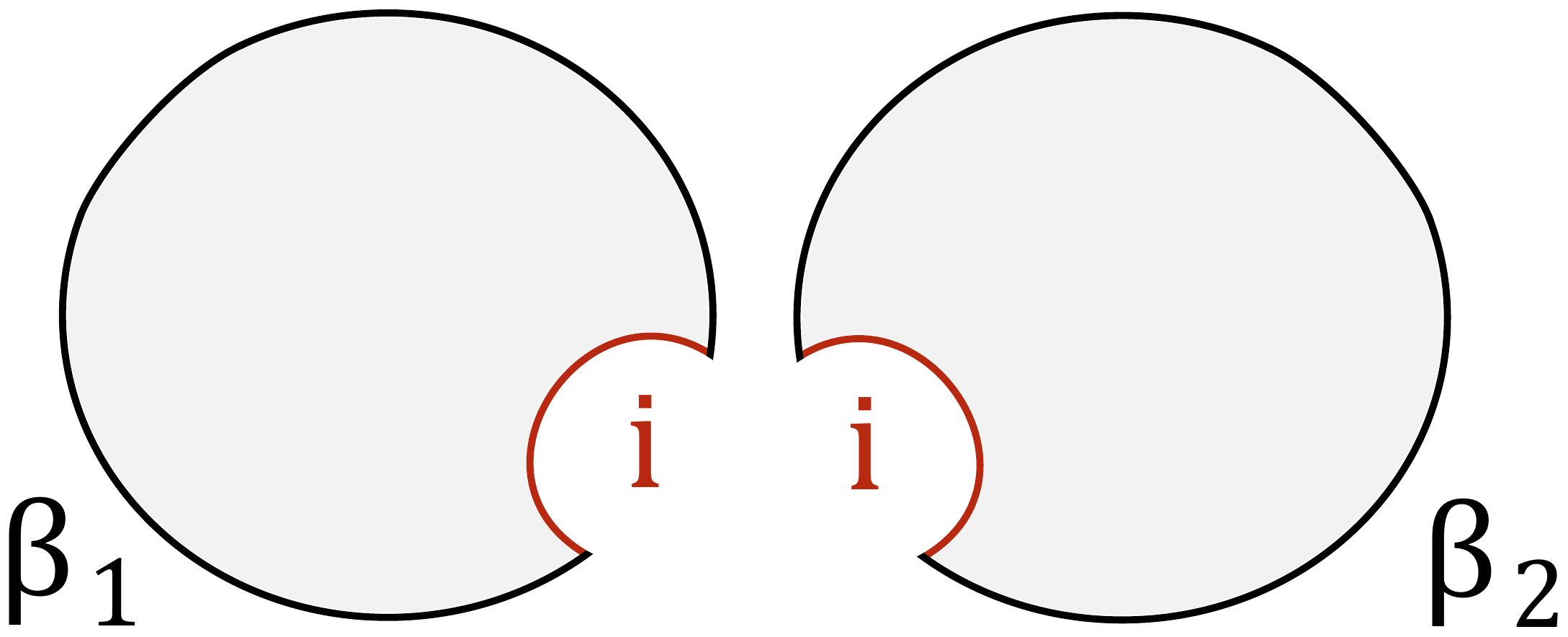}}\quad+\quad \raisebox{-7mm}{\includegraphics[width=44mm]{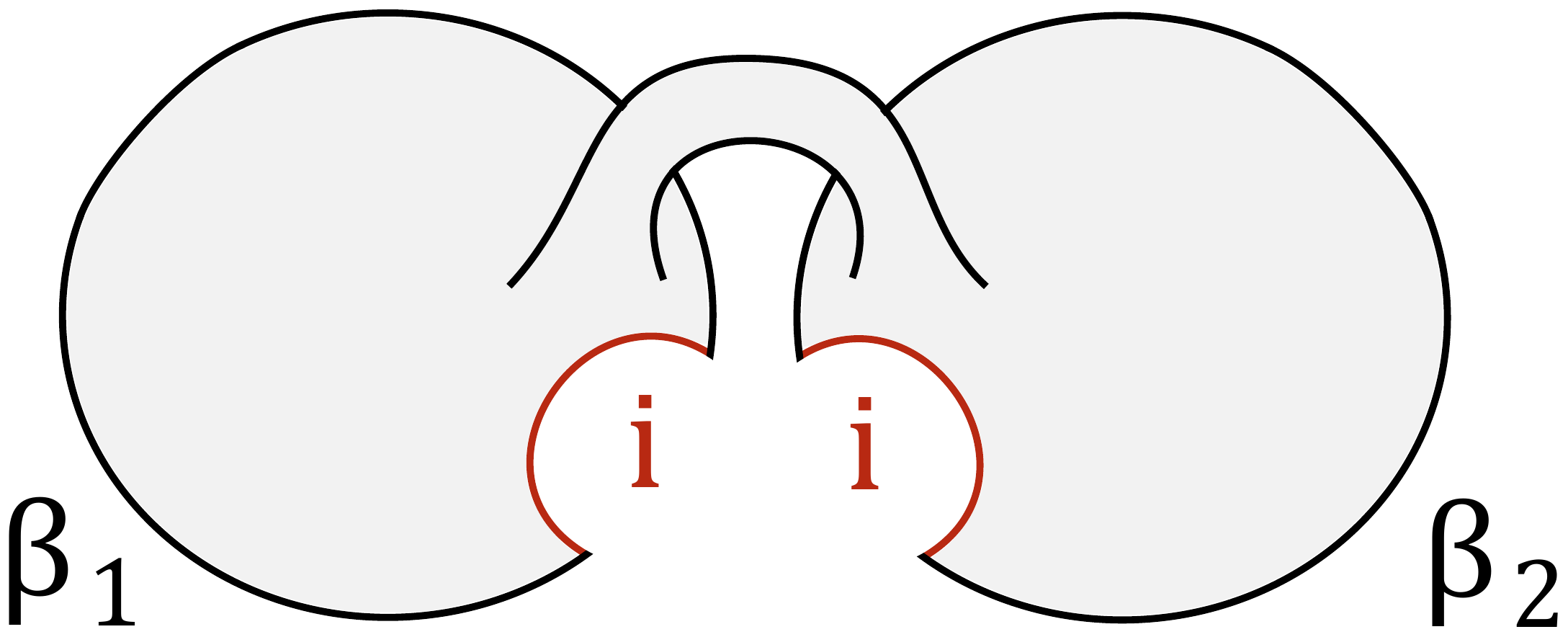}}\quad+\dots\label{136}
\end{align}
The contributions on the second line would factorize if not for geometries connecting the two boundaries. These suffer from the same lack of factorization as does $Z(\beta_1)Z(\beta_2)$ associated with statistical correlations between energy eigenvalues. The connected geometries on the first line arise due to the sum over contractions between bras and kets. These terms embody statistical correlations between eigenstate components on which eigenbranes have no effect.

For this to factorize we must include additional branes of eigenvector boundaries in the definition of the gravitational path integral. For matrix integrals this was achieved by including branes of the type
\begin{equation}
    \psi(P_i)\psi(P_i^*)=\exp\bigg(i\sum_{a=1}^L P_{i a} \bra{i}\ket{a}\bigg)\exp\bigg(i\sum_{b=1}^L P_{i a}^* \bra{a}\ket{i}\bigg)\,.\label{4.59}
\end{equation}
We first ignore the contributions of eigenbranes to the gravitational path integral such that energies can take continuous values. Let us introduce the operators $\bra{i}\ket{E}$ as
\begin{equation}
    \bra{i}e^{-\beta H}\ket{i}=\int_\cas d E\,e^{-\beta E}\,\bra{i}\ket{E}\rho(E)\bra{E}\ket{i}\,.\label{4.59bis}
\end{equation}
We claim that the matrix integral branes \eqref{4.59} generalize to the continuum theory as
\begin{align}
&\psi(P_i)\psi(P_i^*)\nonumber\\&=\exp\bigg(i\int_\cas d M_1\,\sqrt{\rho(M_1)}\,P_i(E)\,\bra{i}\ket{E}\bigg)\exp\bigg(i\int_\cas d M_2\,\sqrt{\rho(M_2)}\,P_i(E)^*\,\bra{E}\ket{i}\bigg)\,. \label{4.60}
\end{align}
The half fixed energy boundaries combine into full boundaries in the naive way. Notice the energy dependent brane coupling for the bras and the kets in the matrix elements. 

Examples clarify how to interpret these branes. As a first example we compute the brane partition function. By following the rules on how to contract bras and kets we find
\begin{align}
    \average{\psi(P_i)\psi(P_i^*)}&=1-\int_\cas d E_1\,\rho(E_1)\,\frac{\rvert P_i(E_1)\rvert^2}{L} +\frac{1}{2}\int_\cas d E_1\,d E_2\,\rho(E_1,E_2)\,\frac{\rvert P_i(E_1)\rvert^2}{L}\,\frac{\rvert P_i(E_2)\rvert^2}{L}+\dots\nonumber\\&=\exp(-\int_\cas d E\,\rho(E)\,\frac{\rvert P_i(E)\rvert^2}{L})\,.\label{142}
\end{align}
This is just a Gaussian functional of the couplings. In terms of geometries we can compare to \eqref{133} and \eqref{4.57}. The terms in the brane partition function are associated to similar geometries where the length of the piece of wiggly boundary is taken to zero. For example
\begin{equation}
    \quad \raisebox{-7mm}{\includegraphics[width=13mm]{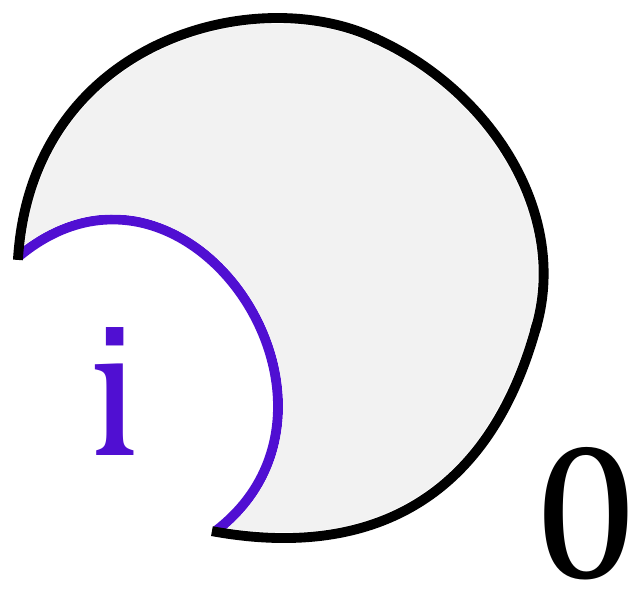}}\quad + \quad \raisebox{-7mm}{\includegraphics[width=13mm]{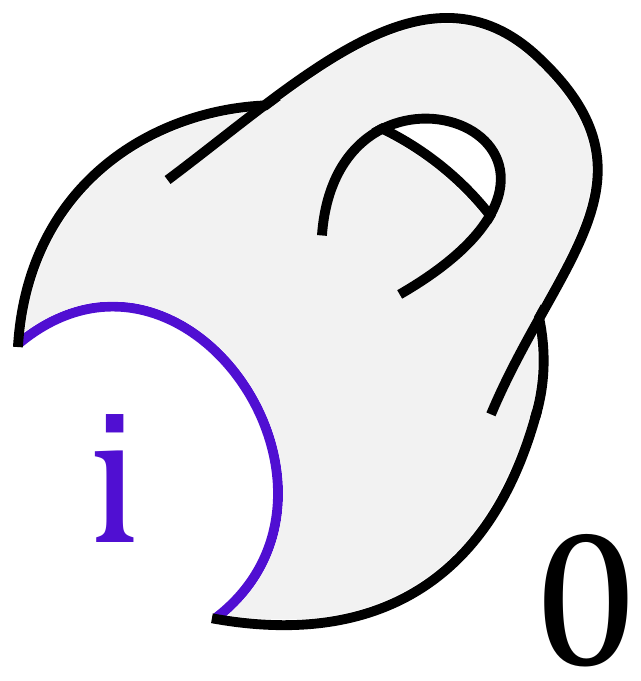}}\quad + \dots =-\int_\cas d E_1\,\rho(E_1)\,\frac{\rvert P_i(E_1)\rvert^2}{L}.
\end{equation}
The brane coupling supplies an energy dependent weight factor for the eigenvector boundary segments (blue) that can be interpreted as setting a boundary wavefunction $\psi(E)=\rvert P_i(E)\rvert^2$. The third term in \eqref{142} is due to two copies of these boundaries
\begin{align}
   \quad &\raisebox{-7mm}{\includegraphics[width=27.5mm]{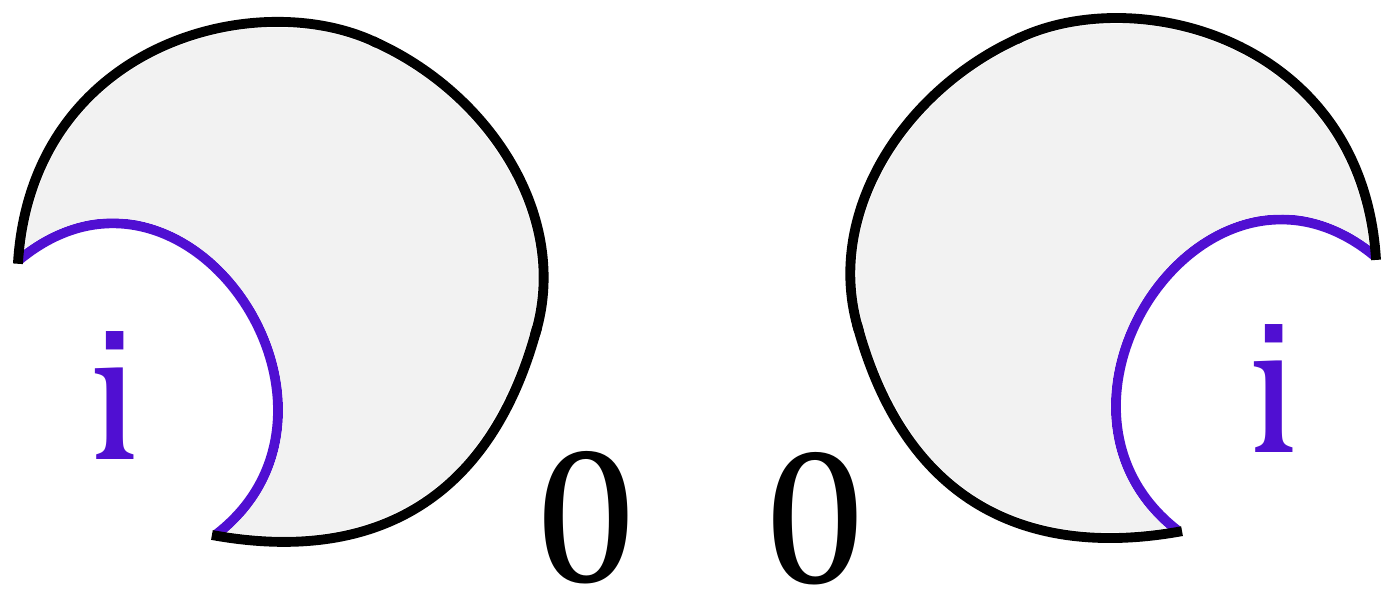}}\quad +\quad \raisebox{-7mm}{\includegraphics[width=27.5mm]{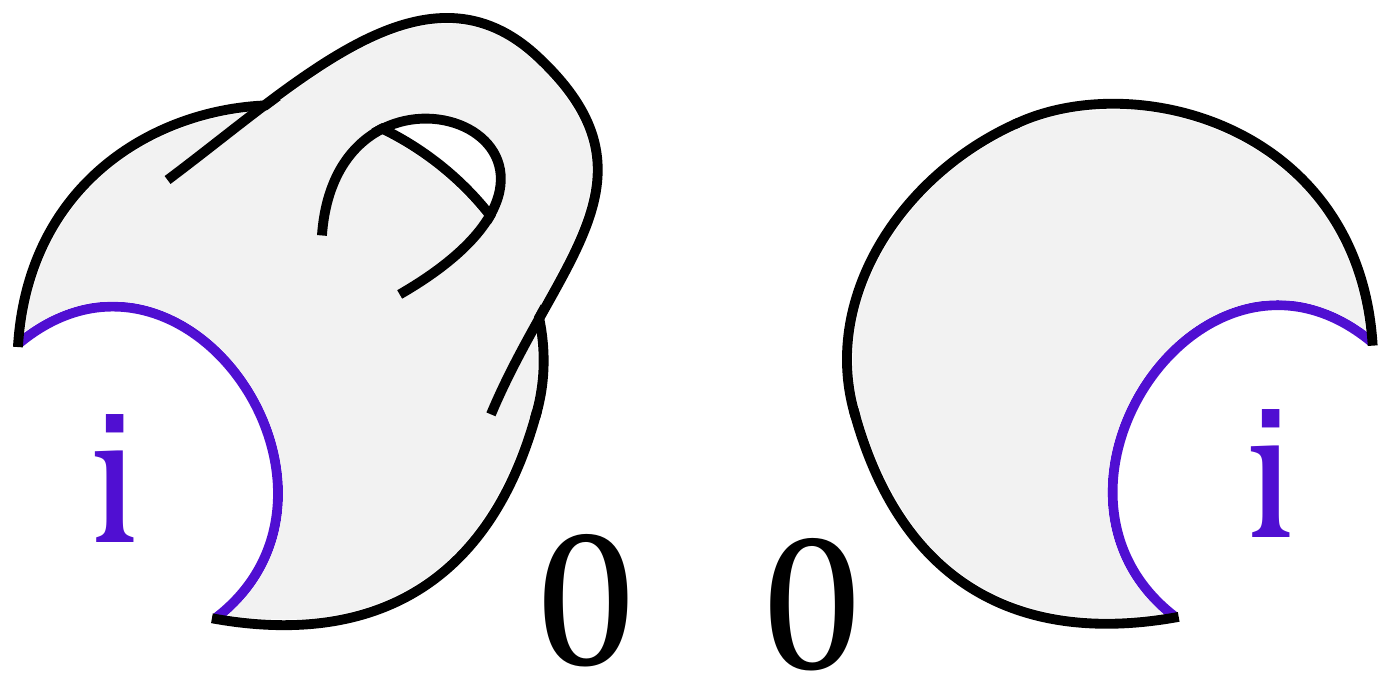}}\quad + \dots \\\nonumber &\qquad\quad+\quad \raisebox{-7mm}{\includegraphics[width=27.5mm]{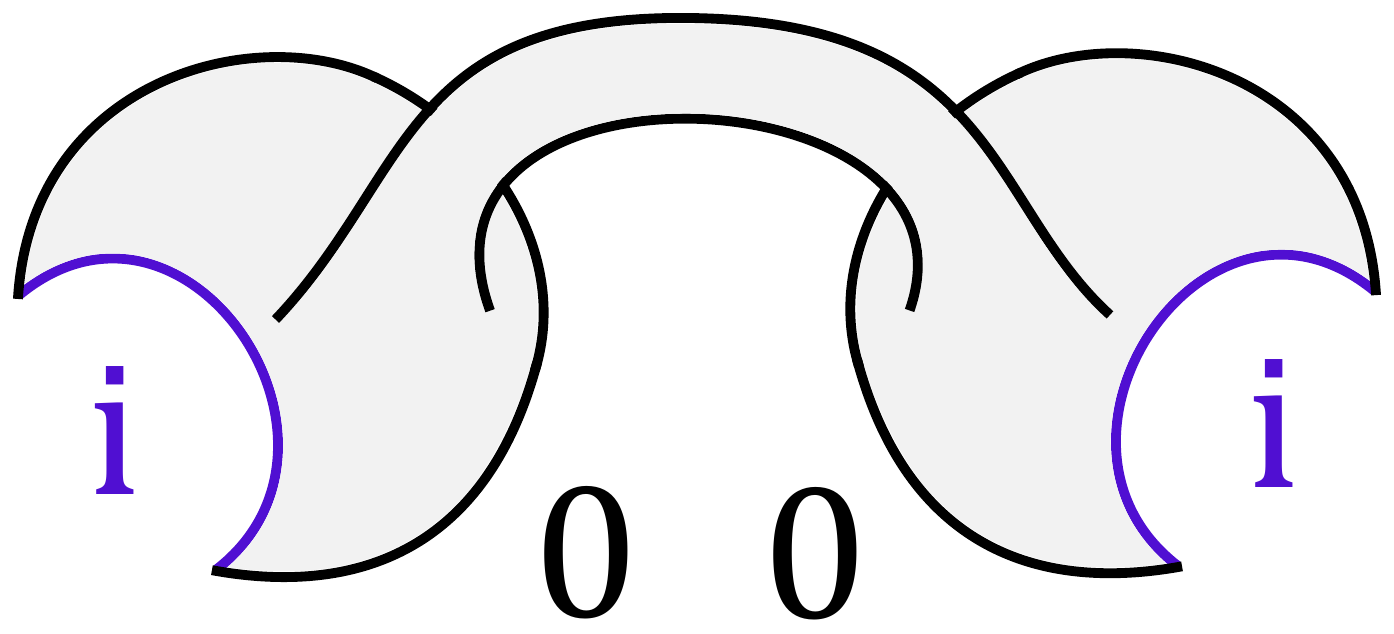}}\quad +\dots\quad = \int_\cas d E_1\,d E_2\,\rho(E_1,E_2)\,\frac{\rvert P_i(E_1)\rvert^2}{L}\,\frac{\rvert P_i(E_2)\rvert^2}{L}\,.
\end{align}
From formula \eqref{4.60} we see that there are no contributions to the brane partition function due to boundaries with multiple eigenstate segments such as those in the first line of \eqref{136}. We can compute the brane partition function to arbitrary precision using formula (167) of \cite{sss2} but we will not need the explicit answer here.

As a second example we consider the matrix element $\bra{i}e^{-\beta H}\ket{i}$ in a gravitational theory that includes these eigenstate branes. There are contributions where additional eigenvector boundaries (blue) are present besides the probe eigenvector boundaries (red)
\begin{align}
    \average{\bra{i}e^{-\beta H}\ket{i}\psi(P_i)\psi(P_i^*)}=&\quad \raisebox{-7mm}{\includegraphics[width=22mm]{fac2.pdf}}\quad + \quad \raisebox{-7mm}{\includegraphics[width=22mm]{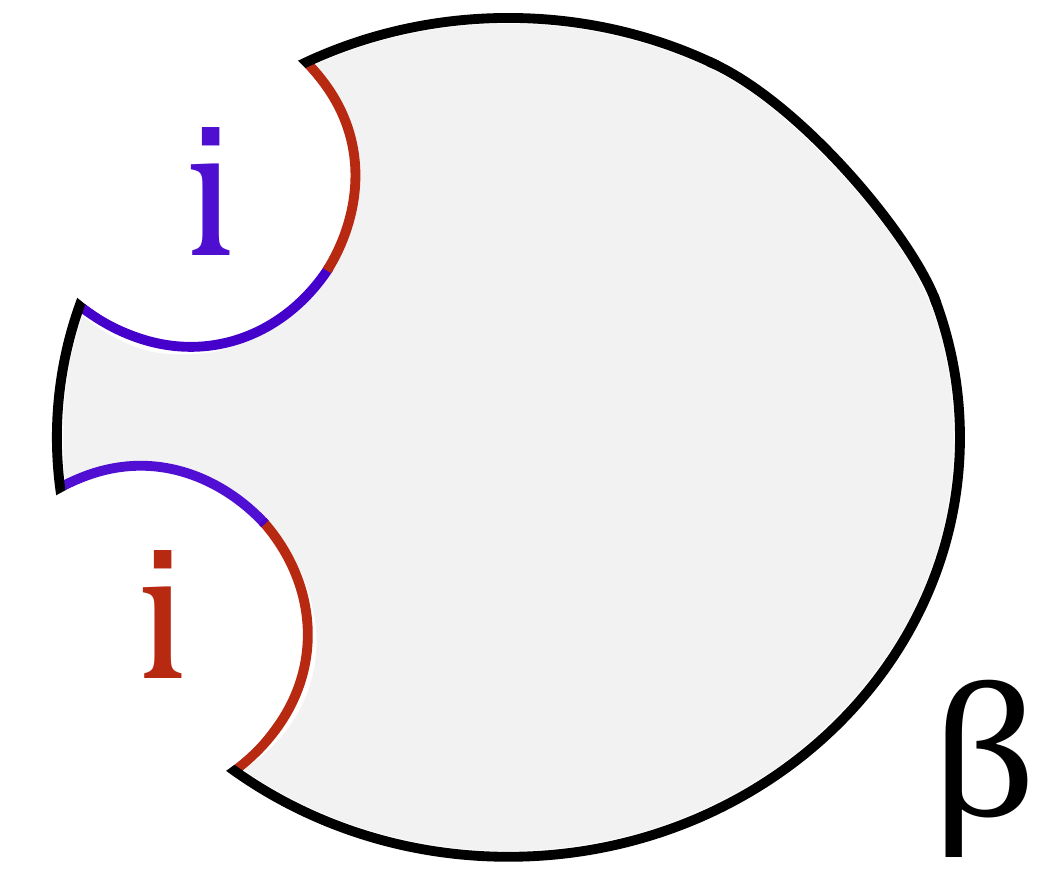}}\quad\nonumber\\&\qquad\qquad+  \quad \raisebox{-7mm}{\includegraphics[width=38mm]{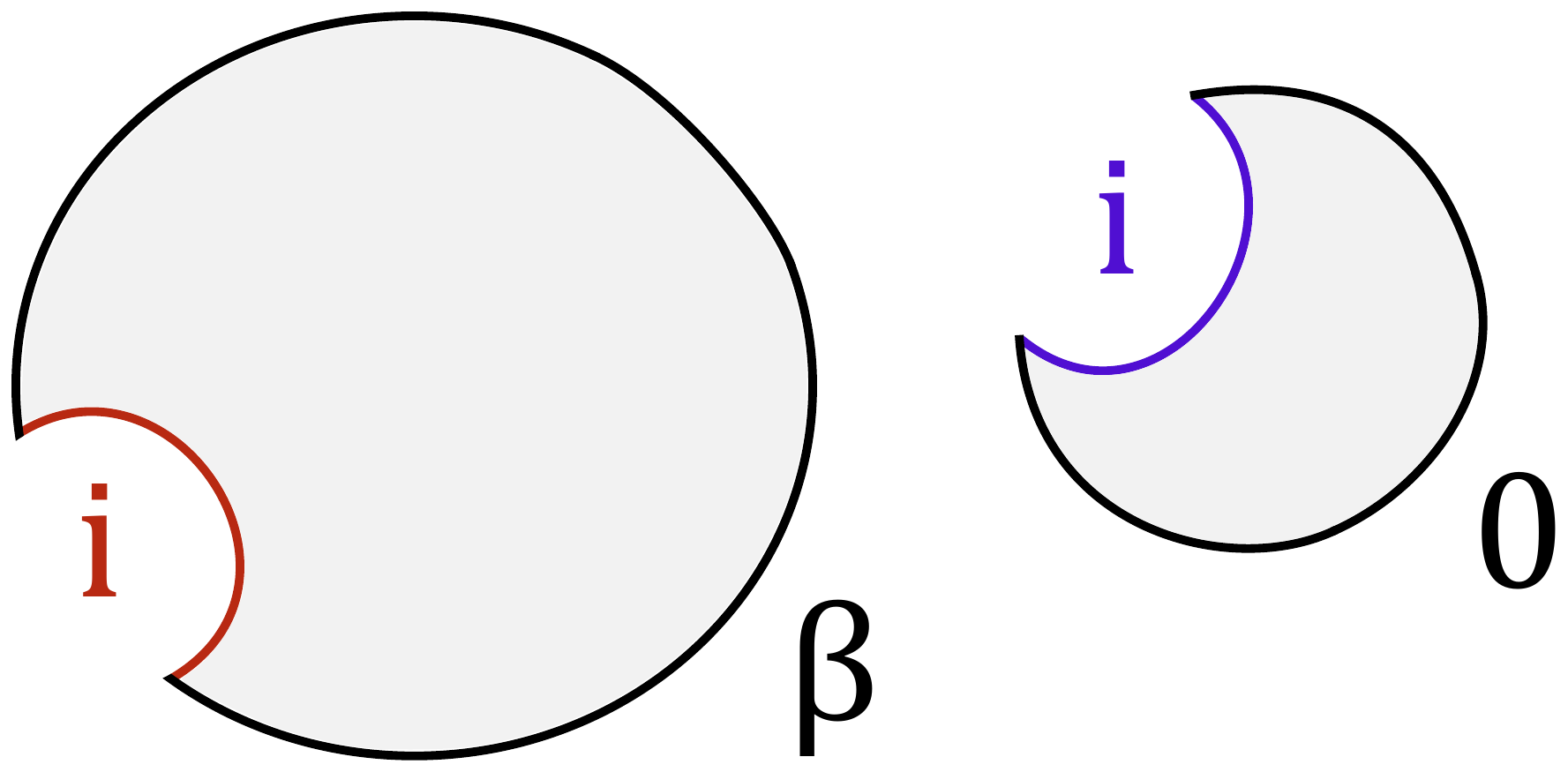}}\quad+\dots \label{137}
\end{align}
Higher genus geometries can connect to each of the disk shaped components. They can for example connect the two disks in the contribution on the second line. Summing over such geometries we find for the second term on the first line
\begin{equation}
    \raisebox{-7mm}{\includegraphics[width=22mm]{correct3.pdf}}\quad +\quad \raisebox{-7mm}{\includegraphics[width=22mm]{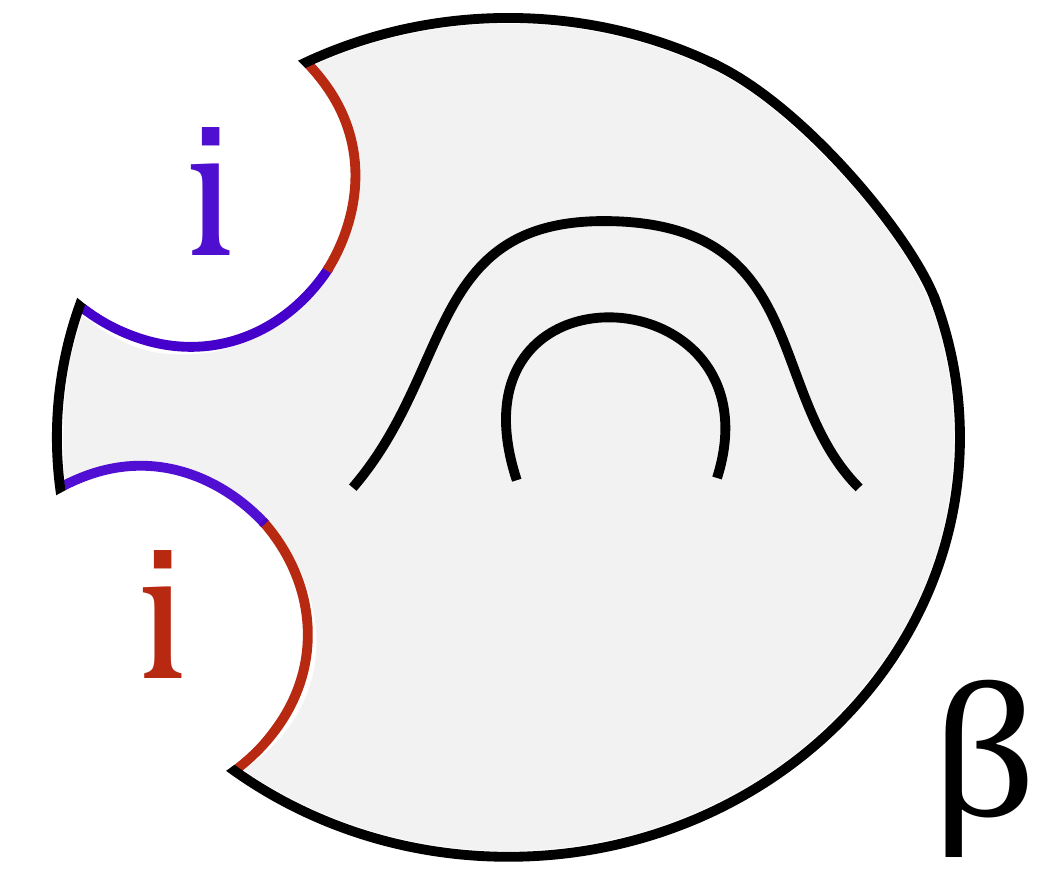}}\quad+\dots  \quad =  -\int_\cas d E \,e^{-\beta E}\,\rho(E)\,\frac{\rvert P_i(E)\rvert^2}{L^2}\,.\label{4.62}
\end{equation}
Repeating this path integral calculation for each of the terms we find
\begin{align}
    \average{\bra{i}e^{-\beta H}\ket{i}\psi(P_i)\psi(P_i^*)}&=\int_\cas d E_1\,e^{-\beta E_1}\,\rho(E_1)\,\frac{1}{L}-\int_\cas d E_1\,e^{-\beta E_1}\,\rho(E_1)\,\frac{\rvert P_i(E_1)\rvert^2}{L^2} \nonumber \\&\quad-\int_\cas d E_1\,d E_2\,e^{-\beta E_1}\,\rho(E_1,E_2)\, \frac{\rvert P_i(E_2)\rvert^2}{L^2}\,+\dots \label{4.66bis}
\end{align}

Notice that the brane partition function is a generating functional of matrix element correlation functions. Taking functional derivatives of either \eqref{4.60} or \eqref{142} we find
\begin{equation}
    \average{\bra{i}e^{-\beta H}\ket{i}\psi(P_i)\psi(P_i^*)}=-\int_\cas d E\, e^{-\beta E}\, \frac{\delta}{\delta P_i(E)}\,\frac{\delta}{\delta P_i(E)^*}\,\average{\psi(P_i)\psi(P_i^*)}\,.\label{4.65}
\end{equation}
When acting with the functional derivatives on \eqref{142} we need to be careful with diagonal terms. It is convenient to write the third term in the brane partition function \eqref{142} as
\begin{align}
    \frac{1}{2}\int_\cas d E_1&\,d E_2\,\rho(E_1,E_2)\,\frac{\rvert P_i(E_1)\rvert^2}{L}\,\frac{\rvert P_i(E_2)\rvert^2}{L}\\&=\frac{1}{2}\int_\cas d E_1\,d E_2\,\rho(E_1,E_2)_\text{smooth}\,\frac{\rvert P_i(E_1)\rvert^2}{L}\,\frac{\rvert P_i(E_2)\rvert^2}{L}+\frac{1}{2}\int_\cas d E_1\,\rho(E_1)\,\frac{\rvert P_i(E_1)\rvert^4}{L^2}\,.\nonumber
\end{align}
The smooth term vanishes when $E_1=E_2$ and the second term is due to the contact term in the two level spectral density \eqref{224}. The functional derivatives act on $\rvert P_i(E)\rvert^4$ in the second term whereas they act on $\rvert P_i(E)\rvert^2$ in the first term. This generates a relative factor of two and a surplus of contact term which becomes the second term on the first line of \eqref{4.66bis}. Using similar care one checks that
\begin{align}
    &\average{\bra{i}e^{-\beta_1 H}\ket{i}\bra{i}e^{-\beta_2 H}\ket{i}\psi(P_i)\psi(P_i^*)}\nonumber\\&\qquad=\int_\cas d E_1\,d E_2\,e^{-\beta_1 E_1}\,e^{-\beta_2 E_2}\,\frac{\delta}{\delta P_i(E_1)}\,\frac{\delta}{\delta P_i(E_1)^*}\,\frac{\delta}{\delta P_i(E_2)}\,\frac{\delta}{\delta P_i(E_2)^*}\,\average{\psi(P_i)\psi(P_i^*)}.\label{4.64}
\end{align}
For example we recognize all terms in formula \eqref{136} by taking these functional derivatives on the third term in the perturbative expansion of the brane partition function \eqref{142}.

To dissect the ensemble we are led to consider Fourier transformed branes
\begin{equation}
    \psi(Z_i)=\dots\int [\dpi P_i]\exp(i \int d M\,\sqrt{\rho(M)}\,Z_i(M)\,P_i(M))\,\psi(P_i)\,.\label{4.66}
\end{equation}
The normalization prefactor $\dots$ is introduced to cancel the one loop determinant. It is not essential but simplifies the presentation. The brane partition function with fixed couplings is just a Gaussian \eqref{142} so we find
\begin{equation}
    \average{\psi(Z_i)\psi(Z_i^*)}=\exp(- L \int d E\, \abs{Z_i(E)}^2)\,.
\end{equation}
This is just a number from the point of view of the gravitational path integral. By applying partial functional integration we see that the Fourier transformed branes are eigenfunctionals of the operators $\bra{i}\ket{E}$ and $\bra{E}\ket{i}$. This is the analogous statement to \eqref{4.40} and so these Fourier transformed branes play the role of alpha states in gravity. 

Normalizing by the brane partition function we find an elegant answer for the expectation value of a thermal matrix element in the presence of Fourier transformed branes
\begin{equation}
    \average{\bra{i}e^{-\beta H}\ket{i}}_Z =\int_\cas d E\,e^{-\beta E}\,\rho(E)\,\rvert Z_i(E)\rvert^2\,.
\end{equation}
A similar answer follows for the product of two such matrix elements
\begin{align}
     &\average{\bra{i}e^{-\beta_1 H}\ket{i}\bra{i}e^{-\beta_2 H}\ket{i}}_Z=\int_\cas d E_1\,d E_2\,e^{-\beta_1 E_1}\,e^{-\beta_2 E_2}\,\rho(E_1,E_2)\,\rvert Z_i(E_1)\rvert^2\,\rvert Z_i(E_2)\rvert^2\,.
\end{align}
We no longer see traces of connected contributions due to statistical correlations of eigenstate components. The sum over contractions is no more. The data $Z$ of the Fourier transformed branes can be read as implementing boundary conditions in the form of boundary wavefunctions $\psi(E)=L\,\rvert Z_i(E)\rvert^2$ for the eigenstate boundaries (orange)
\begin{align}
\displaybreak[0]
    &\nonumber \average{\bra{i}e^{-\beta_1 H}\ket{i}\bra{i}e^{-\beta_2 H}\ket{i}}_Z
    =\\&\qquad\qquad \quad \raisebox{-7mm}{\includegraphics[width=44mm]{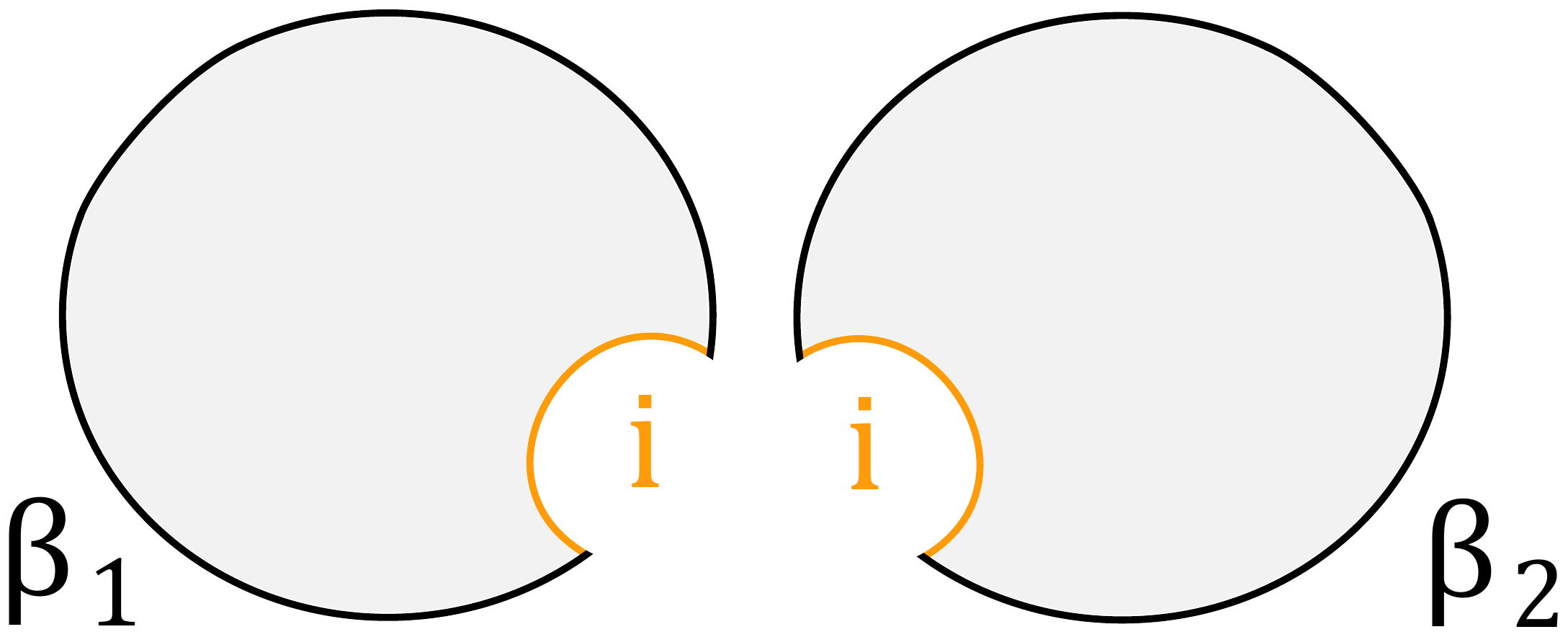}}\quad+\quad \raisebox{-7mm}{\includegraphics[width=44mm]{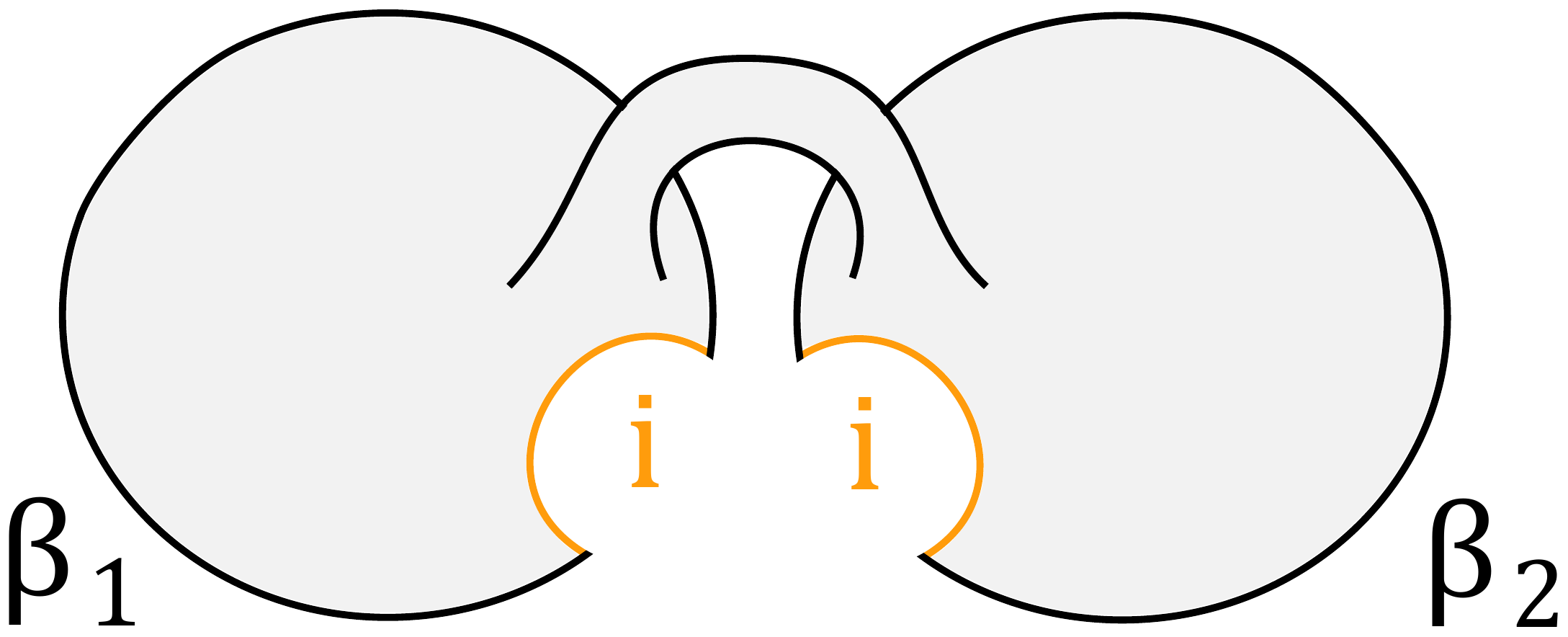}}\quad+\dots
\end{align}
The remaining lack of factorization in this observable is due to eigenvalue correlations in $\rho(E_1,E_2)$. The statistical ensemble of random eigenvalues can be collapsed by furthermore including eigenbranes to which geometries may connect \cite{paper6}. Including eigenbranes as well as eigenvector branes one finds a perfectly factorizing answer
\begin{align}
     \average{\bra{i}e^{-\beta_1 H}\ket{i} \bra{i}e^{-\beta_2 H}\ket{i}}_{Z\,E_1\dots E_L}&=\sum_{a=1}^L e^{-\beta_1 E_a}\,\rvert Z_{i a}\rvert^2 \sum_{b=1}^L e^{-\beta_2 E_b}\,\rvert Z_{i b}\rvert^2\,.
\end{align}
This means we have reproduced the matrix integral answer \eqref{4.42} from a gravity calculation. We have learned that alpha states correspond geometrically to eigenbranes and eigenvector branes. Observables factorize because geometries can connect to these branes
\begin{equation}
    \average{\bra{i}e^{-\beta H}\ket{i}\psi(P_i)\psi(P_i^*)\,\rho(E_1)\dots \rho(E_L)}\supset \quad \raisebox{-7mm}{\includegraphics[width=64mm]{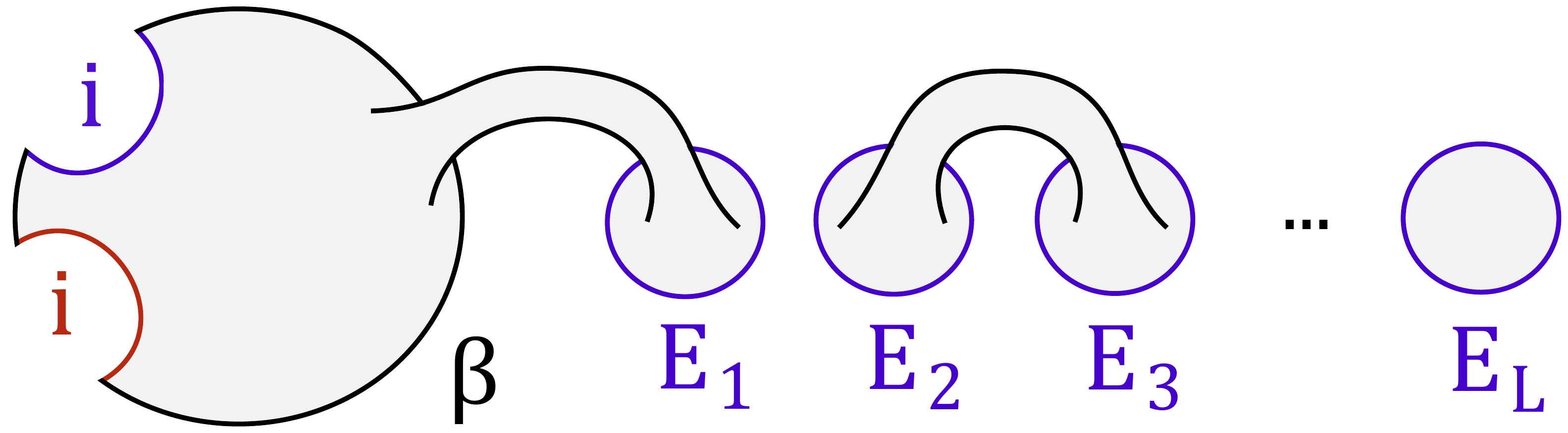}}\quad .\nonumber
\end{equation}

It is straightforward to include eigenvector branes for each of the states so that we can understand factorization of more general products of matrix elements.

In section \ref{sect:2} we learned that gravity is best described by averaging over $e^{S_0}$ dimensional Haar random unitaries for each block of $e^{S_0}$ neighbouring energy levels. We were further led to approximate that Haar random ensemble by considering the matrix elements of the blocks to be independent Gaussian random complex variables with variance $e^{-S_0}$. One could modify the contents on this section to that more realistic ensemble and wield this to study correlation functions. In the end one recovers the erratic oscillations at large distances and late time as expected of a finite entropy system.
%%%%%%%%%%%
% SECTION %
%%%%%%%%%%%
\section{Discussion}\label{sect:disc}
In section \ref{sect:2} we gave a prescription to compute higher genus corrections to Lorentzian bulk observables in JT gravity. We wrote bulk observables in terms of boundary observables by using bulk operator reconstruction and then included higher genus corrections to the latter. There is ample recent evidence \cite{rw1,rw2,sss2,maxfieldmarolf} suggesting it is in general sensible to write observables in terms of boundary conditions for the Euclidean gravitational path integral and to sum over all geometries consistent with these boundary conditions. The boundary correlators discussed in section \ref{sect:2} are an application of this where the boundary conditions define a graph consisting of pieces of fixed length boundary and pieces of geodesic boundary.

The rule book for what to count and what not to count in the gravitational path integral is yet incomplete. The best we can do is make an ansatz and find out if it results in desirable physical output. Recent progress strongly suggests to include summing over all geometries consistent with given boundary conditions in the rule book. The output is very desirable as it reproduces the Page curve of an evaporating black hole \cite{rw1,rw2} and the late time averaged behavior of boundary correlators \cite{sss,sss2,paper5,phil,paper6}. 

In similar spirit we see the results of section \ref{sect:4} as evidence that our prescription for bulk correlators is sensible. We expect and confirmed that large distance bulk correlators in finite entropy gravitational systems do not decay to zero. One further incentive that this is on the right track are the results of \cite{paper201}.

The results of section \ref{sect:5} suggest that gravitational branes play an important role at late times and long distances in describing the microstructure of gravitational systems. However it is not yet clear to what degree this important role of branes and wormholes generalizes to higher dimensional models. We comment on this below. 

The elephant in the room is the lack of a satisfactory Lorentzian interpretation of these Euclidean wormholes and branes even in these simple models of two dimensional quantum gravity \cite{sss2,paper6,maxfieldmarolf}. We cannot claim to know how quantum gravity works without having a dynamical understanding of these instanton (not solutions) corrections to the path integral.
\\~\\
We end this work with various small comments.
\\~\\
\textbf{\emph{Random matrices and factorization}}
\\~\\
We summarize the logic on alpha states in random matrix ensembles which we view as discrete models of quantum gravity. We can isolate one single member of the ensemble by inserting an appropriate set of branes $\psi(H_0)$ in the gravitational path integral. This set is chosen wisely to conspire to a delta in the matrix integral $\psi(H_0)=\delta(H-H_0)$. Observables in the resulting theory of gravity with branes are calculated as
\begin{equation}
    \average{\mo}_{H_0}=\frac{\average{\mo\, \psi(H_0)}}{\average{\psi(H_0)}}=\mo(H_0)\quad,\quad \average{\psi(H_0)}=\frac{1}{\mathcal{Z}}\int d H\,P(H)\,\delta(H-H_0)\,.\label{121}
\end{equation}
This is trivial from the matrix integral perspective but less obvious from the geometrical point of view. For example the branes explain how observables factorize in gravity. There are factorization issues due to nontrivial correlations between eigenvalues and eigenvector components in the ensemble averaged description of JT gravity \cite{sss2}.
\begin{enumerate}
    \item There is a sum over contractions between bras and kets due to statistical correlations of eigenstate components. For example there are two contractions in
    \begin{equation}
        \average{\bra{i}e^{-\beta_1 H}\ket{i} \bra{i}e^{-\beta_2 H}\ket{i}}=\frac{1}{L^2}\average{Z(\beta_1+\beta_2)}+\frac{1}{L^2}\average{Z(\beta_1)Z(\beta_2)}.
    \end{equation}
    The first connected factor is manifestly impeding factorization. This particular issue is resolved by including Fourier transformed eigenvector branes
    \begin{equation}
        \average{\bra{i}e^{-\beta_1 H}\ket{i} \bra{i}e^{-\beta_2 H}\ket{i}}_Z=\int_\cas d E_1\,d E_2\,e^{-\beta_1 E_1}\,e^{-\beta_2 E_2}\,\rho(E_1,E_2)\,\rvert Z_i(E_1)\rvert^2\,\rvert Z_i(E_2)\rvert^2\,.\nonumber
    \end{equation}
    These collapse the ensemble of random eigenvector components.
    \item There are wormhole geometries contributing to $\rho(E_1,E_2)$. These wormholes embody statistical correlations of eigenvalues. We can deal with this by including eigenbranes
     \begin{equation}
        \average{\bra{i}e^{-\beta_1 H}\ket{j}\bra{j}e^{-\beta_2 H}\ket{i}}_{Z\,E_1\dots E_L}=\sum_{a=1}^L e^{-\beta_1 E_a}\,\rvert Z_{i a}\rvert^2\sum_{b=1}^L e^{-\beta E_b}\,\rvert Z_{i b}\rvert^2\,.
    \end{equation}
    The eigenbranes do not necessarily destroy the wormholes but instead they include off diagonal terms to allow for factorization. Nevertheless it remains unclear if we should consider wormholes to be real objects or if they rather are just an effective description of more microscopic physics.
\end{enumerate}

We emphasize that JT gravity plus branes describes just one quantum mechanical system and not an ensemble. These branes are expected to play a role in understanding dynamical unitary in terms of geometry. Most late time observables are not self averaging and require an understanding of the microscopic degrees of freedom of bulk geometry. The Page curve is an exception that remains self averaging at exponentially late times.

So we need to include these exotic branes to capture the microstructure of a dual theory. The arrow of implication is that we take some dual theory and ask for a bulk pure JT gravity description. This bulk description includes branes.
\\~\\
\textbf{\emph{Haar random unitaries}}
\\~\\
Throughout most of this work we have considered matrix elements of $U$ to be independent Gaussian random complex variables. From a physical point of view we would be inclined to instead consider $U$ to be Haar random unitaries. This would be preferable because now $\ket{i}$ and $\ket{a}$ would both be orthogonal bases for any choice of $U$. There are two reasons why in this work we favored independent Gaussian random complex matrix elements over Haar random unitaries.  
\begin{enumerate}
    \item The leading contribution to correlators in the Haar random ensemble are due to Wick contractions which are reproduced by assuming the matrix elements are independent Gaussian random complex variables \cite{wein}.
    \item In the gravitational sum over particle trajectories and topologies in section \ref{sect:2} we found no traces of contributions that could be associated to subleading contributions due to the Weingarten functions. Such contributions are the difference between correlators in the two ensembles under consideration.
\end{enumerate}
This leaves open two options. One option is that JT gravity really is dual to a theory where the matrix elements of $U$ are independent Gaussian random complex variables. Alternatively it might be that there are additional configurations which contribute to correlators that we did not think of. One of the simplest setups in which to probe this would be in the ensemble averaged gravitational calculation of the purity in the context of \cite{rw1}
\begin{equation}
    \sum_{i\,j}\average{\bra{i}e^{-\beta H}\ket{j}\bra{j}e^{-\beta H}\ket{i}}.
\end{equation}
If the matrix elements are independent Gaussian random complex variables then we get
\begin{equation}
    \sum_{i\,j}\average{\bra{i}e^{-\beta_1 H}\ket{j}\bra{j}e^{-\beta_2 H}\ket{i}}=\average{Z(2\beta)}+\frac{1}{L}\average{Z(\beta)Z(\beta)}\,.\label{5.5}
\end{equation}
But if we consider Haar random unitaries then we only recover the first of these terms. Are there additional configurations that contribute to the purity besides those considered in \cite{rw1} which end up cancelling the second term?
\\~\\
\textbf{\emph{Speculation on higher dimensional ensembles}}
\\~\\
JT gravity is a subsector of AdS$_3$ gravity \cite{origins,paper3,cotlerads3} so these wormholes and branes must descend from similar gravitational structures in AdS$_3$ gravity. Therefore we expect that in some form or another the results of this work can be generalized to the AdS$_3$ setup.

As wormholes we could imagine path integrating over all geometries of the form $\Sigma\times S_1$ that end on a torus. Depending on ones attachment to modular invariance as opposed to covariance we can include a sum of modular images or cycles. The JT gravity case suggest that this restricted sum over three manifolds might already have an ensemble averaged dual description. For recent evidence see \cite{cotlerrandom}. 

As branes we could include exponentials of toric boundaries with some boundary conditions. Given that one could calculate the path integral on $\Sigma\times S_1$ it should be straightforward to include such branes and see if they have the desired effect of collapsing the ensemble.
\\~\\
It has been argued that the Hilbert space of baby universes must be trivial in higher dimensions \cite{maxfieldmarolf,vafa}. We comment on three reasons for this assumption. 
\begin{enumerate}
    \item In examples of higher dimensional holography there are no signs of alpha states in the bulk. Note however that it is quite difficult to detect the microstructure provided by the alpha states. The erratic oscillations only show at exponentially large separations. Most likely the checks on holography in higher dimensional examples have not probed these wild parametric regimes as such precision checks would require such a theory to have been solved to the same level of accuracy as our two dimensional examples.
    \item There is at first sight no natural data to average over in the boundary dual. In this sense the recent examples of averaging over Narain moduli space provide new hope \cite{randomcft1,randomcft2}. Who knows what we could average over. The question is if we should. Clearly averaging is not necessarily an improvement. However the hope is that certain suitably averaged boundary theories are dual to a simpler effective theory of quantum gravity in which to get traction. The lower dimensional models suggest this might be possible.
    \item String theory or quantum gravity in general has naively no free parameters that could be associated to alpha states. Note that this is also naively true for JT gravity where the only free parameter is the string coupling. We see no logical contradiction why the bulk dual to a given quantum field theory with fixed data could not be something like string theory plus branes where the coupling constants or multiplicities of the branes could label alpha states.
\end{enumerate}
It might be that the Hilbert space of baby universes in higher dimensions is trivial. However there does not seem to be a strict reason why it absolutely must be trivial. Given that the only available exactly soluble models of gravity have a nontrivial such Hilbert space we find it too soon to discard these examples as exceptions. The more likely option of the two might be that higher dimensional ensembles are real.
%%%%%%%%%%%%%%%%%%%%%%%%%%%%%%%%%%%
\section*{Acknowledgements}
I thank Jordan Cotler, Thomas Mertens and Shunyu Yao for discussions and I acknowledge financial support from FWO Vlaanderen.
%%%%%%%%%%%%%%%%%%%%%%%%
\appendix
%%%%%%%%%%%
% SECTION %
%%%%%%%%%%%

%%%%%%%%%%%
% SECTION %
%%%%%%%%%%%
%%%%%%%%%%%
% SECTION %
%%%%%%%%%%%

\end{document}